\documentclass[12pt]{article}
\usepackage{amsfonts}
\usepackage{float}
\usepackage{amsmath}
\usepackage{epsfig, graphicx}
\usepackage{amssymb}
\usepackage{float}
\usepackage{epsfig, graphicx}
\usepackage{amsmath}

\newtheorem{Defi}{Definition}
\newtheorem{theorem}{Theorem}
\newtheorem{lem}{Lemma}
\newtheorem{conj}{Conjecture}
\newtheorem{prop}{Proposition}
\newtheorem{coro}{Corollary}
\font\bb=msbm10 at 12pt
\font\bbbis=msbm10 at 10pt

\def\rR{\hbox{\bb R}}
\def\nN{\hbox{\bb N}}

\def\nN{\hbox{\bb N}}

\def\sS{\hbox{\bb S}}

\def\QED{\quad\hbox{\hskip 4pt\vrule width 5pt height 6pt depth 1.5pt}}

\font\bb=msbm10 at 12pt
\font\bbbis=msbm10 at 10pt

\font\bb=msbm10 at 12pt
\font\bbbis=msbm10 at 10pt

\def\rR{\hbox{\bb R}}

\def\nN{\hbox{\bb N}}

\def\nN{\hbox{\bb N}}

\def\sS{\hbox{\bb S}}

\def\QED{\quad\hbox{\hskip 4pt\vrule width 5pt height 6pt depth 1.5pt}}

\newcommand{\p}{\partial}

\begin{document}

\title{ Semi-classical limits  of the first eigenfunction and concentration
on the recurrent sets of a dynamical system}
\author{David Holcman \thanks{
Weizmann Institute of Science, department of Mathematics, Rehovot
76100, Israel. D. H. is incumbent of the Madeleine Haas Russel
Career Development Chair. D.H thanks the Ashoff family for their
hospitality at Ukiah, during the elaboration of this work.}
\thanks{D\'epartement de Math\'ematiques et de Biologie, Ecole
Normale Sup\'erieure, 46 rue d'Ulm 75005 Paris, France. D. H. is
supported by the program ``Chaire d'Excellence''.} \and Ivan
Kupka\thanks{Department of Mathematics, University of Toronto, 40
St. George St. Toronto, Ontario, Canada M5S 2E4. I.K. thanks the
Weizmann Institute of Science for its hospitality during the
preparation of this paper.}}
\date{}
\maketitle

\begin{abstract}
We study the semi-classical limits of the first eigenfunction of a
positive second order operator on a compact Riemannian manifold,
when the diffusion constant $\epsilon$ goes to zero. If the drift of
the diffusion is given by a Morse-Smale vector field $b$, the limits
of the eigenfunctions concentrate on the recurrent set of $b$.  A
blow-up analysis enables us to find the main properties of the limit
measures on a recurrent set.

We consider generalized Morse-Smale vector fields, the recurrent set
of which is composed of hyperbolic critical points, limit cycles and
two dimensional torii. Under some compatibility conditions between
the flow of b and the Riemannian metric $g$ along each of these
components, we prove that the support of a limit lies on those
recurrent components of maximal dimension, where the topological
pressure is achieved. Also, the restriction of the limit measure to
 either a cycle or a torus is absolutely continuous with respect to
the unique invariant probability measure of the restriction of b to
the cycle or the torus. When the torii are not charged, the
restriction of the limit measure is absolutely continuous with
respect to the arclength on the cycle and we have determined the
corresponding density. Finally, the support of the limit measures
and the support of the measures selected by the variational
formulation of the topological pressure (TP) are identical.

\end{abstract}

\pagebreak

 \tableofcontents

\pagebreak

\section{ Introduction}

A random particle can escape in finite time from the basin of
attraction of a dynamical system \cite{FW,Schuss}. Indeed a large
fluctuation will ultimately push the stochastic particle outside of
the basin even if the drift points inside. On the other hand, on a
compact manifold,  the analysis of the motion of a stochastic
particle is quite different. In particular the particle can wander
inside the manifold from one basin of attraction to another. We are
interested here in finding the behavior of the random particle for
large time and small noise.

The answer to this problem depends on two main data, at least. First
the geometrical space which is a Riemannian manifold and second a
dynamical system. It is legitimate to think that the recurrent set
of the field plays a crucial role since they are visited repeatedly
by the deterministic particle. One can ask, for example, if certain
recurrent sets are visited more often than others. To address this
question, one would try to obtain an explicit solution of the
Fokker-Planck Equation (FPE) and derive large time and small noise
asymptotics. We shall study the FPE and in particular the ground
state (first eigenfunction) when the noise is small.
\subsubsection{Formulation of the first eigenfunction problem}
On compact Riemannian manifolds, the behavior of the first
eigenvalue of singular elliptic perturbations of first order
operators has been the topic of many studies (among others
\cite{Day,FW}). As the viscosity parameter goes to
zero, the first eigenvalue tends to a quantity called the
topological pressure. We shall discussed briefly the concept
developed in \cite{Kifer90} and link it with the limits of the first
eigenfunctions.

In a previous study \cite{HK1}, when the first order term is a
Morse-Smale field and a killing potential $c$ is added to the
dynamic, some properties of the limit measures of the first
eigenfunction were obtained. It was shown that the limit of a
normalized eigenfunction is concentrated on a subset of the
recurrent set (see \cite{HK1}). There, we have left open the
characterizations of the limit measures. The purpose of the present
work is to complete and extend this previous study \cite{HK1} and to
describe in some cases the properties of these measures. In order to
make our exposition self contained, we shall recall briefly the
links between the dynamics of a random particle and its survival
probability distribution.

If $\ X_{\epsilon }(t)$ denotes the position at time t of a random particle
on a compact Riemannian manifold $(V$,$g)$ its motion is described by the
following stochastic equation:
\begin{equation*}
dX_{\epsilon }(t)=b(X_{\epsilon }(t))dt+\sigma (X_{\epsilon }(t))d\mathbf{w}
\end{equation*}%
where $\mathbf{w}$ is the N-dimensional classical Brownian process and $%
\sigma $ is a vector bundle homomorphism from the trivial vector bundle $%
V\times \mathbb{R}^{N}$ into the tangent bundle $T_{V}$ such that $\sigma
\sigma ^{\ast }=\widetilde{g}^{-1}$ where $\sigma ^{\ast }:T^{\ast }{V}%
\rightarrow V\times \hbox{\bb R}^{N}$ is the homomorphism dual to
$\sigma $ and $\widetilde{g}:T$ $V\longrightarrow T^{\ast }$ $V$ is
the canonical fiber bundle isomorphism induced by the metric $g$. A
potential $c$ representing a killing term is added. Physical
interpretations of $c$ and computation of the survival probability
can be found in \cite{Schuss,HMS}. The survival probability
distribution of a particle $X_{\epsilon }(t)$ located in the region
$y+dy$ at time t, conditioned by the initial condition $X_{\epsilon
}(0)=x$ (see \cite{Schuss}) is given by $p_{\epsilon
}(t,x,y)dy=Pr(X_{\epsilon }(t)\in y+dy|X_{\epsilon }(0)=x)$ and
satisfies the backward Fokker-Planck equation (FPE):
\begin{eqnarray*}
\frac{\partial p_{\epsilon }(t,x,y)}{\partial t} &=&\varepsilon \Delta
_{g}p_{\epsilon }(t,x,y)+<b(x),\nabla p_{\epsilon }(t,x,y)>+c(x)p_{\epsilon
}(t,x,y) \\
p_{\epsilon }(0,.,y) &=&\delta _{y}.
\end{eqnarray*}%
We consider the Fokker Planck operator defined by
\begin{equation}
L_{\epsilon }=\epsilon \Delta _{g}+\theta (b)+c,  \label{fpo}
\end{equation}%
where $\Delta _{g}$ is the Laplace-Beltrami operator and $\theta (b)$ is the
Lie derivative in the direction $b$. The function $c$ is chosen such that $%
L_{\epsilon }$ is positive. $L_{\epsilon }$ cannot be conjugated to a
self-adjoint operator by scalar multiplication. Since the operator is
positive and compact, by the Krein-Rutman theorem the first eigenvalue $%
\lambda _{\epsilon }$ is real positive and associated to a unique positive
normalized eigenfunction $u_{\epsilon }$ (see for example
\cite{Pinsky}).

The probability density function $p_{\epsilon }$ can be expanded using the
eigenfunction of $L_{\epsilon }$
\begin{equation*}
p_{\epsilon }(t,x,y)=e^{-\lambda _{\epsilon }t}u_{\epsilon }(x)u_{\epsilon
}^{\ast }(y)+R_{\epsilon }(t,x,y),
\end{equation*}%
where the second term $R_{\epsilon }(t,x,y)$ decreases exponentially
faster than the first as t goes to infinity, and $u_\epsilon^*$ is
the first
positive formalized eigenfunction associated to the adjoint operator $%
L_{\epsilon }^{\ast }$. The first normalized positive eigenfunction $%
u_{\epsilon }>0$, is a solution of
\begin{eqnarray}  \label{edpfdt}
\epsilon \Delta _{g}u_{\epsilon }+(b,\nabla u_{\epsilon })+cu_{\epsilon }
&=&\lambda _{\epsilon }u_{\epsilon } \\
\int_{V_{n}}u_{\epsilon }^{2}dV_{g} &=&1.  \notag
\end{eqnarray}
Equation (\ref{edpfdt}) has a long story in the literature, starting
from the work in the 70's in $\hbox{\bb R}^{n}$ to more recent
endeavors, to find a quantum analog of the weak KAM theory (see for
example \cite{Evans1}). In that case, when the field b is
Hamiltonian, the measure $u_{\epsilon }u_{\epsilon }^{\ast }dV_{g} $
as $\epsilon$ goes to zero concentrates on a specific set.

\subsubsection{Limit of the first eigenvalue and the topological pressure}

Under hyperbolicity assumptions on the field $b$, the limit of the
sequence $\lambda _{\epsilon }$ has been determined in
\cite{DoV,Kifer90}. It was based on the fact that $\lambda _{\epsilon }$
is given by the following variational formula (see \cite{DoV}),
\begin{equation*}
\lambda _{\epsilon }=\sup_{\mu \in P(V)}(\int_{V}cd\mu +\inf_{u>0}\left[
-\int_{V}\frac{L_{\epsilon }(u)}{u}d\mu \right] ),
\end{equation*}%
where $\mu $ belongs to $P(V)$, the space of probability measure on $V$. It
is proved in \cite{Kifer90} that when the recurrent set K of the field $b$
is a finite union of isolated components K$_{1}$,K$_{2}$,...,K$_{N}$ which
are hyperbolic invariant sets, the limit of the first eigenvalue $\lambda
_{\epsilon }$ as $\epsilon $ goes to zero is equal to topological pressure,
denoted by $TP$ and defined as follows:
\begin{eqnarray*}
TP(K_{j}) &=&\sup_{\mu }\{h_{\mu }+\int_{V}(c-\frac{d\det D\phi _{t}^{u}}{dt}%
|_{t=0})d\mu \mid \mu \in P(V),\\& &\hbox{ support }\mu \subset K_{j}\,,\ \mu \,\ \phi
_{t}-invariant\}.
\end{eqnarray*}
For any union U of $K_j$'s,
\begin{eqnarray*}
TP(U) &=&\underset{K_j \in U}{\sup }TP(K_{j}),\hbox{ where }
\end{eqnarray*}%
$h_{\mu }$ is the metric entropy (see \cite{Robin}), $\phi _{t}$ is
the flow of the vector field $b$ and $D\phi _{t}^{u}$ is the tangent
mapping of $\phi $ restricted to the unstable bundle
$T_{K_{j}}^{u}V$ of $K_{j}$ and $\det D\phi _{t}^{u}$ is the
determinant of $\ D\phi _{t}^{u}$ with respect to the metric $g.$ A
$\phi _{t}-invariant$ measure $\mu _{j}\in P(V)$ with support in
$K_{j}$ such that
\begin{equation*}
TP(K_{j})=h_{\mu _{j}}+\int_{V}(c-\frac{d\det D\phi _{t}^{u}}{dt}|_{t=0})d\mu _{j}
\end{equation*}
is called an equilibrium state associated to $K_{j}$ (see \cite%
{Kifer90},\cite{DoV} for existence). Up to a reordering of  $%
K_{j}$ we can assume that $TP(K_{j})=TP(\cup_i K_{i})$ if 1$\leq j\leq l$, and $%
TP(K_{j})<TP(\cup_i K_{i})$ \ if $j>l.$

A measure $\mu =\sum_{j=1}^{l}p_{j}\mu _{K_{j}}$ where $\mu
_{K_{j}}$ is an equilibrium state associated with the set $K_{j}$ ,
$p_{j}\geq 0,\sum_{j=1}^{l}p_{j}=1,$ is called an equilibrium
measure (see Theorem 3.4 in \cite{Kifer80} p.20). Unfortunately,
neither these results nor the methods used in
\cite{Kifer90}, Donsker-Varadhan \cite{DoV}- \cite{DoV2}-\cite{DoV3}
and many others do give us any information about the first
eigenfunction $u_{\epsilon }$ as $\epsilon $ goes to zero.


\subsubsection{Limit of the first eigenfunction} \label{definitionintro}

Let us introduce the following notations,
\begin{eqnarray}
b=\Omega +\nabla \mathcal{L}\, &,&v_{\epsilon }=e^{-\frac{\mathcal{L}}{%
2\epsilon }}u_{\epsilon },  \label{gauge} \\
c_{\epsilon } &=&\epsilon (c+\frac{\Delta _{g}\mathcal{L}}{2})+\Psi _{%
\mathcal{L}},  \notag \\
\Psi _{\mathcal{L}} &=&\frac{1}{4}(||{\nabla \mathcal{L}||}^{2}+2(\nabla
\mathcal{L},\Omega )),  \notag
\end{eqnarray}%
where the function $\mathcal{L}$ is a special Lyapunov function (see
appendix I of \cite{HK1}, for a construction of $\mathcal{L}$
associated with a Morse-Smale vector field on a compact Riemannian
manifold). The introduction of the function  $\mathcal{L}$ is
natural because in the neighborhood of any recurrent sets of the
field b, it coincides to the third order with the solution of the
associated Hamilton-Jacobi equation $(\Psi _{\mathcal{L}} =0)$. This
solution has already been introduced in the classical text books by
Courant-Hilbert to find approximations to the solutions of the Wave
equation, and later on by Schuss \cite{Schuss}, Kamin
\cite{Kamin,Kamin2,Kamin3}, Friedman \cite{Fried1}, and many other, in the
context of the diffusion equation. The exponential of the Lyapunov
function plays the role of a convergence factor, which is a well
established technic in analysis especially in the study of divergent
series and integrals.

Equation (\ref{edpfdt}) is transformed into
\begin{eqnarray}
L_{\epsilon }(v_{\epsilon })=\epsilon ^{2}\Delta _{g}v_{\epsilon
}+\epsilon (\Omega ,\nabla v_{\epsilon })+c_{\epsilon }v_{\epsilon
}=\epsilon \lambda _{\epsilon }v_{\epsilon },  \label{edpfdt1}
\end{eqnarray}%
and we impose the normalization condition
\begin{eqnarray*}
\int_{V_{n}}v_{\epsilon }^{2}dV_{g}=1.
\end{eqnarray*}
In equation (\ref{edpfdt1}), $\epsilon \lambda _{\epsilon }$ is the
first eigenvalue of the operator
\begin{eqnarray*}
{L^{\prime }}_{\epsilon }=\epsilon ^{2}\Delta _{g}+\epsilon \theta
(\Omega )+c_{\epsilon }
\end{eqnarray*}%
and $v_{\epsilon }$ is the associated positive \cite{Pinsky}
eigenfunction. It has been proved in \cite{HK1} that the weak limits
of the normalized measures $v_{\epsilon }^{2}$ $dV_{g}$ are
supported by the limit sets of the field $b$. In order to obtain a
precise description of these limits additional assumptions will be
made on the behavior of the vector field $b$
near the hyperbolic recurrent sets. Usually the first eigenfunction $u_{\epsilon }$ will not have any limits as $%
\epsilon $ goes to zero in any of the classical "strong" topologies. On the
other hand we will characterize the limits of the measure
$v^2_{\varepsilon }dV_g$ if we assume that $b$ is a generalized
Morse-smale field, the only recurrent components of which are
hyperbolic points, cycles, two dimensional torii and the hyperbolic
structure satisfies some compatibility conditions with the metric.
The main result of this work can be summarized as follow:

\bigskip

{\bf \emph{\textquotedblleft\   On a Riemannian manifold, for any
choice of a special Lyapunov function $\mathcal{L}$, vanishing at
order 2 on the recurrent sets of the field, the limits as
}$\varepsilon $ \emph{tends to 0, of the normalized measures
$e^{-\frac{\mathcal{L}}{2\epsilon }}u_{\epsilon }^{2}dV_{g},$ are
concentrated on the components of the recurrent sets which are of
maximal dimension and where the topological pressure is achieved."}}


\subsection{Notations and Assumptions}\label{notat}

We shall make the following assumptions on the field $b$:

\textrm{I)} The recurrent set is a finite union of stationary
points, limit cycles and two dimensional torii.

\textrm{II)} The stationary points are hyperbolic and for each such $P$ the
stable and unstable manifolds belonging to $P$ are orthogonal at $P$ with
respect to the metric $g.$

\textrm{III)} Any limit cycle $S$ has a tubular neighborhood $T^{S}$
equipped with a covering map $\Phi :\mathbb{R}^{m-1}\times \mathbb{R}%
\longrightarrow T^{S}$ having the following properties:

(a) for all $(x^{\prime },\theta )\in \mathbb{R}^{m-1}\times \mathbb{R},$
\begin{eqnarray*}\Phi
^{-1}\circ \Phi (x^{\prime },\theta )=\{(x^{\prime },\theta +nT_{S})|n\in
\mathbb{Z}\}
\end{eqnarray*}
where $T_{S}$ is the minimal period of the the cycle $S$.

(b) at any point ($0,\theta )\in \mathbb{R}^{m-1}\times \mathbb{R}$,\
\begin{eqnarray*}
\left( \Phi \right) ^{\ast }g_{(0,\theta )}=\sum_{n=1}^{m-1}dx_{n}^{2} +g^{m
m }(\theta )d\theta ^{2},
\end{eqnarray*}
 where $\theta=x^m$.

(c) at any point ($0$,$\theta )\in \mathbb{R}^{m-1}\times \mathbb{R}$ ,\
\begin{equation*}
\left( \Phi \right) _{\ast }b=\frac{\partial }{\partial \theta }%
+\sum_{i,j=1}^{m-1}B_{ij}x_{j}\frac{\partial }{\partial x_{i}},
\end{equation*}%
up to term of order two in $x^{\prime }=(x_{1},..x_{m-1}),$ canonical
coordinates on $\mathbb{R}^{m-1}.$

(d) the $\left( m-1\right) \times (m-1)$ matrix $B=$\{$B_{ij}|1\leq
i,j\leq m-1$\} is hyperbolic and its stable and unstable spaces are
orthogonal with
respect to the Euclidean metric $\sum_{n=1}^{m-1}x_{n}^{2}$ on $\mathbb{R}%
^{m-1}$. Denote by $B^{s}$(resp. $B^{u}$) the restriction of B to
the stable (resp unstable) space.

\textrm{IV)} Any 2-dimensional torus $R$ has a tubular neighborhood
$T^{R}$ equipped with a diffeomorphism $\Phi :\mathbb{R}^{m-2}\times \mathbb{R}%
^{2}\longrightarrow T^{R}$ having the following properties:

(a) at any point ($0$,$\theta )\in $ $\mathbb{R}^{m-2}\times
\mathbb{R}^{2}$, $\theta =(\theta _{1},\theta _{2})$ $\theta
_{1},\theta _{2}$ cyclic coordinates, with period one.

\begin{eqnarray*}
\left( \Phi ^{-1}\right) ^{\ast }g_{(0,\theta
)}=\sum_{n=1}^{m-2}dx_{n}^{2}+a(\theta )d\theta _{1}^{2}+2b(\theta )d\theta
_{1}d\theta _{2}+c(\theta )d\theta _{2}^{2}.
\end{eqnarray*}

(b)
at any point ($0$,$\theta )\in \mathbb{R}^{m-2}\times \mathbb{R}^{2}$ ,%
\begin{equation*}
\ \left( \Phi \right) _{\ast }\left( b\right) (0,\theta )=\ \ k_{1}\frac{%
\partial }{\partial \theta ^{1}}+k_{2} \frac{\partial }{\partial \theta ^{2}}%
+\sum_{i,j=1}^{m-2}B_{ij}x_{j}\frac{\partial }{\partial x_{i}},
\end{equation*}%
where:

(i) the $\left( m-2\right) \times (m-2)$ matrix $B=$\{$B_{ij}|1\leq i,j\leq
m-2$\} is hyperbolic and its stable and instable spaces are orthogonal with
respect to the Euclidean metric $\sum_{n=1}^{m-2}x_{n}^{2}$ on $\mathbb{R}%
^{m-2}$,

(ii) $k_1,k_2 \in \hbox{\bb R}$ and $\frac{k_1}{k_2} \in \hbox{\bb R}-%
\hbox{\bb Q}$.
A torus with such a flow will be called an irrational torus.

(iii) there exist constants $C>0,\alpha>0$ such that for all $m_{1},m_{2}\in %
\hbox{\bb N}$,
\begin{equation}
|m_{1}k_{1}+m_{2}k_{2}|>C(m_{1}^{2}+m_{2}^{2})^{\alpha }  \label{sd}
\end{equation}%
This is usually called the small divisor condition. We call
assumption (i) in all the previous cases the {\bf \em  Orthogonality
Assumption}.

\subsubsection*{Definition and construction of a special Lyapunov
function} Given a Riemannian manifold (V, g), of dimension m and a
vector field b on V, having as recurrent components stationary
points, limit cycles and two-dimensional torii satisfying
hyperbolicity conditions and compatibility conditions with g, stated
as in paragraph \ref{notat}, then there exists a Lyapunov function
$\mathcal{L}$ satisfying the following properties
\begin{enumerate}
\item Outside the recurrent sets, $d\mathcal{L}(b)<0$.
\item In the neighborhood of any recurrent elements S (points, cycles and
torii), in the coordinate system defined in \ref{notat}, the first
nonzero term of the Taylor expansion of $\mathcal{L}$ is a quadratic
form $\mathcal{L}(x) = <A(S)x,x>_{\mathbb{R}^{m-\sigma}}
+O(||x||^3)$. $\sigma \in \{0,1,2\}$, is the dimension of the
recurrent components.
\item In the splitting of $\rR^{m-\sigma}=\rR^{m_s} \bigoplus \rR^{m_u}$,
$\sigma \in \{0,1,2\}$, the matrix $A(S)$, splits into $A_s(S)$ and
$A_u(S)$. They have the
 form:
in the system $(U,x_{1},..,x_{m})$.
\begin{eqnarray*}
A^{-1}_{s}(S)&=&-\int_{0}^{+\infty }e^{tB_{s}}\Pi
_{s}e^{tB_{s}^{\ast }}dt,\\
A^{-1}_{u}(S)&=&\int_{0}^{+\infty }e^{-tB_{u}}\Pi
_{u}e^{-tB_{u}^{\ast }}dt,
\end{eqnarray*}
where $\Pi _{s},\Pi _{u}$ \ are positive definite. For later
purposes, we take  $\Pi_u>>2Id_{m_u}$ and $\Pi_s>>2Id_{m_s}$. The
symbol $>>$ denote the canonical order on the set of symmetric
matrices.
\item
\begin{eqnarray*}
\Psi(L)&=&\frac{1}{4}({\ ||\nabla\mathcal{L}||}%
_{g}^{2}+2<\nabla\mathcal{L}, \Omega>_{g})=\\
&&\frac{1}{4}({\ -||\nabla\mathcal{L}||}%
_{g}^{2}+2<\nabla\mathcal{L}, b>_{g}) \geq0,
\end{eqnarray*}
where equality occurs only on the recurrent sets. Moreover,
\begin{eqnarray*}
\Psi(L)=\frac{1}{2}<B^*A(S)+ A(S)B
-2A(S)^2x,x>_{\mathbb{R}^{m-\sigma}} +O(||x||^3),
\end{eqnarray*}
where the conditions on $\Pi _{s},\Pi _{u}$ given in 3 implies that
\begin{eqnarray*}
B^*A(S)+ A(S)B -2A(S)^2>>0.
\end{eqnarray*}
\end{enumerate}

\noindent \textbf{Remark.}
From the conditions above, it may be surprising that all we need for our study is the knowledge
of g and b up to the first order along the recurrent set.

\subsubsection{General Notations}

\begin{align*}
<,>_{g}& :=\text{ scalar product associated to g} \\
d_{g}& :V\times
V\rightarrow \mathbb{R}_{+}:=\text{ distance associated to }g
\\
\exp _{x}& :T_{x}V\longmapsto V\text{:=exponential map of \ }g\text{
with
pole }x \\
dV_{g}& :=\text{volume element associated to }g \\
L^{2}(V)& \text{:=}L^{2}-\text{space associated to } dV_g \\
H_{1}([0,t];V)& \text{:= the space of all $H_{1}$ curve from [0,t] to V} \\
\Delta _{g}& :=\text{ negative Laplacian associated to the metric }g \\
b& :=\text{ vector field on }V \\
\theta (b)& :=\text{ Lie derivation operator associated to }b \\
\nabla & :=\text{ gradient associated to }g \\
P(V)& :=\text{ space of all probability measures on }V \\
C^{\infty }-\text{topology}& \text{:=uniform convergence of all the
derivatives on compact sets}\\
TP& :=\text{Topological Pressure} \\
g& :=\sum_{ij=1}^{m}g_{ij}dx_{i}dx_{j} \\
\Delta _{g}& :=-\frac{1}{\sqrt{\det (g)}}\sum_{ij=1}^{m}\frac{\partial }{%
\partial x_{i}}\sqrt{\det (g)}g^{ij}\frac{\partial }{\partial x_{j}} \\
g^{ij}& :=\text{ inverse matrix of }g_{ij} \\
\Delta _{E}^{m-1}& :=-\sum_{i=1}^{m-1}\frac{\partial ^{2}}{\partial x_{i}^{2}} \\
\det (g)& :=\text{det (}g_{ij}) \\
\Gamma _{ij}^{k}& :=\text{Christoffel symbols of }g_{ij} \\
\Gamma _{ij}^{k}& =\frac{1}{2}g^{kl}\left( \frac{\partial g_{il}}{\partial
x_{j}}+\frac{\partial g_{jl}}{\partial x_{i}}-\frac{\partial g_{ij}}{%
\partial x_{l}}\right) \\
R_{ijk\cdot }^{\cdot \cdot \cdot l}& :=\frac{\partial \Gamma _{jk}^{l}}{%
\partial x_{i}}-\frac{\partial \Gamma _{ik}^{l}}{\partial x_{j}}%
+\sum_{n=1}^{m}\left[ \Gamma _{in}^{l}\Gamma _{jk}^{n}-\Gamma
_{jn}^{l}\Gamma _{ik}^{n}\right] \\
R_{ijkl}& :=\sum_{n=1}^{m}g_{\ln }R_{ijk\cdot }^{\cdot \cdot \cdot n} \\
Ric_{kl}& :=\sum_{j=1}^{m}R_{jklj} \\
R& :=\sum_{j=1}^{m}Ric_{jj}=\sum_{i,j=1}^{m}R_{ijji} \\
\theta (b)& :=\sum_{i=1}^{m}b^{i}\frac{\partial }{\partial x^{i}} \\
dx^{i}(\nabla f)& :=\sum_{i=1}^{m}g^{ij}\frac{\partial f}{\partial x^{j}}%
\text{,}1\leq i\leq m
\end{align*}
\noindent Along a limit cycle $S$, parametrized by the trajectory $x_{\ast }:\rR
\rightarrow V$%
, we denote by $<g>_{S}$ the average of any regular function $g$,
\begin{equation*}
<g>_{S}=\frac{1}{T_{S}}\int_{0}^{T_{S}}g(x_{\ast }(\theta ))d\theta,
\end{equation*}%
where $T_S$ is the minimal period of the limit cycle. Finally we
denote
\begin{equation*}
\bar{g}=g-<g>_{S}.
\end{equation*}%


\subsubsection{Expression of the Topological Pressure in some cases.}

\noindent The topological pressure of
the recurrent components $\omega $ of $b$ are given by the following
formula
\begin{eqnarray*}
TP(\omega)=\pi_{\omega }+R(\omega ).
\end{eqnarray*}
where
\begin{eqnarray*}
\pi _{\omega}= \sum_{i}\min (0,Re \lambda_{i}(\omega)),
\end{eqnarray*}
where $\lambda_{i}(\omega)$ are the eigenvalues of the matrix B (see paragraph \ref{notat}) and
\begin{itemize}
\item (i) if $\omega $ stationary point, $R(\omega )=c(\omega ),$
\item (ii) if $\omega $ is a cycle parametrized by $\theta $ (notation \ref{notat}),%
\begin{equation*}
R(\omega )=\frac{1}{T_{\omega}}\int_{0}^{T_{\omega
}}c(\theta)d\theta.
\end{equation*}
\item (iii) if $\omega $ is a 2-dimensional irrational torus, in the coordinate system
defined in the paragraphs \ref{notat},
\begin{eqnarray*}
R(\omega )=\frac{1}{k_1k_2}\int_{\mathbb{T}^{2}}c(\theta_1,\theta_2)d%
\theta_1d\theta_2.
\end{eqnarray*}
where $c(\theta_1,\theta_2)$ is the restriction of c to $\omega$.
\end{itemize}
The topological pressure of the field b is
\begin{eqnarray*}
TP=\max \{\pi _{\omega }+R(\omega )| \omega \hbox{ recurrent components of } b\}.
\end{eqnarray*}


\subsection{Definition of Concentration Phenomena}

We shall say that a limit of a measure $\frac{v_{\epsilon }^{2}dV_{g}}{%
\int_V v_{\epsilon}^{2}dV_{g}}$
is concentrated on a set S if S has an open neighborhood $U$ such that the
support of the restriction of a limit measure to U is S. The total mass of
the restriction is called the concentration coefficient.

\begin{Defi}
For all $\delta $ small enough fixed, $B_{P}(\delta )$ is the geodesic ball
of radius $\delta $, centered at $P$, the concentration coefficient $c_{P}$
is,
\begin{equation*}
c_{P}={\lim_{n\rightarrow \infty }}
\frac{\int_{B_{P}(\delta )} v_{\epsilon_{n}}^{2}dV_g}{\int_V v^2_{\epsilon_{n} }dV_g}.
\end{equation*}%
There exists $r>0$ such that the restriction of the limit measure to the
ball $B_P(r)$ is $c_P \delta_P$.
\end{Defi}
This coefficient characterizes the concentration measure at point $P$.
Similarly, for a set $S$, which can be a cycle $\Gamma $ or torus $%
\hbox{\bbbis T}$, we have
\begin{Defi}
\label{concencoeffgeneral} If $\mu$ is a limit measure for a small enough $%
\delta $, such that $supp \mu \cap T^{S}(\delta ) = S$, ($T^{S}(\delta)$ the
tubular neighborhood of $S$) and the concentration coefficient $c_{S}(\mu)$ is
defined by:
\begin{equation*}
\hbox{ if } v_{\epsilon_{n}}^{2}dV_g \rightarrow \mu \hbox{ then }
c_{S}(\mu)= \mu(T^S(\delta)) = {\lim_{n\rightarrow \infty }}\frac{\int_{T^{S}(\delta )}%
{v_{\epsilon_{n}}^{2}}dV_g}{\int_V v^2_{\epsilon_{n} } dV_g} \, .
\end{equation*}%
It depends on the limit $\mu$, but not on $\delta $ if small enough.
\end{Defi}
We will need the following additional definitions.
Let us denote by $Q^{U_S}_\epsilon$ a maximum point of $v_{\epsilon}$ in $U_S$.
\begin{Defi}\label{Charged}
A sequence $v_{\epsilon_n}$ where $\epsilon_n$ tends to zero is said
to charge a set S if the following limit exists and
\begin{eqnarray*}
\gamma _{S}=\lim_{ \epsilon_n\rightarrow 0 } \frac{v_{\epsilon_n}(Q^{U_S}_{\epsilon_n})%
}{\max_{V} v_{\epsilon_n }}>0.
\end{eqnarray*}
$\gamma _{S}$ is called a modulating coefficient. Note that this coefficient depends on the subsequence.
Such sequence is called a charging sequence. Finally, a sequence $v_{\epsilon_n}$
where $\epsilon_n$ tends to zero is said to maximally charge a set S if
\begin{eqnarray*}
\gamma _{S}=1.
\end{eqnarray*}
We remark that the definition of $\gamma _{S}$ does not depend on the open set $U_S$, small enough .
\end{Defi}



\section{Main Results}


\begin{theorem}\label{th1}
On a compact Riemannian manifold $(M,g)$, let $b$ be a
Morse-Smale vector field and $\mathcal{L}$ be a special Lyapunov function
for $b$. Consider the normalized positive eigenfunction $u_{\epsilon }>0$ of
the operator $L_{\epsilon }=\epsilon \Delta +\theta (b)+c$, associated to
the first eigenvalue $\lambda _{\epsilon }$.
\begin{enumerate}
\item The recurrent set $R$ of $b$ is a union of a finite set of stationary points $R^s$,
a finite set of periodic orbits $R^p$ and a finite set of two
dimensional irrational torii $R^t$. The limit set of a  normalized
measure
\begin{eqnarray*}
\frac{ u_{\epsilon }^{2}e^{-\mathcal{L}/\epsilon }dV_{g}}
{\int_{V_{n}}u_{\epsilon}^{2}e^{-\mathcal{L}/\epsilon }dV_{g}},
\end{eqnarray*}
is contained in the set of probability measure $\mu$ of the form
\begin{eqnarray*}
\mu =\sum_{P\in R^s_{tp}}c_{P}\delta _{P}+\sum_{\Gamma \in
R^p_{tp}}a_{\Gamma }\delta _{\Gamma }+\sum_{\hbox{\bbbis T}\in R^t_{tp}}b_{%
\hbox{\bbbis T}}\delta_{\hbox{\bbbis T} }
\end{eqnarray*}
where $R^s_{tp}$ (resp.$R^p_{tp},R^t_{tp}$) is the subset of $R^s$
(resp.$R^p,R^t$), where the topological pressure is attained.
$\delta _{P}$ is the Dirac measure at P.

For $\Gamma \in R^p$ and h $\in C(V )$,
\begin{eqnarray} \label{delta_g}
 \delta _{\Gamma }(h)= \int_{0}^{T_{\Gamma }} f_\Gamma(\theta) h(\Gamma
(\theta )) d\theta,
\end{eqnarray}
where $\theta \in \mathbb{R\longrightarrow }\Gamma (\theta )\in V$
is a
solution of $b$ representing $\Gamma$ 
(see the notations for the precise definition of $\theta$). The periodic function
 $f_\Gamma$ is given by
\begin{eqnarray*}
f_\Gamma(\theta) = \exp \{  -{\int_0^\theta{ c(\Gamma(s)) } ds} +
\frac{\theta}{T_\Gamma}\int_0^{T_\Gamma} c(\Gamma(s)) ds \}
\end{eqnarray*}
and $T_\Gamma$ is the minimal period of $\Gamma$.

$\hbox{ For {\bbbis T}} \in R^t \hbox{ and }\,  h \in C(V ),$

\begin{eqnarray}\label{delta_t}
\delta_{\hbox{\bbbis T} }(h) =\int_{\hbox{\bbbis T}}h(\theta
_{1},\theta _{2}) f_{\hbox{\bbbis T}}(\theta _{1},\theta
_{2})dS_{\hbox{\bbbis T}},
\end{eqnarray}
where $dS_{\hbox{\bbbis T}}$ is the unique probability measure on
$\hbox{\bbbis T}$
 invariant under the action of the field b and $f_{\hbox{\bbbis T}}$ is the unique
 solution of maximum 1, of the equation
\begin{eqnarray*}
k_1 \frac{\partial f}{\partial \theta_1}+k_2 \frac{\partial f}{\partial
\theta_2} +cf = \mu_2 f \hbox{ where } \mu_2=\int_{\hbox{\bbbis T}} c dS_{\hbox{\bbbis T}}.
\end{eqnarray*}
\item The coefficients $c_{P}, a_{\Gamma}, b_{\hbox{\bbbis T}}$ obey the following rule:
If at least one coefficient $b_{\hbox{\bbbis T}}>0$, then for all cycle $a_{\Gamma }=0$ and
 all points $c_{P}=0$. If all coefficients  $b_{\hbox{\bbbis T}}=0$, and at least one coefficient $a_{\Gamma }>0$ then all $c_{P}=0$.
\item The limit measures of the first eigenfunction are
concentrated on the recurrent set $\mathcal{R}$ of $\Omega $, where the
topological pressure is attained. If $\omega \in \mathcal{R}$, and $TP
(\omega )\neq TP(\mathcal{R})$, where $TP (\omega )$ is the topological pressure
 at $\omega $, then the limit measure associated with the normalized
eigenfunction $\frac{e^{-\mathcal{L}/\epsilon }u_{\epsilon }^{2}dV_{g}}{%
\int_{V_{n}}e^{-\mathcal{L}/\epsilon }u_{\epsilon }^{2}dV_{g}}$ has no
contribution on $\omega $.
\end{enumerate}
\end{theorem}

{\noindent \textbf{Remarks.}}

\noindent In this last theorem, we consider only two dimensional torii such
that the restriction of the vector field to the torus is diffeomorphic to an
irrational flow. We do not known what should be
the equivalent theorem for a field where the recurrent set contains a two
dimensional surface $\Sigma$ of arbitrary genus. If the recurrent set
contains a two dimensional sphere, any field contains critical points. Since
there is no ergodic field on $S^{2}$, this suggests that the concentration
of the first eigensequence cannot be absolutely continuous with respect to
the surface of the sphere but should occur on some subsets $S^{2}$, such as
the critical points. For a general Riemannian surface, it is not know how to
construct an ergodic field on it and how the concentration occurs on such a
surface.

When a recurrent set is a Riemannian surface $\Sigma$ which contains a limit
cycle, it may be that the concentration occurs on the limit cycle rather
that on the entire surface. Using a stability argument, we expect the
concentration to occur on a minimal recurrent set of maximum dimension. We
have not made any investigation in that fascinating direction.

It is a very interesting problem to determine the limiting weight
$c_{P}, a_{\Gamma}, b_{\hbox{\bbbis T}}$ of each component.
Actually, even the uniqueness of the weight is still unknown. It is
indeed a very difficult problem and no much results about this
problem are mentioned in literature. The results presented in this
paper constitute a first contribution toward the solution of that
problem.


\subsection{ Description of the method}

We prove the main results in several steps. First by using gauge
transformation involving a special Lyapunov function $\mathcal{L}$,
the first order term in the partial differential operator given by
expression (\ref{fpo}) is transformed into one which becomes
arbitrarily small as $\epsilon $ goes to zero.

The second step is the blow-up analysis of the eigenfunction. In \cite{HK1}
 it has been proved that the concentration occurs on the recurrent sets of
the field, and that the supports of the limit measures are contained in the
set where the function $\Psi _{\mathcal{L}}$, defined by equation (\ref{gauge})
(paragraph \ref{definitionintro}), vanishes. Under appropriate assumptions we show that if a component of the
recurrent set is a cycle or an irrational 2-dimensional torus, the
restriction of any limit measure to it is absolutely continuous with respect
to the unique measure invariant under the flow generated by b on the
component (in the case of a torus, the measure is unique because our assumptions imply
that the flow is ergodic). Moreover we give explicit formulas for the densities of
these limit measures.

In the last step, we compute precise decay estimates for the eigenfunction and
for that purpose, we use the Feynman-Kac formula for the eigenfunction.


\subsection{ Induction Principle for the localization of concentration}

As stated in the main theorem, the selection of the recurrent set
depends on the topological pressure only. On the other hand, the
limit measures depend on the total jets of the field $b$  and the
potential $c$ along the recurrent.
The TP is not sufficient to determine the final support of the limit
measures, because it involves only the first term in the expansion
of the eigenvalue $\lambda_{\epsilon}$ in power of $\epsilon^{1/2}$.
To obtain a more precise localization, all the expansion has to be
used. This question is similar to the double-well potential problem:
what are the coefficients of the limit measure when there are two
recurrent sets where the TP is achieved (see \cite{HK1})?


\subsection*{Remark}

In this section we study the concentration of
a limit measures on the critical points of the field b. Because $\Omega $
is not a gradient, we cannot use variational methods in our study
(equation (\ref{edpfdt1}) can not be conjugated in general to a
variational equations). Using the special Lyapunov
function, constructed in \cite{HK1}, equation (\ref{edpfdt1}) is transformed into
\begin{eqnarray}
\epsilon ^{2}\Delta _{g}v_{\epsilon }+\epsilon \theta (\Omega
)v_{\epsilon }+c_{\epsilon }v_{\epsilon }=\epsilon \lambda
_{\epsilon }v_{\epsilon } \notag
\end{eqnarray}%
\begin{eqnarray*}
c_{\epsilon }=\epsilon (c+\frac{\Delta _{g}\mathcal{L}}{2})+\Psi _{\mathcal{L%
}}
\end{eqnarray*}%
where $c_{\epsilon }$ and $\Psi _{\mathcal{L}}$ have been computed in (\ref%
{gauge}). This transformation is crucial in our analysis to obtain interesting results.

Weighting $v^2_{\epsilon }$ by $e^{-\frac{\mathcal{L}}{\epsilon}}$
enables us to extract the main features of the limit measures.
Another advantage of using the transformed
 equation is that the first order term tends to zero with $\epsilon $.
Moreover, since the choice of the  special Lyapunov function
$\mathcal{L}$ is not unique, one would expect that the limits depend
on this  choice. In fact, it turns out that according to our
construction of the Lyapunov function, the second order term of
which satisfies the linearized Hamilton-Jacobi equation,  the limit
does not depend on this choice.

\section{ The case of critical points}


\subsection{Rate of convergence of the maximum points}

Here we study the behavior of the maximum points of $v_{\epsilon
}=u_{\varepsilon }\exp -\frac{\mathcal{L}}{2\varepsilon }$. This is usefull
when we are going to normalize the eigenfunctions. Let $\mathcal{M}%
_{\varepsilon }$ be the set of maximum points of $v_{\epsilon}.$ $\mathcal{R}
$ denoted the recurrent set of $b$.

\begin{lem}\label{lemfdt}
There exists a constant $C>0$ such that $\sup \{d(P,\mathcal{R})|P\in
\mathcal{M} _{\varepsilon }\} \leq C\sqrt{\epsilon}$.
\end{lem}
\textbf{\noindent Remark.} This situation is similar to the
variational case \cite{HK1} where the sequence $P_{\epsilon }$ of
global (also local) maximum points converges to a point of the
recurrent set of the field. The rate of convergence depends on the
order of the vanishing of $\Psi _{\mathcal{L}}$ on the recurrent
sets.

\bigskip

\noindent \textbf{Proof:} \noindent The proof is an immediate consequence of
the Maximum Principle. Indeed, at a maximum point (or local maximum) $P\in
\mathcal{M}_{\varepsilon }$, $\Delta _{g}v_{\epsilon }(P)\geq 0$ and $\left(
\theta (\Omega )v_{\epsilon }\right) (P)=0$. Thus, using equation (\ref%
{edpfdt1}), because the sequence $v_{\epsilon }$ is positive:
\begin{eqnarray*}
c_{\epsilon }(P)\leq \epsilon \lambda _{\epsilon }.
\end{eqnarray*}%
Because $c_{\epsilon }=\epsilon (c+\frac{\Delta _{g}\mathcal{L}}{2})+\Psi _{%
\mathcal{L}}$ , we obtain the following estimate
\begin{eqnarray*}
0\leq \Psi _{\mathcal{L}}(P)\leq \epsilon (\lambda _{\epsilon }+\underset{V}{%
\max }|c+\frac{\Delta _{g}\mathcal{L}}{2}|).
\end{eqnarray*}
The definition of $\Psi _{\mathcal{L}}$ implies that there exists a
constant
$C_{1}>0$ such that%
\begin{eqnarray*}
\Psi _{\mathcal{L}}(P)\geq \frac{1}{C_{1}}d(P,\mathcal{R} )^{2}.
\end{eqnarray*}
\bigskip Hence for some constant C and all P $\in \mathcal{M}_{\varepsilon }$
\begin{eqnarray*}
d(P,\mathcal{R} )^{2}\leq C\epsilon.
\end{eqnarray*}%
\subsection{Weak limits of the eigenfunctions w$_{\protect\varepsilon }$: case of points}

Let ($U,x_{1},...x_{m})$ be a coordinate system at P, as defined in section %
\ref{notat}(I). The blown up function $w_{\epsilon }$ is defined on the
subset $\frac{1}{\sqrt{\varepsilon }}x_{1}\times ...\times x_{m}(U)$ of $%
\mathbb{R}^{m}$ by
\begin{equation*}
w_{\epsilon }(x)=\frac{v_{\epsilon }(\sqrt{\epsilon }x)}{\bar{v_{\epsilon }}}%
,
\end{equation*}%
where $x=(x_{1},...x_{m})$. As $\epsilon $ tends to 0 the behavior of $%
w_{\epsilon }$ is described in the following theorem where $\Delta
_{E}=-\sum_{n=1}^{m}\frac{\partial ^{2}}{\partial x_{n}^{2}}.$

\begin{theorem}\label{blowup} Let $P$ be a critical point of $b$ and let $%
(U,x_{1},...,x_{m})$ be a normal coordinate system centered at $P$, as in
section (I) of ( \ref{notat}).
\begin{itemize}
\item Any weak limit  $w$ of $w_{\epsilon }$ as $\varepsilon \rightarrow 0$ ,
is in $C^{\infty }$  and is a solution of the equation
\begin{eqnarray}
\Delta _{E}w+\sum_{i,j=1}^{m}\Omega _{ij}(P)x_{j}\frac{\partial
w}{\partial x_{i}}+[c(P)+\frac{\Delta _{E}\mathcal{L}}{2}(P)+\psi
_{2}(x)]w=\lambda w, \label{bua}
\end{eqnarray}%
\begin{eqnarray*}
0<w\leq 1,
\end{eqnarray*}%
where $\sum_{i,j=1}^{m}\Omega _{ij}(P)x_{j}\frac{\partial }{\partial
x_{i}}$ is the linear part of the field $\Omega $ at P, $\psi
_{2}(x)$ is the quadratic form representing the terms of order two
in the Taylor development of $\Psi _{\mathcal{L}}$ at P and $\lambda
\geq 0$ is equal to the topological pressure:
\begin{eqnarray*}
\Pi (P)=c(P)+\frac{\Delta _{E}\mathcal{L}}{2}(P)+\sum \{\max (0,
Re\sigma (\Omega(P))) | \sigma (\Omega (P))\hbox{ eigenvalue of
}\Omega (P)\}\},
\end{eqnarray*}%
where the sum in the right hand-side is the sum of all real parts of
the eigenvalues of the matrix $(\Omega _{ij}(P)|1\leq i,j\leq m)$,
each counted according to its multiplicity.

\item There exits a sequence $w_{\epsilon _{n}}$ such that $\varepsilon
_{n}\rightarrow 0$ as n goes to infinity, which converges to $w$ in the $%
C^{\infty }$ topology.
\item Either w is identically zero or $0<w \leq 1$. In addition,
if the sequence is maximally charging at the point P  then $w(P) = 1$.
\end{itemize}
\end{theorem}

\bigskip {\noindent \textbf{Proof. }}$w_{\epsilon }$ satisfies the equation:%
\begin{equation}
\Delta _{g_{\epsilon }}w_{\epsilon }(x)+\sum_{i,j=1}^{m}\frac{\Omega ^{i}(x%
\sqrt{\epsilon })}{\sqrt{\epsilon }}\frac{\partial w_{\epsilon }(x)}{%
\partial x^{i}}+\left[ \frac{\Psi _{\mathcal{L}}(x\sqrt{\epsilon })}{%
\epsilon }+c+\frac{\Delta _{g_{\epsilon }}\mathcal{L(}x\mathcal{)}}{2}\right]
w_{\epsilon }(x)=\lambda _{\epsilon }w_{\epsilon }(x)  \label{bin}
\end{equation}%
where $g_{\epsilon }(x)=g(\sqrt{\epsilon }x)$. The assumptions on $b$ and $%
\mathcal{L}$ at the point $P$ imply that the functions
$\widehat{\Omega}$ and $\widehat{\Psi _{\mathcal{L}}}$ of the
variables $(x,\epsilon)$ defined for $1\leq i\leq m$ by

\begin{eqnarray*}
\widehat{\Omega ^{i}}(x,\sqrt{\varepsilon }) &=&\frac{\Omega ^{i}(\sqrt{\epsilon }x)}{\sqrt{%
\epsilon }}, \hbox{ for }\varepsilon>0  \\
&=& \sum_{j=1}^{m}\Omega _{j}^{i}(P)x_{j}, \hbox{ for }
\varepsilon=0
\end{eqnarray*}
and
\begin{eqnarray*}
\widehat{\Psi _{\mathcal{L}}}(x,%
\sqrt{\varepsilon }) &=& \frac{\Psi _{\mathcal{L}}(\sqrt{\epsilon
}x)}{\epsilon }\hbox{ for }\varepsilon>0
\\ \widehat{\Psi _{\mathcal{L}}}(x,0)&=&\psi_2(x)=
\frac{1}{2}<A(P)x,x>_{\mathbb{R}^{m}} ,
\end{eqnarray*}
where $<A(P)x,x>_{\mathbb{R}^{m}} $ is the initial term of the Taylor expansion of $\Psi_\mathcal{L}$ at P. Let
us call $L_{\varepsilon }$ the operator%
\begin{eqnarray*}
L_{\varepsilon }=\Delta _{g_{\epsilon }}+\sum_{i,j=1}^{m}\widehat{\Omega ^{i}%
}(x,\sqrt{\epsilon })\frac{\partial }{\partial x^{i}}+\widehat{\Psi }_{%
\mathcal{L}}(x,\sqrt{\epsilon })+c+\frac{\Delta _{g_{\epsilon }}\mathcal{L(}x%
\mathcal{)}}{2}.
\end{eqnarray*}
Because the second order operator $L_{\varepsilon }$ is uniformly
elliptic for $\varepsilon \in \lbrack 0,1]$ ($0$ included!) on every
compact set and its coefficients are C$^{\infty }$(in the variables
x and $\sqrt{\varepsilon})$,
classical interior elliptic estimates (see \cite{GT}) imply that $%
w_{\varepsilon }$ is bounded on every compact set in the C$^{\infty
}$ topology uniformly in ${\varepsilon} \in \lbrack 0,1],$
$w_{\epsilon }$ being
bounded by 1. Ascoli-Arzela's theorem implies w is the limit of a sequence $%
\{w_{\epsilon _{n}}|n\in \mathbb{N},\varepsilon _{n}\},\varepsilon
_{n}\rightarrow 0,$ which converges to $w$ in the C$^{\infty }$ topology.
 $w$ is a classical solution of the elliptic equation%
\begin{eqnarray*}
\Delta _{E}w(x)+\sum_{i,j=1}^{m}\Omega _{j}^{i}(P)x_{j}\frac{\partial w}{%
\partial x_{i}}+[\psi _{2}(x)+(c+\Delta _{E}\mathcal{L}/2)(0)]w(x)=\lambda
w(x)\hbox{ on }\hbox{\bb R}^{m}.
\end{eqnarray*}

We want to determine the function $w$. With that goal in mind we
shall transform  equation (\ref{bua}) into a well known one.\ Set
\begin{eqnarray}
z=w\exp \frac{<A(P)x,x>_{\mathbb{R}^{m}}}{2}.
\end{eqnarray}
 Then
\begin{eqnarray*}
\Delta _{E}z+\sum_{i,j=1}^{m}(\Omega _{j}^{i}(P)+A_{j}^{i}(P))x_{j}\frac{%
\partial z}{\partial x_{i}}=[\lambda -c(0)-trA]z
\end{eqnarray*}%
If $\ \sum_{ij=1}^{m}B_{ij}(P)x_{j}\frac{\partial }{\partial x_{j}}$
denotes
the linear part of $b$ at $P$ in the coordinate system $(U,x_{1},..,x_{n})$:%
\begin{eqnarray*}
\Omega _{j}^{i}(P)+A_{j}^{i}(P)=B_{j}^{i}(P)
\end{eqnarray*}%
\bigskip The operator $\Delta _{E}+\sum_{i,j=1}^{m}B_{j}^{i}(P)x_{j}\frac{%
\partial }{\partial x_{i}}$ is well known: it is the celebrated
Ornstein-Uhlenbeck operator. $z$ is a solution of
\begin{eqnarray}
\Delta _{E}z+\sum_{i,j=1}^{m}B_{j}^{i}(P)x_{j}\frac{\partial
z}{\partial x_{i}}=[\lambda -c(0)-trA]z  \label{eqfdtb}
\end{eqnarray}
If $\sigma _{1},\sigma _{2},...,\sigma _{m}$ are the eigenvalues of
$B(P)$, each appearing a number of times equal to its multiplicity,
using the result of \cite{Kifer80}, we have
\begin{eqnarray*}
\lambda =\Pi (P)=c(0)+TrA-\sum_{n=1}^{m}\min [0,Re\sigma _{n}],
\end{eqnarray*}
equivalently
\begin{eqnarray*}
\lambda -c(0)-trA=-TrB^{s}.
\end{eqnarray*}

\subsection{The Ornstein-Uhlenbeck model }

Clearly the function $\zeta :\mathbb{R\times R}^{m}\rightarrow \mathbb{R}$, $%
\zeta (t,x)=z(x)\exp (tTrB^{s})$ is a solution of the parabolic equation%
\begin{equation}
\frac{\partial \zeta }{\partial t}+\Delta _{E}\zeta
+\sum_{ij=1}^{m}B_{j}^{i}(P)x_{j}\frac{\partial \zeta }{\partial x_{i}}=0
\label{parou}
\end{equation}
We are going to show that Kolmogorov's integral%
\begin{eqnarray*}
T_{t}(z)(x)&=&\frac{1}{(4\pi )^{m/2}(detQ_{t})^{1/2}}\int_{\mathbb{R}%
^{m}}e^{-<Q_{t}^{-1}(e^{-tB}x-y),(e^{-tB}x-y)>_{\mathbb{R}^{m}}/4}z(y)dy, \\
&=&\frac{1}{(4\pi )^{m/2}(\det Q_{t})^{1/2}}\int_{\hbox{\bb
R}^{m}}e^{-<Q_{t}^{-1}y,y>/4}z(e^{tB}x-y)dy \nonumber
\label{kor}
\end{eqnarray*}
is a $C^{\infty }$ function satisfying the equation (\ref{parou}). Here
\begin{eqnarray*}
Q_{t}=\int_{0}^{t}e^{-sB}e^{-sB^{\ast }}ds.
\end{eqnarray*}
Because $z=w\exp \frac{%
<A(P)x,x>_{\mathbb{R}^{m}}}{2},$
\begin{eqnarray}
T_{t}(z)(x)=\frac{1}{(4\pi )^{m/2}(detQ_{t})^{1/2}}\int_{\mathbb{R}%
^{m}}w(y)e^{-q(x,y,t)}dy  \label{i}
\end{eqnarray}
where%
\begin{equation*}
q(x,y,t)=\frac{1}{4}<Q_{t}^{-1}(e^{-tB}x-y),(e^{-tB}x-y)>_{\mathbb{R}^m}-%
\frac{<A(P)y,y>_{\mathbb{R}^{m}}}{2},
\end{equation*}
\begin{eqnarray}\label{iv}
q(x,y,t)=-\frac{1}{4}<U_{t}x,x>_{\mathbb{R}^{m}}+\frac{1}{4}%
||R_{s,t}y_{s}-P_{s,t}x_{s}||_{\mathbb{R}^{m}}^{2}+\frac{1}{4}%
||R_{u,t}y_{u}-P_{u,t}y_{u}||_{\mathbb{R}^{m}}^{2} ,
\end{eqnarray}
where $R_{s,t},R_{u,t}$ are the unique positive definite operators such
that:
\begin{equation*}
R_{s,t}^{2}=Q_{s,t}^{-1}-2A_{s},\text{ }R_{u,t}^{2}=Q_{u,t}^{-1}-2A_{u}.
\end{equation*}
\begin{equation*}
P_{s,t}=R_{s,t}^{-1}Q_{s,t}^{-1}e^{-B_{s}},\text{ }%
P_{u,t}=R_{u,t}^{-1}Q_{u,t}^{-1}e^{-tB_{u}}.
\end{equation*}
\begin{equation*}
U_{t}=U_{s,t}\oplus U_{u,t}
\end{equation*}
where
\begin{equation*}
U_{s,t}=e^{-tB_{s}^{\ast }}\left( Q_{s,t}^{-1}\
-Q_{s,t}^{-1}R_{s,t}^{-2}Q_{s,t}^{-1}\right) e^{-tB_{s}},\text{ }%
U_{u,t}=e^{-tB_{u}^{\ast }}\left( Q_{u,t}^{-1}\
-Q_{u,t}^{-1}R_{u,t}^{-2}Q_{u,t}^{-1}\right) e^{-tB_{u}}
\end{equation*}
$U_{s,t}$ is negative definite and $U_{u,t}$ positive definite.
\begin{equation*}
T_{t}(z)(x)=\frac{e^{\frac{1}{4}<U_{t}x,x>_{\mathbb{R}^{m}}}}{(4\pi
)^{m/2}(detQ_{t})^{1/2}}\int_{\mathbb{R}^{m}}w(y)\exp \left( -\frac{1}{4}%
||R_{s,t}y_{s}-P_{s,t}x_{s}||_{\mathbb{R}^{m}}^{2}-\frac{1}{4}
||R_{u,t}y_{u}-P_{u,t}y_{u}||_{\mathbb{R}^{m}}^{2}\right)dy
\end{equation*}

\begin{lem}
\label{inequ} The operators just defined have the following properties:

(i)For small $t>0:$
\begin{enumerate}
\item $Q_{s,t}=t\left[ Id_{s}-t(\frac{B_{s}+B_{s}^{\ast }}{2})+O(t^{2})\right] $
and $Q_{s,t}^{-1}=\frac{1}{t}Id_{s}+\frac{B_{s}+B_{s}^{\ast }}{2}+O(t).$

\item $Q_{u,t}=t\left[ Id_{u}-t(\frac{B_{u}+B_{u}^{\ast }}{2})+O(t^{2})\right] \
$and\ $Q_{u,t}^{-1}=\frac{1}{t}Id_{u}+\frac{B_{u}+B_{u}^{\ast }}{2}+O(t).$

\item $R_{s,t}^{2}=\frac{1}{t}Id_{s}+\frac{B_{s}+B_{s}^{\ast }}{2}-2A_{s}+O(t)$
and $R_{s,t}^{-2}=t\left[ Id_{s}-t(\frac{B_{s}+B_{s}^{\ast }}{2}%
-2A_{s})+O(t^{2})\right] $

\item $R_{u,t}^{2}=\frac{1}{t}Id_{u}+\frac{B_{u}+B_{u}^{\ast }}{2}-2A_{u}+O(t)$
and \ $R_{u,t}^{-2}=t\left[ Id_{u}-t(\frac{B_{u}+B_{u}^{\ast }}{2}%
-2A_{u})+O(t^{2})\right] $

\item $U_{s,t}=-2A_{s}+O(t)$ and \ $\ \ \ \ \ U_{u,t}=-2A_{u}+O(t).$

\item $R_{s,t}^{-2}Q_{s,t}^{-1}e^{-tB_{s}}=Id_{s}+\ O(t)$ and$\
R_{u,t}^{-2}Q_{u,t}^{-1}e^{-tB_{u}}=Id_{u}+O(t)$

\item $\det Q_{t}=t^{m-1}(1+O(t)).$
\end{enumerate}

(ii) When $t\rightarrow +\infty $:

\begin{enumerate}
\item $Q_{s,t}\rightarrow \infty $ and $%
Q_{u,t}\rightarrow \int_{0}^{+\infty }e^{-sB_{u}}e^{-sB_{u}^{\ast }}ds$
\item $R_{s,t}\rightarrow R_{s,\infty }=\sqrt{-2A_{s}}$ and $R_{u,t}\rightarrow R_{u,\infty }=%
\sqrt{Q_{u,\infty }^{-1}-2A_{u}}>>0$
\item $Q_{s,t}^{-1}e^{-tB_{s}}\rightarrow 0$ and hence $P_{s,t}=$ $%
R_{s,t}^{-2}Q_{s,t}^{-1}e^{-tB_{s}}\rightarrow 0,U_{s,t}\rightarrow \left(
\int_{0}^{+\infty }e^{\tau B_{s}}e^{\tau B_{s}^{\ast }}d\tau \right) ^{-1}$
\item $Q_{u,t}\rightarrow Q_{u,\infty }=\int_{0}^{+\infty
}e^{-tB_{u}}e^{-tB_{u}^{\ast }}dt,$  P$_{u,t}=
R_{u,t}^{-2}Q_{u,t}^{-1}e^{-tB_{u}}\rightarrow 0,U_{u,t}\rightarrow
0.$
\end{enumerate}

\bigskip

(iii) $e^{2tTrB_{s}}\det Q_{s,t}\rightarrow \det \int_{0}^{+\infty }e^{\tau
B_{s}}e^{\tau B_{s}^{\ast }}d\tau $
\end{lem}

\bigskip

\textbf{Proof.} Most of the statements of the lemma are trivial. Let us prove
(ii) 3).

\begin{eqnarray*}
\left( Q_{s,t}^{-1}e^{-tB_{s}}\right)
^{-1}&=&e^{tB_{s}}Q_{s,t}=e^{tB_{s}}\int_{0}^{t}e^{-\tau
B_{s}}e^{-\tau B_{s}^{\ast }}d\tau =\left(\int_{0}^{t}e^{(t-\tau
)B_{s}}e^{(t-\tau )B_{s}^{\ast
}}d\tau\right) e^{-tB_{s}^{\ast }}\\
&=&\left(\int_{0}^{t}e^{\sigma B_{s}}e^{\sigma B_{s}^{\ast
}}d\sigma\right) e^{-tB_{s}^{\ast }}.
\end{eqnarray*}
Hence $e^{tB_{s}}Q_{s,t}\rightarrow \infty ,$ $%
Q_{s,t}^{-1}e^{-tB_{s}}\rightarrow 0.$ We now compute
\begin{eqnarray*}
 U_{s,t}=e^{-tB_{s}^{\ast
}}\left( Q_{s,t}^{-1}\ -Q_{s,t}^{-1}R_{s,t}^{-2}Q_{s,t}^{-1}\right)
e^{-tB_{s}}=e^{-tB_{s}^{\ast }}Q_{s,t}^{-1}e^{-tB_{s}}-e^{-tB_{s}^{\ast
}}Q_{s,t}^{-1}R_{s,t}^{-2}Q_{s,t}^{-1}e^{-tB_{s}}.
\end{eqnarray*}
 Hence
\begin{eqnarray*}
& e^{-tB_{s}^{\ast }}Q_{s,t}^{-1}=\left( Q_{s,t}^{-1}e^{-tB_{s}}\right)
^{\ast }\rightarrow 0 \\
& e^{-tB_{s}^{\ast
}}Q_{s,t}^{-1}R_{s,t}^{-2}Q_{s,t}^{-1}e^{-tB_{s}}\rightarrow 0.
\end{eqnarray*}
Moreover
\begin{eqnarray*}
e^{-tB_{s}^{\ast }}Q_{s,t}^{-1}e^{-tB_{s}}=\left(
e^{tB_{s}}Q_{s,t}e^{tB_{s}^{\ast }}\right)
^{-1}.e^{tB_{s}}Q_{s,t}e^{tB_{s}^{\ast }}=e^{tB_{s}}\int_{0}^{t}e^{-\tau
B_{s}}e^{-\tau B_{s}^{\ast }}d\tau e^{tB_{s}^{\ast }}.
\end{eqnarray*}
Thus
\begin{eqnarray*}e^{tB_{s}}Q_{s,t}e^{tB_{s}^{\ast }}=\int_{0}^{t}e^{(t-\tau
)B_{s}}e^{(t-\tau )B_{s}^{\ast }}d\tau =\int_{0}^{t}e^{\sigma
B_{s}}e^{\sigma B_{s}^{\ast }}d\sigma.
\end{eqnarray*}
Hence as $t\rightarrow +\infty ,$
\begin{eqnarray*}
 e^{-tB_{s}^{\ast }}Q_{s,t}^{-1}e^{-tB_{s}} \rightarrow
\left( \int_{0}^{+\infty }e^{\sigma B_{s}}e^{\sigma B_{s}^{\ast }}d\sigma
\right) ^{-1}.
\end{eqnarray*}
 and
\begin{eqnarray*}
 U_{s,t}\rightarrow \left( \int_{0}^{+\infty }e^{\tau
B_{s}}e^{\tau B_{s}^{\ast }}d\tau \right) ^{-1}.
\end{eqnarray*}

\underline{To prove (ii) 4)} note that

\begin{eqnarray*}
U_{u,t}=e^{-t B_{u}^{\ast }}(Q_{u,t}^{-1}\
-Q_{u,t}^{-1}R_{u,t}^{-2}Q_{u,t}^{-1})e^{-tB_{u}}.
\end{eqnarray*}
Because
\begin{eqnarray*}
Q_{u,t}^{-1}\ -Q_{u,t}^{-1}R_{u,t}^{-2}Q_{u,t}^{-1}\rightarrow
Q_{u,\infty }^{-1}\ -Q_{u,\infty }^{-1}R_{u,\infty }^{-2}Q_{u,\infty
}^{-1},
\end{eqnarray*}
  $U_{u,t}$ tends to $0.$

\underline{To prove (iii)} note that
\begin{eqnarray*}
e^{tB_{s}}Q_{s,t}e^{tB_{s}^{\ast
}}\rightarrow \int_{0}^{+\infty }e^{\tau B_{s}}e^{\tau B_{s}^{\ast
}}d\tau .
\end{eqnarray*}
Hence
\begin{eqnarray*}
det e^{tB_{s}}\det Q_{s,t}\det e^{tB_{s}^{\ast
}}\rightarrow det \int_{0}^{+\infty }e^{\tau B_{s}}e^{\tau B_{s}^{\ast
}}d\tau .
\end{eqnarray*}
But
\begin{eqnarray*}
\det e^{tB_{s}}=\det e^{tB_{s}^{\ast }}=e^{TrB_{s}}.
\end{eqnarray*}
 So as $t\rightarrow +\infty ,$
\begin{eqnarray*}
  e^{2tTrB_{s}}\det Q_{s,t}\rightarrow \det
\int_{0}^{+\infty }e^{\tau B_{s}}e^{\tau B_{s}^{\ast }}d\tau
\end{eqnarray*}
The formulas (\ref{i}), (\ref{iv}) imply that $T_{t}(z)$ is
a $C^{\infty }$ function of $x$ for $t>0.$ The
integral
\begin{eqnarray}
\frac{e^{-tTrB_{s}}}{(4\pi )^{m/2}(detQ_{t})^{1/2}}\int_{\mathbb{R}%
^{m}}w(y)\exp -\frac{1}{4} \left\{  ||R_{s,t}y_{s}-P_{s,t}x_{s}||_{\mathbb{R}
^{m}}^{2}+\text{ }||R_{u,t}y_{u}-P_{u,t}y_{u}||_{\mathbb{R}^{m}}^{2}\right\}
dy  \label{integ}
\end{eqnarray}%
is bounded by $\underset{\mathbb{R}^{m}}{\sup } w$ for all $t>0,$ because by
Lemma \ref{inequ},(ii) and (iii) 2)
\begin{eqnarray}
e^{tTrB_{s}}\sqrt{\det Q_{s,t}}%
\rightarrow \sqrt{\det \int_{0}^{+\infty }e^{\tau B_{s}}e^{\tau
B_{s}^{\ast }}d\tau }
\end{eqnarray}%
and $R_{s,t}\rightarrow \sqrt{-2A_{s}}$ and $%
R_{u,t}\rightarrow R_{u,\infty }=$ $\sqrt{Q_{u,\infty }^{-1}-2A_{u}}.$
Hence $|T_{t}(z)(x)|$ is bounded by $\ C_{t}\exp -\frac{1}{4}%
<U_{s,t}x_{u},x_{u}>_{\mathbb{R}^{m}},$ where
\begin{eqnarray*}
\underset{t\rightarrow +\infty }{\lim }C_{t}=\frac{\left( \det
\int_{0}^{+\infty }e^{\tau B_{s}}e^{\tau B_{s}^{\ast }}d\tau \right) ^{-%
\frac{1}{2}}}{(4\pi )^{\frac{m}{2}}}\int_{\mathbb{R}^{m}}w(y)e^{\frac{1}{2}%
<A_{s}y_{s},y_{s}>_{\mathbb{R}^{m}}-\frac{1}{4}<\left( Q_{u,\infty
}-2A_{u}\right) y_{u},y_{u}>_{\mathbb{R}^{m}}}dy
\end{eqnarray*}
These properties imply by Tikhonov's theorem \cite{Enc}:
\begin{lem}\label{iden}
\bigskip  for all $t>0,$ $T_{t}(z)=e^{-tTrB_{s}}z.$
\end{lem}
\noindent \textbf{Proof. }It is sufficient to prove that $T_{t}(z)\rightarrow z$ and
that $C_{t}$ $\rightarrow 1$ as $t\rightarrow 0.$U$\sin $g Lemma \ (\ref%
{inequ}) and the representation (\ref{integ}) we can show that:%
\begin{eqnarray*}
T_{t}(z)(x) &=&\frac{\exp +\frac{1+O(t)}{2}<Ax,x>_{\mathbb{R}^{m}}}{\left(
4t\right) ^{\frac{m}{2}}(1+O(t))} \times \\
&&\int_{\mathbb{R}^{m}}w(\eta +(1+O(t))x)e^{-\frac{1+O(t)}{2t}\left(
||\eta _{s}||_{\mathbb{R}^{m_{s}}}^{2}+||\eta
_{u}|_{\mathbb{R}^{m_{u}}}^{2}\right) }d\eta.
\end{eqnarray*}
Because $w$ is bounded at $\infty $, we see
that \bigskip $T_{t}(z)e^{-\frac{1}{2}<Ax,x>_{\mathbb{R}^{m}}}$ and $%
we^{-tTrB_{s}}$ tend to $w$ as $t\rightarrow 0$, uniformly in $x$ and that $%
C_{t}\rightarrow 1.$  The functions $T_{t}(z)e^{-\frac{1}{2}<Ax,x>_{%
\mathbb{R}^{m}}}$ and $w$ are both bounded by quadratic exponentials and
satisfy the same parabolic equation
\begin{eqnarray*}
\Delta _{E}h+\sum_{i,j=1}^{m}\Omega _{j}^{i}(P)x_{j}\frac{\partial h}{%
\partial x_{i}}+[\psi _{2}+(c+\Delta _{E}\mathcal{L}/2)(0)-\lambda ]h+\frac{%
\partial h}{\partial t}=0.
\end{eqnarray*}
and both functions have the same limits as $t\rightarrow 0,$
Tikhonov's theorem
\cite{Enc} implies%
\begin{eqnarray*}
T_{t}(z)=e^{-tTrB_{s}}z.
\end{eqnarray*}

\bigskip

\begin{lem}\label{raq}
The solution z does not depend on the unstable variables $x_{u}$ and
is given explicitly for $x =(x_s,x_u) \in \rR^m$ by
\begin{eqnarray*}
z(x)=C\exp \left( -\frac{1}{4}<\left( \int_{0}^{+\infty
}e^{tB_{s}}e^{tB_{s}^{\ast }}dt\right) ^{-1}x_{s},x_{s}>_{\mathbb{R}%
^{m_{s}}}\right)
\end{eqnarray*}
\end{lem}
where C is a constant.

\noindent \textbf{Proof.} Using lemma (\ref{iden}), Kolmogorov's formula (\ref{kolm}) gives:%
\begin{eqnarray}
z(x)=\frac{e^{-tTrB_{s}+\frac{1}{4}<U_{t}x,x>_{\mathbb{R}^{m}}}}{(4\pi
)^{m/2}(detQ_{t})^{1/2}}\int_{\mathbb{R}^{m}}w(\eta +R_{t}^{-1}P_{t}x)e^{-%
\frac{1}{4}||R_{s,t}\eta _{s}||_{\mathbb{R}^{m_{s}}}^{2}-\frac{1}{4}%
||R_{u,t}\eta _{u}||_{\mathbb{R}^{m_{u}}}^{2}}d\eta  \label{z}
\end{eqnarray}
Letting $t\rightarrow +\infty $ in the expression (\ref{z}) above,
using the preceding estimates in Lemma (\ref{inequ}) and the fact
that $w$ is bounded at $\infty $, we see that:
\begin{eqnarray*}
z(x)=\frac{\exp -\left[ \frac{1}{4}\left( \int_{0}^{+\infty
}e^{tB_{s}}e^{tB_{s}^{\ast }}dt\right) ^{-1}x_{s},x_{s}>_{\mathbb{R}^{m}}%
\right] }{\left( 4\pi \right) ^{\frac{m}{2}}\sqrt{\det \int_{0}^{+\infty
}e^{tB_{s}}e^{tB_{s}^{\ast }}dt}}\times \int_{\mathbb{R}^{m}}w(\eta )e^{-%
\frac{1}{4}||R_{s,\infty }\eta _{s}||_{\mathbb{R}^{m_{s}}}^{2}-\frac{1}{4}%
||R_{u,\infty }\eta _{u}||_{\mathbb{R}^{m_{u}}}^{2}}d\eta .
\end{eqnarray*}

\begin{eqnarray*}
z(x)=\frac{\exp -\left[ \frac{1}{4}\left( \int_{0}^{+\infty
}e^{tB_{s}}e^{tB_{s}^{\ast }}dt\right) ^{-1}x_{s},x_{s}>_{\mathbb{R}^{m}}%
\right] }{\left( 4\pi \right) ^{\frac{m}{2}}\sqrt{\det \int_{0}^{+\infty
}e^{tB_{s}}e^{tB_{s}^{\ast }}dt}}\times \int_{\mathbb{R}^{m}}w(\eta )e^{%
\frac{1}{2}<A_{s}\eta _{s},\eta _{s}>_{\mathbb{R}^{m_{s}}}^{2}-\frac{1}{4}%
<(Q_{u,\infty }^{-1}-2A_{u})\eta _{u},\eta _{u}>_{\mathbb{R}%
^{m_{u}}}^{2}}d\eta
\end{eqnarray*}

\bigskip

\noindent Let us summarize in the following theorem, the results obtained so far
\begin{theorem} \label{thcy}
 Let $(V_{m},g)$ be a Riemannian compact manifold, and $b$ be
a Morse-Smale vector field  the recurrent set $\mathcal{R}$ of which
is a finite number of points. Let $\mathcal{L}$ be a special
Lyapunov function. The limits $\mu$ of the probability measures
associated to the eigenfunction $u_{\epsilon }$ of $L_{\epsilon }$,
defined by
\begin{eqnarray*}
\frac{e^{-\mathcal{L}/\epsilon }u_{\epsilon }^{2}dV_{g}}{\int_{V_{m}}{e^{-%
\mathcal{L}/\epsilon }u_{\epsilon }^{2}dV_{g}}}
\end{eqnarray*}%
are of the form
\begin{eqnarray*}
\mu =\sum_{P\in S_{tp}}c_{P}\delta _{P},
\end{eqnarray*}
where the set $\mathcal{R}_{tp}$ is the subset of $%
\mathcal{R}$ where the topological pressure is achieved and $\sum_{P\in
\mathcal{R}_{tp}}c_{P}=1,c_{P}\geq 0$.

\begin{enumerate}
\item Wherever the topological pressure is achieved, the solution $w$ of the
blow up equation (\ref{bua}) satisfies,
\begin{eqnarray*}
\Delta _{E}w+(\Omega _{ij}x^{i},\nabla _{j}w)+[\Psi _{\mathcal{L}%
}(x)+(c+\Delta _{E}\mathcal{L}/2)(0)]w &=&\lambda w\hbox{ on }\hbox{\bb R}%
^{m} \\
0 &\leq &w\leq 1.
\end{eqnarray*}%
w is $C^{\infty }$.It attains its maximum and decays quadratically
exponentially fast at $\infty :$ there exists a constant $C>0$ such
that
\begin{eqnarray*}
\forall x\in \hbox{\bb R}^{m},\,w(x)\leq
e^{-C||x||_{\mathbb{R}^{m}}^{2}}.
\end{eqnarray*}

\item There exists a constant $C(b,g)$ such that
\begin{eqnarray*}
\lim_{\epsilon \rightarrow 0}\epsilon ^{m/2}\bar{v}_{\epsilon
}^{2}=C(b,g),
\end{eqnarray*}%
where $\bar{v}_{\epsilon }=\sup_{P\in V}v_{\epsilon }(P)$.

\item The coefficient $c_{P}$ can be computed from $C(b,g).$ The
function $w_{P}$ is equal to the blow up solution $w$ at the point P
and the concentration coefficient is computed using the modulating
factor $\gamma _{P}$ and is given
\begin{equation*}
c_{P}=C(b,g)\gamma _{P}^{2}\int_{\hbox{\bb R}^{m}}w_{P}^{2}.
\end{equation*}

\item Let us call $\mathcal{CR}$ the set of recurrent points, which are
charged ($\gamma _{P}>0$ for all $P\in \mathcal{CR}$), then
\begin{equation*}
C(b,g)=\frac{1}{\sum_{P\in \mathcal{CR}}\gamma _{P}^{2}\int_{\hbox{\bb R}^{m}}w_{P}^{2}}%
.
\end{equation*}
\end{enumerate}
\end{theorem}

{\textbf{Proof.}} The proof is a consequence of Theorem \ref{blowup}
and the previous lemmas. Using these results, the proofs of the
statements of theorem 3 follow the same steps as in theorem 4 of
\cite{HK1}. Anyway, the proof of the parts 2-3-4 will be given in
the section \ref{lcs} about limit cycles. \quad \QED

\section{Concentration near limit cycles \label{lcs}}

When the recurrent set of the MS vector field $\Omega $ contains
limit cycles, the limits of the probability measures associated to
the eigenfunctions can have components located on  the limit cycles.
Thus it is relevant to study the restrictions of the limit measures
to the cycles. We shall prove here that they are absolutely
continuous with respect to the length induced by the metric $g$
along the cycle.

Moreover, we will show that the limit measures are concentrated on
the cycles or critical points where the topological pressure is
attained. Actually if some limit cycles are charged, then no
critical point is charged. The reason for this is: Once a limit
cycle is charged, then the speed with which the maximum of the
eigenfunction tends to infinity when $\epsilon $ goes to zero is
determined by the local behavior near the cycle.
\subsection{Statement of the main results \ {\protect\Huge \ }}
This section is devoted to the proof of the second part of theorem
\ref{th1} and auxiliary propositions.
\begin{prop}\label{prop1}
Let $S$ be a limit cycle,  the restriction of any limit of the
eigenfunction measures to a sufficiently small tubular neighborhood
of $S$ is carried by $S$ and has the following form $c_S f_{\Gamma }\delta _{S}$, where $%
f_{\Gamma }:S\rightarrow \hbox{\bb R}$ is the unique normalized (maximum
equal one) periodic solution of
\begin{equation*}
\frac{\partial {f}}{\partial \theta }+c(S(\theta ))f=\lambda f(\theta )\text{
on }S
\end{equation*}%
where
\begin{equation*}
\lambda =\frac{1}{T_{S}}\int_{0}^{T_{S}}c(S(\theta ))d\theta .
\end{equation*}%
$c_S$  is the concentration coefficient, $S(\theta )$ the parametrization of a cycle and $T_{S}$ the minimal
period of $S$ .
\end{prop}
\textbf{Remarks.}

The next two propositions describe the behavior of the renormalized
sequence of eigenfunctions near the limit cycles. The notations are the same as
before, $sup_{V_{m}}{v}_{\epsilon }=v(P_{\epsilon })=\bar{v}_{\epsilon }$
and $P_{\epsilon }$ is a maximum point of ${v}_{\epsilon }$. Finally $%
\Delta _{E}^{m-1}$ denotes opposite the Laplacian operator in $\hbox{\bb R}%
^{m-1}$.

\bigskip

\begin{lem}
\begin{enumerate}
\item If S is charged, it contains limits of sequences $P_{\epsilon_n}$,
where $\epsilon_n$ tends to zero (see lemma 1).
\item Let $P_{\epsilon_n }, n \in \nN,$ where $\epsilon_n$ tends to zero,
be a sequence of maximum points ($v_{\epsilon_n }(P_{\epsilon_n })=
\sup_{V_n}v_{\epsilon_n })$, which converges to a point $P_{0} \in
S$. Then by Lemma 1, since $\frac{d(P_{\epsilon
},P_{0})}{\sqrt{\epsilon }}\leq Cte$, passing to a subsequence if
necessary, we assume that $\frac{d(P_{\epsilon
},P_{0})}{\sqrt{\epsilon }}$ converges to a finite limit and
$\frac{1}{\sqrt{\epsilon_n }}(\delta_{P_{\epsilon_n}}-\delta_{P_{0}}
)$ to $\Theta(T)$, where $T \in T_{P_0}(V)$ and $\Theta(T)$ the Lie
derivative associated to T.
\end{enumerate}
\end{lem}
The proof is a consequence of lemma \ref{lemfdt}.

\begin{prop}\label{BLOWUP}
Any sequence of $\epsilon$ converging to zero contains a sub-sequence $\epsilon_n$ such that
the blow-up sequence defined by
\begin{eqnarray*}
w_{\epsilon_n }(x^{\prime },\theta )=\frac{v_{\epsilon_n }(\sqrt{\varepsilon_n }%
x^{\prime },\theta )}{\overline{v_{\epsilon_n }}}
\end{eqnarray*}%
converges uniformly to a function $w$ on every compact of $\hbox{\bb R}%
^{m-1}\times \mathbb{R}$. $w$ is a periodic solution of period $T_{S}$ in $\theta $ of equation :%
\begin{eqnarray} \label{edpfft2}
\Delta _{E}^{m-1}w+\sum_{i,j=1}^{m-1}\Omega _{j}^{i}x^{j}\frac{\partial w}{%
\partial x^{i}}+\frac{\partial {w}}{\partial \theta }+(c(0,\theta )+\frac{%
\Delta _{g}\mathcal{L}(0,\theta )}{2}+\psi _{2}(x^{\prime }))w=\lambda w\,
\end{eqnarray}%
where $\psi _{2}(x^{\prime })\ $ are the terms of order 2 in the
Taylor expansion of the Lyapunov function $\mathcal{L}$ along $S$.
$1\geq w>0$, and the maximum 1 is attained. In the equation %
(\ref{edpfft2}), $\lambda >0$ equals the topological pressure at
$S$, that is
\begin{eqnarray*}
\lambda =<c_{o}>_{S}-TrB^{s}.
\end{eqnarray*}
\end{prop}
Finally, using the Orthogonality Assumption  with the same notations
as above, we have
\begin{prop}\label{prop3}
Under the orthogonal Assumptions along the cycle , in a coordinate
system defined in \ref{notat}-iii, $x=(x_{u},x_{s},\theta )$, where
$x_{u}$ represents the unstable direction, $x_{s}$ the stable
direction and $\theta $ a parametrization of the cycle, the solution
$w$ is given by
\begin{eqnarray*}
w(x)=z(x^{u},x^{s})f(\theta ),
\end{eqnarray*}%
where $z$ is solution of
\begin{eqnarray}
&&\Delta _{E}^{m-1}z+\sum_{i,j}A_{j}^{i}x^{j}\frac{\partial z}{\partial x^{i}}+[%
\frac{\Delta _{g}\mathcal{L}(P_{0})}{2}+\tilde{Q}(x^{\prime
})]z=\lambda z \hbox{ on } \rR^{n-1},
\notag  \label{edpfft2pp} \\
&&1\geq z>0,  \notag
\end{eqnarray}%
and $f$ is a periodic solution along the cycle of the equation
\begin{equation*}
\frac{\partial f}{\partial \theta }+c(0,\theta )f=<c_{0}>f,
\end{equation*}%
where $<c_{0}>$ is average of c along the cycle.
\end{prop}

Proposition \ref{thcf} given below gives the final information
necessary to characterize the limit measure on the cycle and the
value of the concentration coefficients.

\begin{theorem}\label{thcf}
\begin{enumerate}
\item  When a limit cycle of $\Omega $ is charged, for any charging sequence
there exists a subsequence, \{$\varepsilon _{n}|n\in \mathbb{N}$\} and a
strictly positive constant $C(\mu)>0$, such  that
\begin{equation}
\lim_{n\longrightarrow \infty }\epsilon_{n} ^{m/2-1}\bar{v}_{\epsilon
_{n}}^{2}=C(\mu).
\end{equation}%

\item Suppose that there exists a subsequence converging to a measure $\mu$ such that only the
cycles $S_{1},..,S_{p}$ are charged by this measure. If the cycles are of minimal period $T_{1},..T_{q}$, then
\begin{equation*}
C(\mu)=\frac{1}{\sum_{r=1}^{p}\gamma _{r}^{2}\int_{\hbox{\bb R}%
^{n-1}}z_{r}^{2}(x^{\prime })dx^{\prime
}\int_{0}^{T_{r}}f_{r}^{2}(\theta )d\theta },
\end{equation*}%
where $\gamma _{r}$ is the modulating coefficients.

\item The concentration coefficients $c_{S}$ along a charged cycle $S$
is given by:
\begin{equation*}
c_{S}=\lim_{n\longrightarrow \infty }\int_{T_{S}(\delta )}v_{{\epsilon_n}
}^{2}=C(\mu)\gamma _{S}^{2}\int_{0}^{T_{S}}f_{S}^{2}(\theta )d\theta
\int_{\hbox{\bb R}^{m-1}}z_{S}^{2}(x') dx',
\end{equation*}%
where $T_{S}(\delta )$ is a tubular neighborhood of the cycle,
$T_{S}$ the minimal period and the functions $f_{S},z_{S}$ come from
the blow-up analysis of the previous results (theorem \ref{thcf} and
proposition \ref{prop3}).
\end{enumerate}
\end{theorem}


\subsection{Proofs of the theorems.}

Recall that the normalized eigenfunction $v_{\epsilon }$ satisfies
\begin{eqnarray}
&&\epsilon ^{2}\Delta _{g}v_{\epsilon }+\epsilon (\Omega ,\nabla
v_{\epsilon })+c_{\epsilon }v_{\epsilon }=\epsilon \lambda
_{\epsilon }v_{\epsilon },
\label{recall} \\
&&\int_{V_{m}}v_{\epsilon }^{2}dV_{g}=1  \notag
\end{eqnarray}%

\noindent \textbf{Proof proposition \ref{BLOWUP}  and \ref{prop3}.}

\noindent Consider a coordinate system $x=(x^{\prime },\theta )$ on the
universal covering of a tubular neighborhood of the cycle $S$ as defined in (%
\ref{notat}) \textrm{III.} The blown-up function is defined by
\begin{equation*}
w_{\varepsilon }(x^{\prime },\theta )=\frac{v_{\epsilon }(\sqrt{\varepsilon }%
x^{\prime },\theta )}{\bar{v}_{\varepsilon }}.
\end{equation*}
\bigskip Consider the equation
\begin{eqnarray} \label{bloupequa}
\Delta _{E}^{m-1}w+\sum_{i,j=1}^{m-1}\Omega _{j}^{i}x^{j}\frac{\partial w}{%
\partial x^{i}}+\frac{\partial {w}}{\partial \theta }+(c(0,\theta )+\frac{%
\Delta _{g}\mathcal{L}(0,\theta )}{2}+\psi _{2}(x^{\prime }))w=\lambda w\,
\end{eqnarray}


\begin{lem}\label{lemma6}
Any sequence of $\epsilon$'s converging to zero contains a
sub-sequence $\epsilon_n$ such that the blown-up sequence
$w_{\epsilon_n}$ converges to a function $w \geq 0$ solution of
equation (\ref{bloupequa}) uniformly on any compact set $%
K\times S^1$.
\end{lem}

\noindent \textbf{Proof.} \noindent\ The blown-up metric is defined by%
\begin{equation*}
g_{\varepsilon}(x^{\prime },\theta )=g(\sqrt{\varepsilon }x^{\prime },\theta ).
\end{equation*}
Define $\Gamma _{\varepsilon ij}^{k}$ by $\Gamma _{\varepsilon
ij}^{k}(x^{\prime },\theta )=\Gamma _{ij}^{k}(\sqrt{\epsilon }x^{\prime
},\theta )$ where the $\Gamma _{ij}^{k}$ are the Christoffel symbols of the
metric $g$ in the coordinates ($x_{1},...,x_{m-1},\theta $). Note that the $%
\Gamma _{\varepsilon ij}^{k}$ are not the Christoffel symbols of the metric $%
g_{\varepsilon }.$ As $\varepsilon $ goes to zero, the sequence $g_{\epsilon
}$ converges uniformly to the metric $g_{E}$ =$%
\sum_{n=1}^{m-1}dx_{n}^{2}+g^{\theta \theta }(\theta )d\theta ^{2}$.

The sequence $w_{\epsilon }$ satisfies the equation%
\begin{equation}  \label{edp1b}
L_{0}w_{\epsilon }+\sqrt{\epsilon }L_{1}w_{\epsilon }=\lambda _{\epsilon
}w_{\epsilon }
\end{equation}
with
\begin{equation*}
L_{1}=D_{1}+\sqrt{\varepsilon }D_{2}
\end{equation*}
where:%
\begin{equation*}
L_{0}=\Delta _{E}^{m-1}+\frac{\partial }{\partial \theta }%
+\sum_{i,j=1}^{m-1}\Omega _{j}^{i}x^{j}\frac{\partial }{\partial x^{i}}%
+c(0,\theta )+\frac{TrA}{2}+<\left( \frac{B^{\ast }A+AB}{2}-A^{2}\right)
x^{\prime },x^{\prime }>_{\mathbb{R}^{m-1}}
\end{equation*}
$L_{0}$ is the limit of the blown-up operator
\begin{equation*}
\varepsilon \Delta _{g}v_{\epsilon }=\Delta _{g_{\epsilon
}}^{m-1}w_{\epsilon }+\sqrt{\varepsilon }D_{1}(w_{\epsilon })+\varepsilon
D_{2}(w_{\epsilon })
\end{equation*}%
where
\begin{equation*}
-\Delta _{g_{\epsilon }}^{m-1}:=\sum_{ij}^{m-1}g_{\varepsilon }^{ij}\frac{%
\partial ^{2}}{\partial x_{i}\partial x_{j}},
\end{equation*}
and
\begin{eqnarray*}
D_{1}:=-\sum_{k=1}^{m-1}\left( \sum_{ij}^{m-1}g_{\varepsilon
}^{ij}\Gamma _{\varepsilon ij}^{k}+g_{\varepsilon }^{\theta \theta
}\Gamma _{\varepsilon
\theta \theta }^{k}\right) \frac{\partial }{\partial x_{k}}%
+2\sum_{i=1}^{m-1}g_{\varepsilon }^{\theta i}(\frac{\partial ^{2}}{\partial
x_{i}\partial \theta }-\sum_{k=1}^{m-1}\Gamma _{\varepsilon \theta i}^{k}%
\frac{\partial }{\partial x_{k}})
\end{eqnarray*}
and
\begin{equation*}
D_{2}:=\left( g_{\varepsilon }^{\theta \theta }(\frac{\partial ^{2}}{%
\partial \theta \partial \theta }-\Gamma _{\varepsilon \theta \theta
}^{\theta }\frac{\partial }{\partial \theta })-\sum_{ij}^{m-1}g_{\epsilon
}^{ij}\Gamma _{\varepsilon ij}^{\theta }\frac{\partial }{\partial \theta }%
-2\sum_{i=1}^{m-1}g_{\varepsilon }^{\theta i}\Gamma _{\varepsilon \theta
i}^{\theta }\frac{\partial }{\partial \theta }\right).
\end{equation*}
Equation (\ref{edp1b}) is considered in the domain
$B^{m-1}(0,\frac{\delta }{\sqrt{\epsilon }}) \subset
\mathbb{R}^{m-1}\times S^1$ and $g_{\epsilon }$ converges to $g_{E}$
uniformly on each compact set, where
$g_{E}=\sum_{i=1}^{m-1}dx_{i}^{2}+g^{\theta \theta }(0,\theta
)d\theta ^{2}$. \bigskip

\textbf{Limit equation.} Any weak limit $w$ of $w_{\epsilon }$ when $%
\varepsilon $ goes to zero, satisfies the equation
\begin{eqnarray}
\Delta _{E}^{m-1}w+\frac{\partial w}{\partial \theta }+\sum_{i,j=1}^{m-1}%
\Omega _{j}^{i}x{}^{j}\frac{\partial w}{\partial x{}^{i}}+(c(0,\theta
)+TrA+<\left( \frac{B^{\ast }A+AB}{2}-A^{2}\right) x^{\prime },x^{\prime }>_{%
\mathbb{R}^{m-1}})w=\lambda w  \label{eeedp}
\end{eqnarray}
where $0\leq w\leq ess\sup w=1$ and $\underset{\varepsilon
\rightharpoondown 0}{\lim }\lambda_{\varepsilon }
=\underset{}{\lambda .\text{ }}$ Actually,  the sequence
$w_{\epsilon }$ converges in the $C^{\infty }$ topology on any
compact set of $\hbox{\bb R}^{m-1}\times S^{1}$, as  proved in the
appendix. \QED

 \bigskip

We now proceed  with the computation of an explicit expression of
the function solution  of equation (\ref{eeedp}). Let us introduce
the simplified notations $c_{0}(\theta )=c(0,\theta )$. We replace
the function $w(x^{\prime },\theta )$ by $\tilde{w}(x^{\prime
},\theta )$,
\begin{eqnarray*}
\tilde{w}(x^{\prime },\theta )=w(x^{\prime },\theta )\exp
\{\int_{0}^{\theta }\left[ c_{0}(t)-\frac{<c_{0}>}{T_{S}}\right]
dt\},
\end{eqnarray*}%
$\tilde{w}$ is bounded, periodic in $\theta $ and solution of
\begin{eqnarray*}
\Delta _{E}^{m-1}\tilde{w}+\sum_{i,j=1}^{m-1}\Omega
_{ij}x^{i}\frac{\partial
\tilde{w}}{\partial x_{i}}+\frac{\partial \tilde{w}}{\partial \theta }+(%
\frac{TrA}{2}+<Ax^{\prime },x^{\prime }>)\tilde{w}=[\lambda -<c_{0}>]\tilde{w}%
=\left( -TrB_{s}\right) \tilde{w}.
\end{eqnarray*}%
A final gauge transformation $\tilde{w}=ze^{-\frac{1}{2}<Ax^{\prime
},x^{\prime }>}$ leads to the equation
\begin{eqnarray*}
\Delta _{E}^{m-1}z+\sum_{i,j=1}^{m-1}(\Omega _{ij}+A_{ij})x_{j}\frac{%
\partial z}{\partial x_{i}}+\frac{\partial z}{\partial \theta }=\left(
-TrB_{s}\right) z.
\end{eqnarray*}%
Going back to the original vector field $b$, we have
\begin{eqnarray}
\Delta _{E}^{m-1}z+\sum_{i,j=1}^{m-1}B_{ij}x_{j}\frac{\partial
z}{\partial x_{i}}+\frac{\partial z}{\partial \theta }=\left(
-TrB^{s}\right) z. \label{eqz}
\end{eqnarray}%
\bigskip
\begin{lem}\label{lemako} The function z is given by Kolmogorov's formula:%
\begin{eqnarray}
z(x^{\prime },\theta )=\frac{e^{-\theta TrB^{s}}}{\left( 4\pi \right) ^{%
\frac{m-1}{2}}\sqrt{\det Q_{\theta }}}\int_{\mathbb{R}^{m-1}}\widetilde{w}%
(y,0)e^{-q(x^{\prime },y^{\prime },\theta )}dy  \label{kom}
\end{eqnarray}%
where $Q_{\theta }=\int_{0}^{\theta }e^{-tB}e^{-tB^{\ast }}dt$ and
\begin{eqnarray}
q(x^{\prime },y^{\prime },\theta )=\frac{1}{4}<Q_{\theta
}^{-1}(e^{-\theta
B}x^{\prime }-y^{\prime }),(e^{-\theta B}x^{\prime }-y^{\prime })>_{\mathbb{R%
}^{m-1}}-\frac{1}{2}<Ay^{\prime },y^{\prime }>_{\mathbb{R}^{m-1}}.
\end{eqnarray}
\end{lem}

{\noindent \textbf{Proof:}}
\begin{eqnarray}
z(x^{\prime },\theta )=\widetilde{w}(x^{\prime },\theta )\exp \left(
\frac{1}{2}<Ax^{\prime },x^{\prime }>_{\mathbb{R}^{m-1}}\right) .
\end{eqnarray}
Because $\widetilde{w}$ is bounded, for some constants
$C_{1}>0,C_{2}>0$, for all $x^{\prime }$,$\theta$,
\begin{eqnarray}
  |z(x^{\prime },\theta )|\leq C_{1}e^{C_{2}||x^{\prime }||_{\mathbb{R}%
^{m-1}}^{2}},
\end{eqnarray}
Thus $z$ belongs to the Tikhonov class and hence is entirely
determined by its value for a given
fixed $\theta $ by Tikhonov's theorem. Let us introduce the function:%
\begin{eqnarray}
v(x^{\prime },\theta )=\frac{e^{-\theta TrB_{s}}}{\left( 4\pi \right) ^{%
\frac{m-1}{2}}\sqrt{\det Q_{\theta }}}\int_{\mathbb{R}^{m-1}}\widetilde{w}%
(y^{\prime },0)e^{-q(x^{\prime },y^{\prime },\theta )}dy^{\prime }.
\label{kolm}
\end{eqnarray}
Lemma \ref{lemma8} completes the proof of lemma  \ref{lemako} and
shows that $v=z$.

\begin{lem} \label{lemma8}
\begin{enumerate}
\item $v$ is well defined.
\item $v$ is a solution of equation (\ref{eqz})
\item  $ v(x^{\prime },0)=z(x^{\prime },0)$
\item  $v$ belongs to Tikhonov's class.
\end{enumerate}
\end{lem}

{\noindent \textbf{Proof. }}Let us estimate $q(x,y,\theta ).$ We
have the decomposition $\mathbb{R}^{m}=W^{s}\times W^u \times \rR$,
 where $W^s$ and $W^u$ are the stable and unstable space respectively.
If $x'\in \rR^{n-1}$, we denote by $x_s$ (resp. $x_u$) the stable
(resp unstable) components. To this decomposition of
$\mathbb{R}^{m}$ corresponds the splitting of matrices:

$B=$\ $%
\begin{vmatrix}
B_{s} & 0 & 0 \\
0 & B_{u} & 0 \\
0 & 0 & 0%
\end{vmatrix}%
$ $Q_{\theta }=\bigskip \left\vert
\begin{array}{ccc}
Q_{s,\theta } & 0 & 0 \\
0 & Q_{u,\theta } & 0 \\
0 & 0 & 0%
\end{array}%
\right\vert $ $R_{\theta }=\left\vert
\begin{array}{ccc}
R_{s,\theta } & 0 & 0 \\
0 & R_{u,\theta } & 0 \\
0 & 0 & 0%
\end{array}%
\right\vert $

 $P_{\theta }=\left\vert
\begin{array}{ccc}
P_{s,\theta } & 0 & 0 \\
0 & P_{u,\theta } & 0 \\
0 & 0 & 0%
\end{array}%
\right\vert $ \ \ \ \ \ $U_{\theta }=\left\vert
\begin{array}{ccc}
U_{s,\theta } & 0 & 0 \\
0 & U_{u,\theta } & 0 \\
0 & 0 & 0%
\end{array}%
\right\vert $ \ \ \ \ \ \ $A_{\theta }=\left\vert
\begin{array}{ccc}
A_{s,\theta } & 0 & 0 \\
0 & A_{u,\theta } & 0 \\
0 & 0 & 0%
\end{array}%
\right\vert ,$ where
\begin{eqnarray}
Q_{s,\theta }=\int_{0}^{\theta }e^{-tB_{s}}e^{-tB_{s}^{\ast
}}dt,Q_{u,\theta }=\int_{0}^{\theta }e^{-tB_{u}}e^{-tB_{u}^{\ast
}}dt,A_{s}^{-1}=-\int_{0}^{+\infty }e^{tB_{s}}\Pi
_{s}e^{tB_{s}^{\ast }}dt,
\end{eqnarray}

\begin{eqnarray}
A_{u}^{-1}=\int_{0}^{+\infty }e^{-tB_{u}}\Pi _{u}e^{-tB_{u}^{\ast
}}dt, \end{eqnarray} and $\Pi _{s}$, $\Pi _{u}$ are
positive-definite and $>>2Id_{s}$ (resp. 2Id$_{u}$) in the natural
order on the symmetric operators. Hence $A_{u}^{-1}>>2Q_{u,\theta
}.$ This implies that $Q_{u,\theta }^{-1}-2A_{u}$ is
positive-definite and obviously so is $Q_{s,\theta }^{-1}-2A_{s}.$ Let $%
R_{s,\theta }$ and $R_{u,\theta }$ be the unique positive-definite symmetric
operators such that $R_{s,\theta }^{2}=Q_{s,\theta }^{-1}-2A_{s}$ and $%
R_{u,\theta }^{2}=Q_{u,\theta }^{-1}-2A_{u}.$ We now need the
following operators to express q. Define

\begin{equation*} P_{s,\theta
}=R_{s,\theta }^{-1}Q_{s,\theta }^{-1}e^{-\theta B_{s}},P_{u,\theta
}=R_{u,\theta }^{-1}Q_{u,\theta }^{-1}e^{-\theta B_{u}}.
\end{equation*}
Then
\begin{eqnarray}
q(x,y,\theta ) &=&\frac{1}{4}\left[ <U_{s,\theta }x_{s},x_{s}>_{\mathbb{R}%
^{m-1}}+<U_{u,\theta }x_{u},x_{u}>_{\mathbb{R}^{m-1}}\right.  \label{kol} \\
&&+\left. ||R_{s,\theta }y_{s}-P_{s,\theta }x_{s}||_{\mathbb{R}%
^{m-1}}^{2}+||R_{u,\theta }y_{u}-P_{u,\theta }y_{u}||_{\mathbb{R}^{m-1}}^{2}%
\right],  \notag
\end{eqnarray}
where
\begin{eqnarray*} U_{s,\theta }=e^{-\theta B_{s}^{\ast }}\left(
Q_{s,\theta }^{-1}\ -Q_{s,\theta }^{-1}R_{s,\theta }^{-2}Q_{s,\theta
}^{-1}\right) e^{-\theta B_{s}},U_{u,\theta }=e^{-\theta B_{u}^{\ast
}}\left( Q_{u,\theta }^{-1}\ -Q_{u,\theta }^{-1}R_{u,\theta
}^{-2}Q_{u,\theta }^{-1}\right) e^{-\theta B_{u}},
\end{eqnarray*}
then
\begin{eqnarray*}
U_{s,\theta }=e^{-\theta B_{s}^{\ast }}Q_{s,\theta }^{-1}\left(
Q_{s,\theta }\ -R_{s,\theta }^{-2}\right) Q_{s,\theta
}^{-1}e^{-\theta B_{s}}.
\end{eqnarray*}
Because $R_{s,\theta }^{2}>>Q_{s,\theta }^{-1},$ $Q_{s,\theta
}>>R_{s,\theta
}^{-2}$ and $U_{s,\theta }$ is positive definite. On the other hand%
\begin{eqnarray*}
U_{u,\theta }=e^{-\theta B_{u}^{\ast }}Q_{u,\theta }^{-1}\left(
Q_{u,\theta }\ -R_{u,\theta }^{-2}\right) Q_{u,\theta
}^{-1}e^{-\theta B_{u}}.
\end{eqnarray*}
Because $2A_{u}$ is positive definite, $Q_{u,\theta
}^{-1}>>R_{u,\theta }^{2} $ and hence $R_{u,\theta
}^{-2}>>Q_{u,\theta }.$ So $U_{u,\theta }$ is negative definite.

Relation (\ref{kol}) shows that $v$ is well defined. Making the change of
variables $y^{\prime}\rightarrow \eta^{\prime},$ $\eta^{\prime}=y^{%
\prime}-R_{\theta }^{-1}P_{\theta }x^{\prime}$ in the integral of equation (%
\ref{kolm}) we get that%
\begin{eqnarray*}
v(x^{\prime },\theta ) &=&\frac{\exp -\left[ \theta TrB_{s}+\frac{1}{4}%
<U_{\theta }x^{\prime },x^{\prime }>_{\mathbb{R}^{m-1}}\right] }{\left( 4\pi
\right) ^{\frac{m-1}{2}}\sqrt{\det Q_{\theta }}}\times \\
&&\int_{\mathbb{R}^{m-1}}\widetilde{w}(\left( \eta ^{\prime
}+R_{\theta }^{-1}P_{\theta }x^{\prime }\right)
,0)e^{-\frac{1}{4}||R_{s,\theta }\eta _{s}^{\prime
}||_{\mathbb{R}^{m-1}}^{2}-\frac{1}{4}||R_{u,\theta }\eta
_{u}^{\prime }||_{\mathbb{R}^{m-1}}^{2}}d\eta ^{\prime }.
\end{eqnarray*}
Because $\widetilde{w}$ is bounded , $v$ is in Tikhonov's class. Now
it can be checked that $v$ satisfies equation (\ref{eqz}). Let us
show that
\begin{equation*}
\underset{\theta \rightarrow 0}{\lim }v(x^{\prime },\theta
)=z(x^{\prime },0).
\end{equation*}
But first let us state a lemma:
\begin{lem}\label{inequ2}
The operators just defined have the following
properties:

(i)For small $\theta >0:$

\begin{enumerate}
\item $Q_{s,\theta }=\theta \left[ Id_{s}-\theta (\frac{B_{s}+B_{s}^{\ast }}{2}%
)+O(\theta ^{2})\right] $ and $Q_{s,\theta }^{-1}=\frac{1}{\theta }Id_{s}+%
\frac{B_{s}+B_{s}^{\ast }}{2}+O(\theta ).$

\item $Q_{u,\theta }=\theta\left[ Id_{u}-\theta (\frac{B_{u}+B_{u}^{\ast }}{2}%
)+O(\theta ^{2})\right] \ $and\ $Q_{u,\theta }^{-1}=\frac{1}{\theta }Id_{u}+%
\frac{B_{u}+B_{u}^{\ast }}{2}+O(\theta ).$

\item $R_{s,\theta }^{2}=\frac{1}{\theta }Id_{s}+\frac{B_{s}+B_{s}^{\ast }}{2}%
-2A_{s}+O(\theta )$ and $R_{s,\theta }^{-2}=\theta \left[ Id_{s}-\theta (%
\frac{B_{s}+B_{s}^{\ast }}{2}-2A_{s})+O(\theta ^{2})\right]. $

\item $R_{u,\theta }^{2}=\frac{1}{\theta }Id_{u}+\frac{B_{u}+B_{u}^{\ast }}{2}%
-2A_{u}+O(\theta )$ and \ $R_{u,\theta }^{-2}=\theta \left[ Id_{u}-\theta (%
\frac{B_{u}+B_{u}^{\ast }}{2}-2A_{u})+O(\theta ^{2})\right]. $

\item $U_{s,\theta }=-2A_{s}+O(\theta )$ and $U_{u,\theta
}=-2A_{u}+O(\theta ).$

\item $R_{s,\theta }^{-2}Q_{s,\theta }^{-1}e^{-\theta B_{s}}=Id_{s}+\
O(\theta )$ and$\ R_{u,\theta }^{-2}Q_{u,\theta }^{-1}e^{-\theta
B_{u}}=Id_{u}+O(\theta ). $

\item $\det Q_{\theta }=\theta ^{m-1}(1+O(\theta )).$
\end{enumerate}

(ii) When $\theta \rightarrow +\infty : $

 1)$Q_{s,\theta
}\rightarrow \infty $ and $Q_{u,\theta }\rightarrow
\int_{0}^{+\infty }e^{-sB_{u}}e^{-sB_{u}^{\ast }}ds$

2)$R_{s,\theta }\rightarrow \sqrt{-2A_{s}}$ and $R_{u,\theta }\rightarrow
R_{u,\infty }=\sqrt{Q_{u,\infty }^{-1}-2A_{u}}>>0$

3)$Q_{s,\theta }^{-1}e^{-\theta B_{s}}\rightarrow 0$, $P_{s,\theta
}=$ $R_{s,\theta }^{-2}Q_{s,\theta }^{-1}e^{-\theta
B_{s}}\rightarrow 0,U_{s,\theta }\rightarrow \left(
\int_{0}^{+\infty }e^{\tau B_{s}}e^{\tau B_{s}^{\ast }}d\tau \right)
^{-1}$

4)$Q_{u,\theta }\rightarrow Q_{u,\infty }=\int_{0}^{+\infty
}e^{-tB_{u}}e^{-tB_{u}^{\ast }}dt$ and \ hence P$_{u,\theta }=$ $R_{u,\theta
}^{-2}Q_{u,\theta }^{-1}e^{-\theta B_{u}}\rightarrow 0,U_{u,\theta
}\rightarrow 0.$

(iii) $e^{2\theta TrB_{s}}\det Q_{s,\theta }\rightarrow \det
\int_{0}^{+\infty }e^{\tau B_{s}}e^{\tau B_{s}^{\ast }}d\tau .$
\bigskip
\end{lem}

\textbf{Proof.} For small
\begin{eqnarray*}
\theta ,Q_{s,\theta }=\theta \left[ 1-\theta (%
\frac{B_{s}+B_{s}^{\ast }}{2})+O(\theta ^{2})\right] ,Q_{s,\theta }^{-1}=%
\frac{1}{\theta }+\frac{B_{s}+B_{s}^{\ast }}{2}+O(\theta ).
\end{eqnarray*}
Similarly $Q_{u,\theta }^{-1}=\frac{1}{\theta }+\frac{B_{u}+B_{u}^{\ast }}{2}%
+O(\theta ).$ From these relations it follows that:

\begin{eqnarray*}R_{s,\theta }^{2}=\frac{1}{\theta }+\frac{B_{s}+B_{s}^{\ast }}{2}%
-2A_{s}+O(\theta ),R_{u,\theta }^{2}=\frac{1}{\theta }+\frac{%
B_{u}+B_{u}^{\ast }}{2}-2A_{u}+O(\theta ),
\end{eqnarray*}
\begin{eqnarray*}
R_{s,\theta }^{-2}=\theta \left[ 1-\theta (\frac{B_{s}+B_{s}^{\ast }}{2}%
-2A_{s})+O(\theta ^{2})\right] ,R_{u,\theta }^{-2}=\theta \left[ 1-\theta (%
\frac{B_{u}+B_{u}^{\ast }}{2}-2A_{u})+O(\theta ^{2})\right]
\end{eqnarray*}
Then:
\begin{eqnarray*}
U_{s,\theta }=-2A_{s}+O(\theta ),U_{u,\theta }=-2A_{u}+O(\theta ).
\end{eqnarray*}%
Also:
\begin{eqnarray*}
R_{s,\theta }^{-2}Q_{s,\theta }^{-1}e^{-\theta B_{s}}=1+\ O(\theta
),\ R_{u,\theta }^{-2}Q_{u,\theta }^{-1}e^{-\theta B_{u}}=1+O(\theta
)
\end{eqnarray*}%
and%
\begin{eqnarray*}
\det Q_{\theta }=\theta ^{m-1}(1+O(\theta )).
\end{eqnarray*}

As $\theta \rightarrow +\infty ,Q_{s,\theta }=\int_{0}^{\theta
}e^{-tB_{s}}e^{-tB_{s}^{\ast }}dt\rightarrow +\infty .$ Hence $Q_{s,\theta
}^{-1}\rightarrow 0$ , $R_{s,\theta }^{2}\rightarrow -2A_{s}$ and \ $%
R_{s,\theta }\rightarrow \sqrt{-2A_{s}}.$ We have:%
\begin{eqnarray*}
e^{\theta B_{s}}Q_{s,\theta }=\left( \int_{0}^{\theta }e^{(\theta
-t)B_{s}}e^{(\theta -t)B_{s}^{\ast }}dt\right) e^{-\theta
B_{s}^{\ast }}
\end{eqnarray*}%
\begin{eqnarray*}
Q_{s,\theta }^{-1}e^{-\theta B_{s}}=e^{\theta B_{s}^{\ast }}\left(
\int_{0}^{\theta }e^{tB_{s}}e^{tB_{s}^{\ast }}dt\right) ^{-1}.
\end{eqnarray*}%
It follows that $Q_{s,\theta }^{-1}e^{-\theta B_{s}}\rightarrow 0$ as $%
\theta \rightarrow +\infty $. Also $R_{s,\theta }^{-2}Q_{s,\theta
}^{-1}e^{-\theta B_{s}}\rightarrow 0$ and $R_{s,\theta
}^{-1}Q_{s,\theta }^{-1}e^{-\theta B_{s}}\rightarrow 0$. It  follows
that $U_{s,\theta }\rightarrow \left( \int_{0}^{+\infty
}e^{tB_{s}}e^{tB_{s}^{\ast }}dt\right) ^{-1}.$ As $\theta
\rightarrow $\ +$\infty ,$ $Q_{u,\theta }\rightarrow
Q_{u,\infty }=\int_{0}^{+\infty }e^{-tB_{u}}e^{-tB_{u}^{\ast }}dt$ and $%
R_{u,\theta }^{2}\rightarrow R_{u,\infty }^{2}=Q_{u,\infty }^{-1}-2A_{u}.$ \
Hence $R_{u,\theta }^{-2}Q_{u,\theta }^{-1}e^{-\theta B_{u}}\rightarrow 0.$
Because $U_{u,\theta }=e^{-\theta B_{u}^{\ast }}(Q_{u,\theta }^{-1}\
-Q_{u,\theta }^{-1}R_{u,\theta }^{-2}Q_{u,\theta }^{-1})e^{-\theta
B_{u}},U_{u,\theta }\rightarrow 0.$ \bigskip

Finally
\begin{eqnarray*} e^{\theta B_{s}}Q_{s,\theta }e^{\theta
B_{s}^{\ast }}=\int_{0}^{\theta }e^{(\theta -t)B_{s}}e^{(\theta
-t)B_{s}^{\ast }}dt=\int_{0}^{\theta }e^{tB_{s}}e^{tB_{s}^{\ast }}dt
\end{eqnarray*}
and
\begin{eqnarray*} e^{\theta B_{s}}Q_{s,\theta }e^{\theta
B_{s}^{\ast }}\rightarrow \int_{0}^{+\infty
}e^{tB_{s}}e^{tB_{s}^{\ast }}dt\\\det e^{\theta B_{s}}\det
Q_{s,\theta }\det e^{\theta B_{s}^{\ast }}\rightarrow \det
\int_{0}^{+\infty }e^{tB_{s}}e^{tB_{s}^{\ast }}dt.
\end{eqnarray*}
But $\det e^{\theta B_{s}}=\det e^{\theta B_{s}^{\ast }}= e^{\theta
TrB_{s}}.$ So
\begin{eqnarray*}
 e^{2\theta TrB_{s}}\det Q_{s,\theta }\rightarrow \det
\int_{0}^{+\infty }e^{tB_{s}}e^{tB_{s}^{\ast }}dt,
\end{eqnarray*}
as $\theta \rightarrow +\infty $. This implies that

\begin{eqnarray*}
\underset{\theta \rightarrow 0}{\lim }v(x^{\prime
},\theta )=\widetilde{w}(x^{\prime },0)\exp -\frac{1}{2}<Ax,x>_{\mathbb{R}%
^{m-1}}=z(x^{\prime },0).
\end{eqnarray*}

\begin{lem} \label{repres}
The solution z does not depend on the unstable variable $%
x_{u}$ and is given explicitly by
\begin{eqnarray*}
z(x^{\prime },\theta )=C\exp \{-\frac{1}{4}<\left( \int_{0}^{+\infty
}e^{tB_{s}}e^{tB_{s}^{\ast }}dt\right) ^{-1}x_{s},x_{s}>\}.
\end{eqnarray*}
\end{lem}
\noindent {\textbf{Proof.}} We have:%
\begin{eqnarray*}
v(x^{\prime },\theta ) &=&\frac{\exp -\left[ \theta TrB_{s}+\frac{1+O(\theta
)}{2}<Ax^{\prime },x^{\prime }>_{\mathbb{R}^{m-1}}\right] }{\left( 4\pi
\theta \right) ^{\frac{m-1}{2}}(1+O(\theta ))} \\
&&\int_{\mathbb{R}^{m-1}}\widetilde{w}(\eta +(1+O(\theta ))x^{\prime },0)e^{-%
\frac{1+O(\theta )}{2\theta }\left( ||\eta _{s}||_{\mathbb{R}%
^{m_{s}}}^{2}+||\eta _{u}||_{\mathbb{R}^{m_{u}}}^{2}\right) }d\eta
^{\prime }.
\end{eqnarray*}
Because z is periodic, for any period $\overline{\theta },$
Kolmogorov's formula (\ref{kolm}) gives:%
\begin{eqnarray}
z(x^{\prime },0) &=&\frac{\exp -\left[  \,\overline{\theta }TrB_{s}+\frac{1}{4}%
<U_{\overline{\theta }}x^{\prime },x^{\prime }>_{\mathbb{R}^{m-1}}\right] }{%
\left( 4\pi \right) ^{\frac{m-1}{2}}\sqrt{\det Q_{\overline{\theta }}}}\times
  \label{KKOlm} \\
&&\int_{\mathbb{R}^{m-1}}\widetilde{w}(\left( \eta ^{\prime }+R_{\overline{%
\theta }}^{-1}P_{\overline{\theta }}x^{\prime }\right) ,0)e^{-\frac{1}{4}%
||R_{s,\overline{\theta }}\eta _{s}^{\prime }||_{\mathbb{R}^{m-1}}^{2}-\frac{%
1}{4}||R_{u,\overline{\theta }}\eta _{u}^{\prime }||_{\mathbb{R}%
^{m-1}}^{2}}d\eta ^{\prime }\notag
\end{eqnarray}
Letting $\overline{\theta }\rightarrow +\infty ,$ in the expression
above and using the estimates of Lemma (\ref{inequ2}), we see that
\begin{eqnarray}z(x^{\prime },0)=\frac{\exp -\left[ \frac{1}{4}\left(
\int_{0}^{+\infty }e^{tB_{s}}e^{tB_{s}^{\ast }}dt\right)
^{-1}x_{s}^{\prime },x_{s}^{\prime
}>_{\mathbb{R}^{m-1}}\right] }{\left( 4\pi \right) ^{\frac{m-1}{2}}\sqrt{%
\det \int_{0}^{+\infty }e^{tB_{s}}e^{tB_{s}^{\ast }}dt}} \nonumber \\
\times \int_{\mathbb{%
R}^{m-1}}\widetilde{w}(\eta ^{\prime
},0)e^{-\frac{1}{4}||R_{s,\infty }\eta _{s}^{\prime
}||_{\mathbb{R}^{m-1}}^{2}-\frac{1}{4}||R_{u,\infty }\eta
_{u}^{\prime }||_{\mathbb{R}^{m-1}}^{2}}d\eta ^{\prime
}.\label{final}
\end{eqnarray}
Now it is easy to check that the function on the right hand side of
(\ref{final}) is a solution of equation (\ref{eqz}). By
Tikhonov's theorem again, this function coincides with z. \quad%
\hbox{\hskip 4pt\vrule
width 5pt height 6pt depth 1.5pt}

\begin{coro} \label{scind}
On a cycle S, we have the following decomposition
\begin{eqnarray}
w_S(x',\theta) = z_S(x') f_S(\theta),
\end{eqnarray}
where
\begin{eqnarray}
z_S(x') &=& Ce^{-\frac{1}{2}<Ax^{\prime },x^{\prime }>}\exp
\{-\frac{1}{4}<\left( \int_{0}^{+\infty
}e^{tB_{s}}e^{tB_{s}^{\ast }}dt\right) ^{-1}x_{s},x_{s}>\}\\
f_S(\theta) &=&\exp\{-\int_{0}^{\theta }\left[
c_{0}(t)-\frac{<c_{0}>}{T_{S}}\right] dt\},
\end{eqnarray}
where C is the normalization constant such that $\sup w_S=1.$
\end{coro}


\subsection{Decay estimate of the blown-up function near recurrent
sets}\label{decyyv}

To compute the concentration coefficients and to study the
convergence properties of the blown-up sequence $w_{\epsilon }$, we
use the Feynman-Kac formula to compute an asymptotic estimate  of
$w_{\epsilon }$. Then we show that $w_{\epsilon }$ converges
strongly in $L^{2}$ to its weak limits.

In particular we prove that the sequence $w_{\epsilon }$ decays
exponentially in the transverse direction of the recurrent set, that
can be a critical point, a limit cycle $S$ or a torus, satisfying
the assumptions of paragraph \ref{notat}. The difficulty in
obtaining such estimates is cause by the fact that the vector field
is not a gradient. In the case of a limit cycle or a torus, the
field has two orthogonal components: first a component along the
recurrent set (which is zero in the case of a point) and second a
transversal one. We shall prove in a coordinate system $(x^{\prime
},\theta )$ defined in a tubular neighborhood $T^{S}$ of $S$,
defined in \ref{notat},  there exist constants $C>0$ and $C_{0}>0,$
such that
\begin{eqnarray*}
w_{\epsilon }(x^{\prime },\theta )=\frac{v_{\epsilon }(\epsilon ^{1/2}x^{\prime },\theta )}{\bar{v_{\epsilon }}}%
\leq Ce^{-C_{0}dist_{g}(x,\mathcal{S})^{2}}\text{ for
}x=(x',\theta)\in T^{S}.
\end{eqnarray*}
The proof involves several steps:
\begin{itemize}
\item An upper bound of the Fokker-Planck solution (see appendix II). This entails
 an estimate of the rate function of a Kac-Feynman integral, representing the
solution.
\item An explicit lower estimate of the rate function $I_{t}(x)$ (see definition
below) in term of the distance of x to the limit cycle.
\end{itemize}


\subsubsection{Optimal trajectories of an auxiliary variational problem} \label{kupka1}
Here we present computations of the optimal trajectories associated
with the rate function $I_{t}(x)$ defined by
\begin{eqnarray}
I_{t}(x)=\inf_{\Gamma _{x,t}}\int_{0}^{t}\left[ \frac{1}{2}||\dot{\gamma}%
(s)+\Omega (\gamma (s))||_{g}^{2}+\Psi _{\mathcal{L}}(\gamma
(s))\right] ds, \label{lisup1}
\end{eqnarray}%
where
\begin{eqnarray}
\Gamma _{x,t}=\{\gamma \in H^{1}([0,t];V)|\gamma (0)=x\,\}.
\end{eqnarray}%
Consider the following functional, defined on $\Gamma _{x,t}$:
\begin{eqnarray*}
I(\gamma )=\int_{0}^{t}\left[ \frac{1}{2}||\dot{\gamma}(s)+\Omega
(\gamma (s))||_{g_{\gamma (s)}}^{2}+\Psi _{\mathcal{L}}(\gamma
(s))\right] ds.
\end{eqnarray*}%
Then for all $x\in V,t>0$:
\begin{eqnarray} \label{Varia}
I_{t}(x)=\inf_{\Gamma _{t,x}}I(\gamma ).
\end{eqnarray}
We state the variational problem in the Hamiltonian formalism.
The associated Hamiltonian function $\mathcal{H}%
:T^{\ast }V\rightarrow \mathbb{R},$ is:
\begin{eqnarray*}
\mathcal{H}(z)=-<\Omega (\pi _{T^{\ast
}V}(z)),z>+\frac{1}{2}||z||_{g^{\ast }}^{2}-\Psi _{\mathcal{L}}(\pi
_{T^{\ast }V}(z)).
\end{eqnarray*}
A curve $z$ is an extremal for the minimization problem
(\ref{Varia}) if it satisfies the first order condition for an
optimum \cite{Giaquinta}. The Hamiltonian function $\mathcal{H}%
:T^{\ast }V\rightarrow \mathbb{R},$ associated to the problem is:
\begin{eqnarray*}
\mathcal{H}(z)=-<\Omega (\pi _{T^{\ast
}V}(z)),z>+\frac{1}{2}||z||_{g^{\ast }}^{2}-\Psi _{\mathcal{L}}(\pi
_{T^{\ast }V}(z)).
\end{eqnarray*}

This can also be expressed as%
\begin{eqnarray*}
\mathcal{H}(z)=\frac{1}{2}||z-\omega (\pi _{T^{\ast }V}(z)||_{g^{\ast }}^{2}-%
\frac{1}{2}||\omega (\pi _{T^{\ast }V}(z)||_{g^{\ast }}^{2}-\Psi _{\mathcal{L%
}}(\pi _{T^{\ast }V}(z)),
\end{eqnarray*}
where $\omega $ is the 1- form $G(\Omega )$ associated to the vector field $%
\Omega $ ($G:TV\rightarrow T^{\ast }V$ is the Legendre transform
associated to the metric $g$). For any trajectory of the Hamiltonian
field $\overrightarrow{\mathcal{H}}$ of $\mathcal{H},$
$z:J\rightarrow T^{\ast }V,J $ open interval, the function $t\in
J\rightarrow \mathcal{H}(z(t)\in
\mathbb{R},$ is constant. Hence for any $\tau \in J:$%
\begin{eqnarray} \label{use}
||z(t)-\omega (\pi _{T^{\ast }V}(z(t))||_{g^{\ast }}^{2}=||\omega
(\pi _{T^{\ast }V}(z(t))||_{g^{\ast }}^{2}+2\Psi _{\mathcal{L}}(\pi
_{T^{\ast }V}(z(t))+2\mathcal{H}(z(\tau )).
\end{eqnarray}%
Let us call $\digamma _{t}$ the flow of
$\overrightarrow{\mathcal{H}}$. In a classical Darboux coordinate
system $z=(x, p)$,  $\digamma _{t}(0,q)=(x^1,..,x^n ,p^1,..,p^n )$,
these functions satisfy the equations: for $ 1\leq n\leq m,$
\begin{eqnarray}
\frac{d x^{n}}{d t}=\frac{\partial \mathcal{H}}{\partial p_{n}}(x ,p
)=-\Omega^{n}(x )+\sum\limits_{k=1}^{m}g^{kn}(x)p_{k}, \label{equa1}
\end{eqnarray}

\begin{eqnarray}
-\frac{d p_{n}}{d t}=\frac{\partial \mathcal{H}}{\partial x_{n}}(x
,p )=
-\sum_{k=1}^{m}p_{k}\frac{\partial \Omega^{k}(x )}{%
\partial x_{n}}-\frac{\partial \Psi _{\mathcal{L}}}{\partial x_{n}}(x )+%
\frac{1}{2}\sum\limits_{i,j=1}^{m}\frac{\partial g^{ij}( x )}{%
\partial x_{n}}p_{i}p_{j},   \label{equa2}
\end{eqnarray}
and $t\rightarrow \digamma _{t}(0,q)$ is an extremal of the
variational problem iff it satisfies the
following boundary condition:%
\begin{eqnarray}\label{equa3bb}
p (T;(0,q))=0.
\end{eqnarray}

\begin{lem} \label{lemnew}
\begin{enumerate}
\item For any  trajectory $z:J\rightarrow T^*(V)$ of $\overrightarrow{\mathcal{H}}$,
and  any $t,\tau \in J$,
\begin{eqnarray*}
||z(t)||^2_{g*} &\leq& K_0 + 4 \mathcal{H}(z(\tau)) \\
||\frac{d\gamma}{dt}(t)||^2_{g} &\leq& K_1 + 4 \mathcal{H}(z(\tau)),
\end{eqnarray*}
where
\begin{eqnarray*}
\gamma &=& \pi_{T^*(V)}(z),\\
 K_0 &=&4\sup_V ||\Omega||^2_{g} +2\sup_V
||\Psi_{\mathcal{L}}||^2_{g}\\
K_1 &=&K_0+2\sup_V ||\Omega||^2_{g}.
\end{eqnarray*}

\item For any $t,\tau \in J$,
\begin{eqnarray*}
|| \overrightarrow{\mathcal{H}}(t)||_g \leq K_2 +K_3
{\mathcal{H}}(z(\tau)).
\end{eqnarray*}
\end{enumerate}
where $K_2,K_3$ depend only on $g, \Omega, \Psi_{\mathcal{L}}$.
\end{lem}
{\noindent \textbf{Proof.}}

The first inequality of the lemma is a consequence of
 expression (\ref{use}) and
\begin{eqnarray}
||z(t)||_{g^{\ast }}^{2}\leq 2||z(t)-\omega(\gamma(t)||_{g^{\ast
}}^{2} +2||\omega(\gamma(t)||_{g^{\ast }}^{2}.
\end{eqnarray}
The second inequality follows from
\begin{eqnarray}
\frac{T\gamma }{dt}=-\Omega (\gamma )+G^{-1}\circ z  \label{p}.
\end{eqnarray}%
Statement 2 follows from the relation (41)-(42) and statement 1 of
the lemma.

\begin{coro}
The field $\overrightarrow{\mathcal{H}}$ is complete.
\end{coro}
{\noindent \textbf{Proof.}}

This follows from the fact that the $\{{\mathcal{H}}\leq R^2\}$ is
compact and invariant by the field $\overrightarrow{\mathcal{H}}$.
\QED
\bigskip

 We  take a $q\in V$ and consider a
geodesic system of coordinates at $q,(U,x^{1},...,x^{m}),U$ being a
geodesic open ball centered at $q$ and of radius $\rho $ such that
$\overline{U}$ , the closure of  $U$ is contained in $V-Cut(x)$.
Because V is compact, we can assume that $\rho$ is independent of
$q$. Let $(T^{\ast
}U,x^{1},...,x^{m},p_{1},...,p_{m})$ be the associated Darboux chart of $%
T^{\ast }V$ . Let us consider the domain
\begin{eqnarray*}
\mathcal{D}_R =\{ z\in T^*(V) | \mathcal{H}(z)\leq R^2 \}.
\end{eqnarray*}
We have
\begin{lem} \label{lemmaest}
\begin{enumerate}
\item For $T\geq 0$ such that $T(K_0+4R^2)<\rho$, then for any $p\in
T_x^*(V)$, for any $t \in[0,T]$,
\begin{eqnarray*}
\digamma _{t}(p) \subset T^*(U).
\end{eqnarray*}
\item There exists a constant $K(R)$ depending only on $g,
\Omega,\Psi_{\mathcal{L}}$ and R (but {\bf not} on x)
 such that in the linear Euclidean structure $T^*(U)$, defined by
 the coordinate system $(x^1,..x^n, p_1,..,p_n)$ ($T^*(U) \simeq U\times \rR^m \subset
 \rR^{2m}$),
 we have for any $z_1,z_2 \in \mathcal{D}_R $, any t such $t(K_0+4R^2)<\rho$
\begin{eqnarray}\label{relation}
||(\digamma _{t}(z_1) -z_1) - (\digamma _{t}(z_2) -z_2)
||_{\rR^{2m}} \leq (e^{K(R)t}-1) ||z_1-z_2||_{\rR^{m}}
\end{eqnarray}
\end{enumerate}
\end{lem}
{\noindent \textbf{Proof.}} The proof is based on Gronwall's lemma
and uses the estimate of lemma 1. \QED
\bigskip

Let us recall (\cite{Giaquinta}):
\begin{lem}\label{lemmGia}
A curve $z:[0,T]\rightarrow T^*(V)$ is an extremal of the
optimization problem \ref{Varia}  if
\begin{enumerate}
\item z is a trajectory of $\mathcal{H}$ and
\item $z(T) \in 0_{T^*(V)}$.
\end{enumerate}
\end{lem}

We now formulate the main proposition of the paragraph
\begin{prop} \label{jeb}
For any $x\in V$, $T\geq 0$ such that
\begin{equation}\label{ineqT}
T < \min\{\frac{\rho}{K_0}, \frac{1}{K(0)}\log(3/2) \},
\end{equation}%
\begin{enumerate}
\item There exists a unique curve $\gamma
_{x}\in H^{1}([0,T];V)$ such that $\gamma _{x}(0)=x$ \ and
\begin{equation*}
I_{T}(x)=I(\gamma _{x}).
\end{equation*}%
\item This curve is located in U.
\item  $\dot{\gamma_x}(T) +\Omega (\gamma_x (T))=0$.
\item  For any $u\in H^{1}([0,T];T_{\gamma
_{x}}V),u(0)=0_{x}\in
T_{x}V,u\neq 0,$%
\begin{equation*}
\begin{array}{c}
\int_{0}^{T}\left\{ ||\nabla _{\dot{\gamma}_{x}(s)}u(s)+\nabla _{u(s)}\Omega
||_{g_{\gamma _{x}(s)}}^{2}+<\dot{\gamma}_{x}(s)+\Omega (\gamma
_{x}(s)),\nabla \nabla \Omega (\gamma _{x}(s))[u(s),u(s)]>_{g_{\gamma
_{x}(s)}}\right. \\
\left. +\nabla d\Psi _{\mathcal{L}}(\gamma _{x}(s))[u(s),u(s)]ds\right\} >0.%
\end{array}%
\end{equation*}
\end{enumerate}
\end{prop}


{\noindent \textbf{Proof.}}

(1) It is easy to see that there exists an
constant $C$ depending only on $g,\Omega $ and $\Psi $ such that for all $T$ $%
>0$ and all $\gamma \in H^{1}([0,T];V)$%
\begin{eqnarray*}
I(\gamma )+CT\geq \frac{1}{4}\int_{0}^{T}\left\Vert \frac{T\gamma }{dt}%
\right\Vert _{g}^{2}dt.
\end{eqnarray*}

The existence of a minimizing curve for the functional $I(\gamma )$ starting at any $x\in V$ and for any $%
T>0 $ is then standard.

2) If a curve $\gamma :[0,T]\rightarrow V$ is optimal, then there
exists an extremal $z:[0,T]\rightarrow T^{\ast }V,$ i.e. a
trajectory of the Hamiltonian field $\overrightarrow{\mathcal{H}}$
of $\mathcal{H},$ lifting $\gamma $ and such that $z(T)=0_{\pi
_{T^{\ast }V}(z(T))}$  (Lemma \ref{lemmGia}).

Because T is small enough (inequality (\ref{ineqT})) and
$z(T)=0_{\pi _{T^{\ast}V}(z(T))}$ (Lemma \ref{lemmGia}),
\begin{eqnarray*}
\mathcal{H}(z(t))=\mathcal{H}(z(T))=-\Psi_{\mathcal{L}}(\gamma(T))
\leq 0,
\end{eqnarray*}
using Lemma \ref{lemmaest}, we conclude that $\gamma([0,T]) \subset
U$.

Let $\gamma_1, \gamma_2: [0,T] \rightarrow V$ be two extremal curves
starting at x. Let $z_1,z_2:[0,T] \rightarrow T^*(V)$ be their
liftings. By the preceding considerations $z_i([0,T]) \subset U$,
for i=1,2. Choose a $R>0$, such that $TK(R)<\log(3/2)$,
$T(K_0+4R^2)<\log(3/2)$,  $z_i(0) \subset \mathcal{D}_0 \subset
\mathcal{D}_R$. By relation (\ref{relation}),
\begin{eqnarray}\label{relationb}
||(\digamma _{t}(z_1(0)) -z_1(0)) - (\digamma _{t}(z_2(0)) -z_2(0))
||_{\rR^{2m}} \leq (e^{K(R)t}-1) ||z_1(0)-z_2(0)||_{\rR^{m}}
\end{eqnarray}
Let us call $\Pi_2:T^*(U)\rightarrow T^*V$ the canonical projection
related to the product structure $T*(U)\simeq U\times \rR^m$.

\begin{eqnarray*}
||\Pi_2( (\digamma _{t}(z_1(0)) -z_1(0)) - (\digamma _{t}(z_2(0))
-z_2(0))) ||_{\rR^{2m}} \leq \\ || \digamma_{t}(z_1(0)) -z_1(0)) -
(\digamma _{t}(z_2(0)) -z_2(0))||_{\rR^{2m}},
\end{eqnarray*}
\begin{eqnarray*}
||\Pi_2( (\digamma _{t}(z_1(0))) -z_1(0)) - \Pi_2((\digamma
_{t}(z_2(0))) -z_2(0))) ||_{\rR^{2m}} \leq \\ \frac{1}{2}|| (z_2(0))
-z_1(0)||_{\rR^{2m}}.
\end{eqnarray*}
Thus
\begin{eqnarray*}
||\Pi_2( (\digamma _{t}(z_1(0)))  - \Pi_2((\digamma
_{t}(z_2(0)))||_{\rR^{2m}} \geq \frac{1}{2}|| (z_2(0))
-z_1(0)||_{\rR^{2m}}.
\end{eqnarray*}
By paragraph 2),$\digamma _{t}(z_i(0))=z_i(T) \in 0_{\pi
_{T^{\ast}V}(z(T))}$, which implies that $\Pi_2( (\digamma
_{t}(z_1(0)))=0_x$ and thus $z_2(0)=z_1(0)$.

\bigskip
\noindent The proofs of part 2) 3) are consequences of Lemma
\ref{lemmaest} and Lemma \ref{lemmGia} respectively.
\bigskip
We now prove Part (4) of proposition \ref{jeb}: There exists a
constant $K_{3}$ depending only on $g,\Omega $ and $\Psi $ such that
for all $s\in \lbrack 0,t],$ all $u\in T_{\gamma (s)}^{\ast }V$
\begin{eqnarray}
\left\vert <\dot{\gamma}_{x}(s)+\Omega (\gamma _{x}(s)),\nabla
\nabla \Omega (\gamma _{x}(s))[u,u]>_{g_{\gamma (s)}}+\nabla d\Psi
_{\mathcal{L}}(\gamma _{x}(s))[u,u]\right\vert \leq
K_{3}||u||_{g_{\gamma _{x}(s)}}^{2} \label{equa3b}
\end{eqnarray}
Also for any $u\in H^{1}([0,t];T_{\gamma _{x}}V)$ such that
$u(0)=0_{x}$
\begin{eqnarray}
\int_{0}^{T}||u(s)||_{g_{\gamma _{x}(s)}}^{2}ds\leq
T^{2}\int_{0}^{T}||\nabla _{\gamma (s)}u(s)||_{g_{\gamma
_{x}(s)}}^{2}ds \label{equa4}
\end{eqnarray}
Suppose that there exists a $\overline{u}\in H^{1}([0,T];T_{\gamma _{x}}V),%
\overline{u}(0) = 0_{x}\in T_{x}V,\overline{u}\neq 0$ and%
\begin{eqnarray}
\begin{array}{c}
\int_{0}^{T}\left\{ ||\nabla _{\dot{\gamma}_{x}(s)}\overline{u}(s)+\nabla _{%
\overline{u}(s)}\Omega ||_{g_{\gamma _{x}(s)}}^{2}+<\dot{\gamma}%
_{x}(s)+\Omega (\gamma _{x}(s)),\nabla \nabla \Omega (\gamma _{x}(s))[%
\overline{u}(s),\overline{u}(s)]>_{g_{\gamma _{x}(s)}}\right. \\
\left. +\nabla d\Psi _{\mathcal{L}}(\gamma _{x}(s))[\overline{u}(s),%
\overline{u}(s)]ds\right\} \leq 0%
\end{array}
\label{cond}
\end{eqnarray}

Then by (\ref{equa3b}),(\ref{equa4}) and (\ref{cond})%
\begin{eqnarray}
\int_{0}^{T}||\nabla _{\dot{\gamma}_{x}(s)}\overline{u}(s)+\nabla _{%
\overline{u}(s)}\Omega ||_{g_{\gamma _{x}(s)}}^{2}ds\leq
K_{3}\int_{0}^{T}||\bar{u}||_{g_{\gamma _{x}(s)}}^{2}ds.
\label{ine1}
\end{eqnarray}
Moreover,
\begin{equation*}
\int_{0}^{T}||\nabla _{\gamma (s)}\overline{u}(s)||_{g_{\gamma
_{x}(s)}}^{2}ds\leq 2\int_{0}^{T}||\nabla _{\dot{\gamma}_{x}(s)}\overline{u}%
(s)+\nabla _{\overline{u}(s)}\Omega ||_{g_{\gamma
_{x}(s)}}^{2}ds+2\int_{0}^{T}||\nabla _{\overline{u}(s)}\Omega
||_{g_{\gamma _{x}(s)}}^{2}ds.
\end{equation*}
There is a constant $K_{4}$ depending only on $g,\Omega $  such that
\begin{equation}
\int_{0}^{T}||\nabla _{\overline{u}(s)}\Omega ||_{g_{\gamma
_{x}(s)}}^{2}ds\leq K_{4}\int_{0}^{T}||\overline{u}||_{g_{\gamma
_{x}(s)}}^{2}ds  \label{ine}.
\end{equation}
Thus from (\ref{equa4}),(\ref{ine1})  and (\ref{ine}) we get
\begin{equation*}
\int_{0}^{T}||\nabla _{\gamma (s)}\overline{u}(s)||_{g_{\gamma
_{x}(s)}}^{2}ds\leq
2(K_{3}+K_{4})\int_{0}^{T}||\overline{u}||_{g_{\gamma
_{x}(s)}}^{2}ds\leq 2(K_{3}+K_{4})T^{2}\int_{0}^{T}||\nabla _{\gamma (s)}%
\overline{u}(s)||_{g_{\gamma _{x}(s)}}^{2}ds.
\end{equation*}
For $T\sqrt{2(K_{3}+K_{4})}<1,$ we get a contradiction. This ends
the proof of proposition \ref{jeb}.\QED

\subsubsection{Explicit decay estimate of the eigenfunction}

\bigskip Let $\Theta_{\varepsilon }:\mathbb{R}_{+ }\times V\rightarrow \mathbb{%
R},$ is the function $e^{-\lambda _{\varepsilon }t}v_{\varepsilon
}(x),$ then
we have%
\begin{eqnarray}
\epsilon \Delta _{g}\Theta_{\epsilon }+\theta (\Omega )\Theta_{\epsilon }+\frac{%
c_{\epsilon }}{\varepsilon }\Theta_{\epsilon }+\frac{\partial \Theta_{\varepsilon }}{%
\partial t}&=&0\text{, on }\mathbb{R}_{+}^{\ast }\times V  \\
\Theta_{\epsilon }\text{ }(0,x)&=&v_{\varepsilon }(x)\text{ for
}x\in V. \nonumber
\end{eqnarray}
Recall the Feynman-Kac formula:%
\begin{eqnarray*}
\Theta_{\varepsilon }(t,x)=E_{x}\left[ v_{\varepsilon
}(X_{\varepsilon
}^{x}(t))\exp -\int_{0}^{t}\frac{c_{\epsilon }(X_{\varepsilon }^{x}(s))}{%
\varepsilon }ds\right],
\end{eqnarray*}%
\begin{eqnarray}
\frac{\Theta_{\varepsilon }(t,x)}{\overline{v}_{\varepsilon }}=E_{x}\left[ \frac{%
v_{\varepsilon }(X_{\varepsilon }^{x}(t))}{\overline{v}_{\varepsilon
}}\exp
\left( -\int_{0}^{t}\left\{ c(X_{\varepsilon }^{x}(s))+\frac{\Delta _{g}%
\mathcal{L(}X_{\varepsilon }^{x}(s)\mathcal{)}}{2}\right\} ds\right) \exp
-\int_{0}^{t}\frac{\Psi _{\mathcal{L}}(X_{\varepsilon }^{x}(s))}{\varepsilon
}ds\right]  \label{feykac}
\end{eqnarray}
$X_{\varepsilon }^{x}$ is the diffusion process on $V$ having
$-\varepsilon \Delta _{g}-\theta (\Omega )$ as generator.

To estimate the right hand side of the relation (\ref{feykac}) when $%
\varepsilon \rightarrow 0,$ we are going to follow the method
presented in the papers (\cite{Azencott,bena}). For $\ 0<$
$t<\overline{T}$ the function $I$ has a unique minimum point on the
set
\begin{eqnarray}
C^{0}([0,t];V,x)=\{f:[0,t]\rightarrow V|f \hbox{ continuous, }
f(0)=x\},
\end{eqnarray}
 namely $\gamma _{x}$, which is non degenerate by Proposition \ref{jeb} (part-3).
 We are now ready to announce and prove the main result of this section:

\begin{theorem} \label{them6}
For \ $\ 0<$ $t<\overline{T}$\ there exists a constant \ $C_{t}$\
depending only on $g,\Omega $ and $\Psi $ such that for
\begin{eqnarray*}
\frac{v_{\varepsilon }(x)}{\overline{v}_{\varepsilon }}\leq
C_{t}\exp \left[ t\lambda _{\varepsilon
}-\frac{I_{t}(x)}{2\varepsilon }\right].
\end{eqnarray*}
\end{theorem}

\bigskip

{\noindent \bf Proof:} Recall that $X_{\varepsilon }(t)$ is a
stochastic process on $V$ having $-\varepsilon \Delta _{g}-\theta
(b)$ as generator, $X_{\varepsilon }^{x}(t)$ the process starting at
$x\in V.$ Choose $x\in V.$
\begin{eqnarray}
\frac{v_{\varepsilon }(x)}{\overline{v}_{\varepsilon }}=E_{x}\left[ \frac{%
v_{\varepsilon }(X_{\varepsilon }^{x}(t))}{\overline{v}_{\varepsilon
}}\exp
\left( -\int_{0}^{t}\left\{ c(X_{\varepsilon }^{x}(s))+\frac{\Delta _{g}%
\mathcal{L(}X_{\varepsilon }^{x}(s)\mathcal{)}}{2}\right\} ds\right)
\exp -\int_{0}^{t}\frac{\Psi _{\mathcal{L}}(X_{\varepsilon
}^{x}(s))}{\varepsilon }ds\right].
\end{eqnarray}
Let us define
\begin{eqnarray}
\tilde{E}_{h}=\{\gamma \in H^{1}([0,t];V,x)|\int_{0}^{t}\frac{1}{2}||\dot{\gamma}%
(s)+\Omega (\gamma (s))||_{g_{\gamma (s)}}^{2}ds\leq h\}.
\end{eqnarray}
We call $\chi _{1,}\chi _{2},\chi _{3}$ the characteristic functions
of the subsets of $C^{0}([0,t];V,x),$
\begin{eqnarray}
 \{\gamma |d_{\infty }(\gamma ,\tilde{E}_{h})>\delta
,d_{\infty }(\gamma ,\gamma _{x})\geq \eta \},
\end{eqnarray}
 \begin{eqnarray}
 \{\gamma |d_{\infty }(\gamma ,\tilde{E}_{h})\leq \delta ,d_{\infty
}(\gamma ,\gamma _{x})\geq \eta \},
\end{eqnarray}
\begin{eqnarray}
S = \{\gamma |d_{\infty }(\gamma ,\gamma _{x})<\eta \}
\end{eqnarray}
respectively and the distance is
\begin{eqnarray}
 d_{\infty
}(\gamma _{1},\gamma _{2})=\underset{[0,t]}{\sup } \, d_{g}(\gamma
_{1}(s),\gamma _{2}(s)).
\end{eqnarray}
Set $F(X_{\varepsilon })=\frac{v_{\varepsilon }(X_{\varepsilon
}^{x}(t))}{\overline{v}_{\varepsilon }}\exp \left(
-\int_{0}^{t}\left\{ c(X_{\varepsilon }^{x}(s))+\frac{\Delta _{g}\mathcal{L(}%
X_{\varepsilon }^{x}(s)\mathcal{)}}{2}\right\} ds\right) \exp -\int_{0}^{t}%
\frac{\Psi _{\mathcal{L}}(X_{\varepsilon }^{x}(s))}{\varepsilon }ds.$ Then:%
\begin{equation}
\frac{v_{\varepsilon }(X_{\varepsilon }^{x}(t))}{\overline{v}_{\varepsilon }}%
=E_{x}\left[ F(X_{\varepsilon })\chi _{1}(X_{\varepsilon })\right] +E_{x}%
\left[ F(X_{\varepsilon })\chi _{2}(X_{\varepsilon })\right]
+E_{x}\left[ F(X_{\varepsilon })\chi _{3}(X_{\varepsilon })\right].
\label{esti}
\end{equation}
Recall that
\[
I:\gamma \in C^{0}([0,t];V)\rightarrow \left\{ {
\begin{array}{l}
 {\int_{0}^{t}\frac{1}{2}||\dot{\gamma}(s)+\Omega (\gamma
(s))||_{g_{\gamma (s)}}^{2} ds \text{ \ if } \gamma \in
H^{1}([0,t];V)\ }  \\ \\
 +\infty   \,\,\,\,\,\, \text{ if \ }\ \gamma \in H^{1}([0,t];V) \\
 \end{array}} \right.
\]
is the rate function for $X_{\varepsilon }.$ Then it follows from
the theory of large deviations that for some constants $M_{1}$,
$l_{1}>0$ and $\varepsilon _{1}>0$: for $0<\varepsilon \leq
\varepsilon _{1},$
\begin{eqnarray*}
\left\vert E_{x}\left[ F(X_{\varepsilon })\chi _{1}(X_{\varepsilon
})\right] \right\vert \leq M_{1}P_{x}(X_{\varepsilon }\in \{\gamma
|d_{\infty }(\gamma ,E_{h})\geq \delta )\leq M_{1}\exp
-\frac{I_{t}(x)+l_{1}}{2\varepsilon t}.
\end{eqnarray*}
The subset $\{\chi _{2}=1\}\ \ of\ $\ \ $C^{0}([0,t];V,x)$ is
compact for the weak topology. On the other hand $I$ is \ lower
semi-continuous for the same topology. Hence it attains its minimum
on the compact  $\{\chi_{2}=1\}.$ Because $\gamma _{x}$ is the only minimum point of\ \ $I$ and \ \ $%
\{\chi _{2}=1\}\subset $\ \ \ $\{\gamma |d_{\infty }(\gamma ,\gamma
_{x})\geq \eta \},$ $\underset{\{\chi _{2}=1\}}{\min }I\geq
I_{t}(x)+l_{2}$ for some constant $l_{2}.$ Then there exists\ \ a
constant $M_{2}>0$ such that
\begin{eqnarray*}
E_{x}\left[ F(X_{\varepsilon })\chi _{2}(X_{\varepsilon })\right]
\leq M_{2}\exp -\frac{I_{t}(x)+l_{2}}{2\varepsilon t}.
\end{eqnarray*}
For these two estimates see \cite{Azencott2}. The study of the last
term on the right hand side of the relation (\ref{esti}) is more
involved. By construction $\gamma _{x}$ is contained in a geodesic
coordinate chart $(U,\xi ^{1},...,\xi ^{m})$\ with pole at $x.$
Choosing $\delta $ sufficiently small we can assume that  $\{\gamma
|d_{\infty }(\gamma ,\gamma _{x})<\eta \}$\ $\subset U.$ Then we can
take advantage of the linear structure induced on $U$ by the
coordinate system. We can consider \ the process $X_{\varepsilon }$
on $U$ defined in the
coordinate system $(U,\xi ^{1},...,\xi ^{m})$ by the stochastic equation:%
\begin{equation*}
dX_{\varepsilon }(t)=-\Omega_{\varepsilon } (X_{\varepsilon
}(t))+\sqrt{\varepsilon }\sigma (X_{\varepsilon }(t))dw(t)
\end{equation*}
where $\Omega_{\varepsilon }(x)=\Omega (x)+\varepsilon
\sum_{i,j}g^{ij}(x)\Gamma _{ij}^{k}(x)e_{k}$ and  $\sigma
:U\rightarrow GL(m;\mathbb{R})$ is a $C^{\infty }$ function such
that $\sigma \sigma ^{\ast }=g$ and $w$ is a standard brownian
motion. Note that there exists a constant $C>0$ such that
\begin{eqnarray*}
 E_{x}\left[ F(X_{\varepsilon })\chi _{3}(X_{\varepsilon
})\right] \leq C E_x \{ \chi _{3}(X_{\varepsilon })
e^{-\frac{\int_0^t\Psi_\mathcal{L}(X_\varepsilon(s))ds}{\varepsilon}}
\}.
\end{eqnarray*}

For the estimate of  $E_x \{ \chi _{3}(X_{\varepsilon })
e^{-\frac{\int_0^t\Psi_L(X_\varepsilon)(s)ds}{\varepsilon}} \}$, we
refer to the appendix II.

\begin{lem}
\bigskip For fixed $t\in ]0,\overline{T}[$, there exists a constant $C_{t}>0$
such that for all $x\in T^{S}$ (the tubular neighborhood of S
defined in paragraph \ref{notat}),
\begin{eqnarray*} I_t(x)\geq
C_{t}||x^{\prime }||_{\mathbb{R}^{m-1}}^{2}.
\end{eqnarray*}
\end{lem}
{\noindent \textbf{Proof.}} To prove this lemma we restate the
variational problem as a optimal control problem and study the
subspace $\mathcal{E}_{t}$
of \ $C^{0}([0,t];T^{\ast }V)$ consisting of the extremal trajectories $%
z:[0,t]\rightarrow T^{\ast }V$ of the problem. Let $T^{S}$ be a
tubular
neighborhood of the limit cycle S endowed with a cyclic coordinate system $%
(x^{\prime },\theta )$ as in paragraph \ref{notat}. Recall that we
use the notations $x^{\prime }=(x^{1},..,x^{m-1}),\theta =x^{m}$ as
in paragraph \ref{notat}. Using this coordinate systems, we shall
identify the tangent and cotangent spaces $TT^{S}$ and $T^{\ast
}T^{S}$ with $T^{S}\times \mathbb{R}^{m}$. An element $z\in T^{\ast
}T^{S}$ will be represented by a couple $(x,p)$ and an element $\tau
\in TT^{S}$ by
a couple $(x,u).$ It is easy to see that $\mathcal{E}_{t}$ is relatively compact in $%
C^{0}([0,t],T^{\ast }V)$ (Because the final point is in the zero
section  which is compact and the derivative of the extremals are
bounded along $[0,T]$). Let $z\in \mathcal{E}_{t}$.

Set $\pi _{T^{\ast }V}$ $\circ z=\gamma .$ \ Define%
\begin{eqnarray}
\widehat{I}(z)=\int_{0}^{t}\left[ \frac{1}{4}||z(s)||_{g^{\ast }}^{2}+\Psi _{%
\mathcal{L}}(\gamma (s))\right] ds  \label{qes}
\end{eqnarray}

The functional $\widehat{I}:\mathcal{E}$ $_{t}\longrightarrow \mathbb{R}$ is
clearly continuous. Let $\mathcal{E}$ $_{t}^{S}$ be the subset of $\mathcal{E%
}$ $_{t}$ of those $z$ such that $z(0)\in $ $T^{\ast }T^{S}$ .

 Define the
functional $In:\mathcal{E}$ $_{t}^{S}\longrightarrow \mathbb{R}$, by setting
$In(z)=||\gamma ^{\prime }(0)||_{\mathbb{R}^{m-1}}^{2}$ where $\gamma
^{\prime }(0)=(x^{1}(z(0)),..x^{m-1}(z(0))).$ The quotient $In\left/ \widehat{%
I}\right. $ is continuous on $\mathcal{E}_{t}^{S}$ $-In^{-1}(0)$ for
the $C^0$ topology.

If this quotient is not bounded on $\mathcal{E}$
$_{t}^{S}-In^{-1}(0)$, then there
exists a sequence \{$z^{k}|k\in \mathbb{N}$\} in $\mathcal{E}$ $%
_{t}^{S}-In^{-1}(0)$ such for each $k\in \mathbb{N}$
\begin{eqnarray}
\frac{In(z^{k})}{\widehat{I}(z^{k})}\geq k.  \label{qos}
\end{eqnarray}%
Recall that $\Psi _{\mathcal{L}}\geq 0.$ It follows from relations
(\ref{qes}) and (\ref{qos}) that for all $k\in \mathbb{N}.$%
\begin{eqnarray}
\int_{0}^{t}\frac{1}{4}||z^{k}(s)||_{g^{\ast }}^{2}ds\leq \frac{1}{k}%
In(z^{k})\text{ , }\int_{0}^{t}\Psi _{\mathcal{L}}(\gamma
^{k}(s))ds\leq \frac{1}{k}In(z^{k}) , \label{jus}
\end{eqnarray}
and
\begin{eqnarray}
|\mathcal{H(}z^{k}(t)\mathcal{)}|=\Psi _{\mathcal{L}}(\gamma
^{k}(t)). \label{huh}
\end{eqnarray}%
For all $s\in \lbrack 0,t]$%
\begin{eqnarray*}
||\gamma ^{\prime k}(0)-\gamma ^{\prime
k}(s)||_{\mathbb{R}^{m-1}}^{2}\leq
s\int_{0}^{s}\left\Vert \frac{d\gamma ^{\prime k}(\sigma )}{d\sigma }%
\right\Vert _{\mathbb{R}^{m-1}}^{2}d\sigma.
\end{eqnarray*}

Using the relation (\ref{p}), noting that $\Omega ^{q}(0,\theta )=0$ if $%
1\leq q\leq m-1,$ for an appropriate constant $C_{2}$ depending only
on $g$ and $\Omega $

\begin{eqnarray*}
||\gamma ^{\prime k}(0)-\gamma ^{\prime
k}(s)||_{\mathbb{R}^{m-1}}^{2}\leq sC_{2}\left(
\int_{0}^{s}\frac{1}{4}||z^{k}(\sigma )||_{g^{\ast
}}^{2}d\sigma +\int_{0}^{s}||\gamma ^{\prime k}(\sigma )||_{\mathbb{R}%
^{m-1}}^{2}d\sigma \right)
\end{eqnarray*}%
for all $s\in \lbrack 0,t].$ By the assumptions on $\Psi
_{\mathcal{L}}$ there exists a constant $C_{3}>0$ depending only on
$\Psi _{\mathcal{L}}$ such that for all $s\in \lbrack 0,t],$ all
$k\in \mathbb{N}$
\begin{eqnarray}
\frac{||\gamma ^{\prime k}(s)||_{\mathbb{R}^{m-1}}^{2}}{C_{3}}\leq \Psi _{%
\mathcal{L}}(\gamma (s))\leq C_{3}||\gamma ^{\prime k}(s)||_{\mathbb{R}%
^{m-1}}^{2}.  \label{psi}
\end{eqnarray}%
Hence for \ $C_{4}=\max (C_{2},C_{2}C_{3}),$ for all $s\in \lbrack 0,t]$%
\begin{eqnarray*}
||\gamma ^{\prime k}(0)-\gamma ^{\prime
k}(s)||_{\mathbb{R}^{m-1}}^{2}\leq sC_{4}\left(
\int_{0}^{s}\frac{1}{4}||z^{k}(\sigma )||_{g^{\ast }}^{2}d\sigma
+\int_{0}^{s}\Psi _{\mathcal{L}}(\gamma ^{k}(\sigma ))d\sigma
\right)
\end{eqnarray*}%
Using equations (\ref{jus}), for all $s\in \lbrack 0,t]$
\begin{eqnarray*}
||\gamma ^{\prime k}(0)-\gamma ^{\prime
k}(s)||_{\mathbb{R}^{m-1}}^{2}\leq sC_{4}\frac{1}{k}In(z^{k}),
\end{eqnarray*}
\begin{eqnarray*}
||\gamma ^{\prime k}(0)||_{\mathbb{R}^{m-1}}^{2}\leq 2||\gamma
^{\prime k}(s)||_{\mathbb{R}^{m-1}}^{2}+2sC_{4}\frac{1}{k}In(z^{k}).
\end{eqnarray*}
Integrating on $[0,t]$
\begin{eqnarray*}
t||\gamma ^{\prime k}(0)||_{\mathbb{R}^{m-1}}^{2}\leq
2\int_{0}^{t}||\gamma ^{\prime
k}(s)||_{\mathbb{R}^{m-1}}^{2}ds+t^{2}C_{4}\frac{1}{k}In(z^{k}).
\end{eqnarray*}
Using the relations (\ref{psi}),(\ref{jus}) for all $k\in \mathbb{N}$%
\begin{eqnarray*}
\int_{0}^{t}||\gamma ^{\prime k}(s)||_{\mathbb{R}^{m-1}}^{2}ds\leq
C_{3}\int_{0}^{t}\Psi _{\mathcal{L}}(\gamma (s))ds\leq C_{3}\frac{1}{k}%
In(z^{k}),
\end{eqnarray*}
\begin{eqnarray*}
t||\gamma ^{\prime k}(0)||_{\mathbb{R}^{m-1}}^{2}\leq (2C_{3}+t^{2}C_{4})%
\frac{1}{k}In(z^{k}),
\end{eqnarray*}
\begin{eqnarray*}
||\gamma ^{\prime k}(0)||_{\mathbb{R}^{m-1}}^{2}\leq (\frac{2C_{3}}{t}%
+tC_{4})\frac{1}{k}In(z^{k}),
\end{eqnarray*}
\begin{eqnarray*}
In(z^{k})\leq (\frac{2C_{3}}{t}+tC_{4})\frac{1}{k}In(z^{k}).
\end{eqnarray*}
\bigskip For $k$ so large  that $(\frac{2C_{3}}{t}+tC_{4})\frac{1%
}{k}<1$ we get a contradiction. This ends the proof of the lemma.
\QED

\noindent Using the result of the lemma, we conclude that
\begin{eqnarray} \label{fondamental}
\frac{v_{\varepsilon }(x)}{\overline{v}_{\varepsilon }} &\leq&
C_{t}\exp \left[ t\lambda _{\varepsilon
}-\frac{I_{t}(x)}{2\varepsilon }\right] \nonumber \\
w_{\varepsilon}(x) &\leq & C_1\exp \left[-C_{2}||x^{\prime
}||_{\mathbb{R}^{m-1}}^{2} \right].
\end{eqnarray}
$C_{1}, C_{2}$ are two positive constants independent of
$\varepsilon$. In particular, we have for a point, a cycle or a
torus belonging to the recurrent set,
\begin{eqnarray}
 w_{\varepsilon}(x) &\leq & C_1\exp
\left[-C_{2}dist_{g}(x,\mathcal{S})^{2}  \right].
\end{eqnarray}
\begin{coro}\label{Coroconverge} Any sequence of $\epsilon$'s converging to zero
contains a subsequence $\varepsilon_n$ such that $w_{\varepsilon_n}$
converges in $L_2$ to the blown up limit function w.
\end{coro}
The proof follows from lemma (\ref{lemma6}) and inequality
(\ref{fondamental}).

\subsection{Absolute continuity of the limit measures on the limit
cycle} \label{limitmeasure}

In this paragraph, as stated at the beginning of section \ref{lcs},
we shall prove that along a limit cycle, the limit measures are
absolutely continuous with respect to the length. The proofs are
based on the decay estimates, established in the previous paragraph
\ref{decyyv}.

When a cycle $S$ is charged (see definition \ref{Charged}), we shall
 prove that:
\begin{prop}
The restriction of a limit measure $\mu $ to a cycle $S$ is
absolutely continuous with respect to the length.
\end{prop}

{\noindent \textbf{Proof.}} Let \{$v_{\varepsilon _{n}}|n\in \mathbb{N\}}$, $%
\varepsilon _{n}\longrightarrow 0$ as $n\longrightarrow \infty ,$ be
a sequence such that the sequence of measures $\ \left[
\frac{v_{\varepsilon _{n}}^{2}dV_{g}}{\int_{V}v_{\varepsilon
_{n}}^{2}dV_{g}}\left\vert n\in \mathbb{N}\right. \right] $\ \
converges weakly to $\mu.$ For simplicity we
shall assume that the $v_{\varepsilon _{n}}$are normalized ($%
\int_{V}v_{\varepsilon _{n}}^{2}dV_{g}=1).$ Let
$(U,x_{1},...x_{m-1},\theta ) $ be an adapted coordinate system for
the cycle $S$ as in paragraph \ref{notat} such that $U\supset
T(\delta $) the image of which  is the set
\{$\sum_{j=1}^{m-1}x_{j}^{2}\leq \delta ^{2}\}$. Let $\varphi$ be a
continuous function defined in the tubular neighborhood $T(\delta )$. Then:%
\begin{eqnarray*}
\int_{S}\varphi d\mu =\lim_{n\rightarrow +\infty }\int_{T(\delta
)}\varphi v_{\epsilon _{n}}^{2}dV_{g}.
\end{eqnarray*}
Take $\ \varphi (x)$ of the form $\psi (x^{\prime })\eta (\theta )$, $%
x^{\prime }=(x_{1},...,x_{m-1})$. Using the blow-up analysis of
theorem \ref{BLOWUP}, for all $\delta >0$, sufficiently small
\begin{eqnarray*}
\int_{T(\delta )}\varphi v_{\epsilon
_{n}}^{2}dV_{g}=\bar{v}_{\epsilon
_{n}}^{2}\varepsilon _{n}^{\frac{m-1}{2}}\int_{0}^{T_S}\int_{B^{m-1}(\delta /%
\sqrt{\epsilon _{n}})\times \{\theta \}}\psi (x^{\prime }\sqrt{\varepsilon
_{n}})\eta (\theta )w_{\epsilon }^{2}(x^{\prime },\theta )d\Sigma
_{g_{\varepsilon _{n}}}d\theta ,
\end{eqnarray*}%
where $d\Sigma _{g_{\varepsilon _{n}}}$ is the measure induced on
each n-1 dimensional ball $B^{m-1}(\delta )$ of radius $\delta $,
transverse to the cycle by the blown-up metric $g_{\varepsilon
_{n}}$.

Since the sequence $\bar{v}_{\epsilon _{n}}^{2}\varepsilon _{n}^{\frac{m-1}{2}}$ converges as $%
\varepsilon_n $ goes to zero (see the next subsection), the blow-up
analysis of theorem \ref{BLOWUP} and the decay estimate along the
cycle show that the integral
\begin{eqnarray*}
\int_{0}^{l}\int_{B^{m-1}(\delta /\sqrt{\varepsilon _{n}})\times
\{\theta \}}\psi (x^{\prime }\sqrt{\varepsilon _{n}})\eta (\theta
)w_{\epsilon _{n}}^{2}(x^{\prime },\theta )d\Sigma _{g_{\varepsilon
_{n}}}d\theta,
\end{eqnarray*}%
converges as $\varepsilon _{n}$ goes to zero. Indeed, $%
w_{\epsilon _{n}}$ converges strongly in $L_{2}(\hbox{\bb
R}^{m-1}\times S^{1})$ to a blown-up function $w_{S}$ and by Fubini
theorem,
\begin{eqnarray*}
\lim_{n\longrightarrow +\infty }\int_{B^{m-1}(\delta /\sqrt{\epsilon }%
)}\varphi w_{\epsilon _{n}}^{2}d\Sigma _{\epsilon } &=&\int_{0}^{T_S}\int_{%
\hbox{\bb R}^{m-1}}w_{S}^{2}(x^{\prime },\theta )\psi (0)\eta (\theta
)dx^{\prime }d\theta \\
&=&\int_{\hbox{\bb R}^{m-1}}\psi (0)\eta (\theta
)\int_{0}^{l}w_{S}^{2}(x^{\prime },\theta )dx^{\prime }d\theta .
\end{eqnarray*}%
Since \ by Proposition \ref{prop3} $w_{S}(x)=z_{S}(x^{\prime })f_{S}(\theta )$ and%
\begin{eqnarray*}
\int_{\hbox{\bb R}^{m-1}}\int_{0}^{l}\psi (0)\eta (\theta
)w_{S}^{2}(x^{\prime },\theta )dx^{n-1}d\theta =\int_{\hbox{\bb R}%
^{m-1}}\psi (0)z_{S}^{2}(x^{\prime
})dx^{n-1}\int_{0}^{T_S}f_{S}^{2}(\theta )\eta (\theta )d\theta .
\end{eqnarray*}%
If $\delta _{x^{\prime }}$ denotes the Dirac measure in the
transverse variable $x^{\prime }$ only, the normalized eigenfunction
sequence converges in the distribution sense and on the cycle $S$,
\begin{eqnarray*}
\underset{n\longrightarrow +\infty }{\lim }v_{\epsilon
_{n}}^{2}dV_{g}=\left( \int_{\hbox{\bb R}^{m-1}}z_{S}^{2}(x^{\prime
})dx^{\prime }{}\right) \delta _{x^{\prime }}\otimes
f_{S}^{2}(\theta )d\theta .
\end{eqnarray*}%
Another representation of the continuous component of the limit
measure is
\begin{eqnarray*}
\frac{d\mu }{d\theta }=f_{S}^{2}(\theta )=\lim_{n\longrightarrow
+\infty }\int_{H_{l}}v_{\epsilon }^{2}d\Sigma _{g}
\end{eqnarray*}%
where $H_{l}$ denotes any hypersurface cutting the orbit at the
point of abscissa $\theta=l$ transversally. The limit is independent
of the choice of the hypersurface. \QED

\subsection{The limits of $\sup_{V_{m}}v_{\protect\epsilon }$}
Let us recall that $\bar{v}_{\varepsilon
_{n}}=\sup_{V_{m}}v_{\epsilon
_{n}}$. We shall now prove that either $\bar{v}_{\epsilon _{n}}^{2}\varepsilon _{n}^{\frac{%
m}{2}}$ converges to zero and then no critical point is charged. In
that case  $\bar{v}_{\epsilon _{n}}^{2}\varepsilon
_{n}^{\frac{m-1}{2}}$ converges to a strictly positive constant and
at at least one cycle is charged.

To prove this statement, we use the $L_2$ normalization condition
\begin{eqnarray}
1=\int_{V_{m}}v_{\epsilon _{n}}^{2} dV_g
\end{eqnarray}%
and we perform the blow-up change of coordinates in the neighborhood
of the recurrent set. Outside any neighborhood of the recurrent set,
the sequence $v_{\epsilon_{n}}$ converges to zero in $L_2$ (see
\cite{HK1}).

\begin{eqnarray}
1=\int_{V_{m}}v_{\epsilon_{n}}^{2}=\bar{v}_{\epsilon
_{n}}^{2}\varepsilon _{n}^{\frac{m-1}{2}}\left(
\sum_{S}\int_{T_{S}(\delta /\epsilon _{n}^{1/2})}w_{\epsilon
_{n}}^{2}(x)dV_{g_{\varepsilon}}+\varepsilon
_{n}^{1/2}\sum_{P}\int_{B_{P}(\delta /\epsilon
_{n}^{1/2})}w_{\epsilon _{n}}^{2}(x)dV_{g_{\varepsilon}} \right)
+\int_{V'_{m}}v_{\epsilon _{n}}^{2} \label{coeff}
\end{eqnarray}%
where
\begin{eqnarray}
V'_{m} = V_{m}-\bigcup_{S} T_{S}(\delta /\epsilon
_{n}^{1/2})-\bigcup_{P}B_{P}(\delta /\epsilon _{n}^{1/2}).
\end{eqnarray}%
The first sum on the right hand side is extended over the limit
cycles and the second over the stationary points.

The last integral in the r.h.s. of equation (\ref{coeff}) tends to
zero, because we integrate over an open set whose closure does not
intersect the recurrent set(see the proof in \cite{HK1}).

After selecting a subsequence if necessary, there exists a component
$C$ of the recurrent set and a sequence $Q_n$ such that
\begin{enumerate}
\item $v_{\varepsilon_{n}} (Q_n) =\bar{v}_{\varepsilon_{n}}$.
\item $Q_n \rightarrow Q_{\infty} \in C$.
\item $\frac{Q_n -Q_{\infty}}{\sqrt{\varepsilon_{n}}}\rightarrow
Q_*$. This difference is taken in the linear structure defined by
the coordinate system introduced in \ref{notat}. $Q_*$ belongs to
the blown-up space $\rR^m $ or $\rR^m\times S^1$.
\end{enumerate}
Let us  show that w is not identically 0. We know that it converges
uniformly on any compact set a to $w_C$ and
\begin{eqnarray}
\lim_{n \rightarrow \infty} w_{\varepsilon_n}(\frac{Q_n
-Q_{\infty}}{\sqrt{\varepsilon_{n}}}) = w_C(Q_*)=1.
\end{eqnarray}%
We conclude that $w_C$ is not identically 0. If C is a point,  using
Corollary \ref{Coroconverge} we have that $w_{\varepsilon_n}$
\begin{eqnarray}\label{e1}
\underset{n\longrightarrow +\infty }{\lim }%
\int_{B_{C}(\delta /\varepsilon _{n}^{1/2})}w_{\varepsilon
_{n}}^{2}(x)dV_{g_{\varepsilon _{n}}}= \int_{\hbox{\bb R}^{m}
}w_C^{2}(x)dx >0.
\end{eqnarray}
If C is a cycle, we have
\begin{eqnarray}\label{e2}
\underset{n\longrightarrow +\infty }{\lim }%
\int_{T_C(\delta /\varepsilon _{n}^{1/2})}w_{\varepsilon
_{n}}^{2}(x)dV_{g_{\varepsilon _{n}}}=  \int_{\hbox{\bb
R}^{m-1}\times \sS^{1}}w_C^{2}(x',\theta)dx'd\theta >0.
\end{eqnarray}
We conclude  using formula (\ref{coeff}) and the previous identities
(\ref{e1}) or (\ref{e2}), that after selecting a subsequence if
needed.

\noindent If C is a point:
\begin{eqnarray}
\underset{%
n\longrightarrow +\infty }{\lim }\bar{v}_{\epsilon
_{n}}^{2}\varepsilon _{n}^{\frac{m}{2}} >0.
\end{eqnarray}
If C is cycle,
\begin{eqnarray}
\underset{%
n\longrightarrow +\infty }{\lim }\bar{v}_{\epsilon
_{n}}^{2}\varepsilon _{n}^{\frac{m-1}{2}} >0.
\end{eqnarray}
When a cycle is charged, it follows from formula (\ref{coeff}) that
no point is charged.

For any component S of the recurrent set, selecting a subsequence if
needed: if S is a cycle,
\begin{eqnarray}
\int_{T_{S}(\delta /\epsilon _{n}^{1/2})}w_{\epsilon
_{n}}^{2}(x)dV_{g_{\varepsilon}} = \mathcal{C}^{-1}(b,g)
\gamma^2_S\int_{\hbox{\bb R}^{m-1}\times \sS^1
}w_C^{2}(x,\theta)dxd\theta,
\end{eqnarray}
\begin{equation*}
\lim_{\epsilon \rightarrow 0}\epsilon ^{(m-1)/2}\bar{v}_{\epsilon
}^{2}=\mathcal{C}(b,g).
\end{equation*}
If S is a point, we have
\begin{eqnarray}
\int_{B_{S}(\delta /\epsilon _{n}^{1/2})}w_{\epsilon
_{n}}^{2}(x)dV_{g_{\varepsilon}} = \mathcal{C}^{-1}(b,g)
\gamma^2_S\int_{\hbox{\bb R}^{m} }w_C^{2}(x)dx,
\end{eqnarray}
\begin{equation*}
\lim_{\epsilon \rightarrow 0}\epsilon ^{(m-1)/2}\bar{v}_{\epsilon
}^{2}=\mathcal{C}(b,g).
\end{equation*}
If no cycle is charged, then
\begin{equation*}
\mathcal{C}(b,g) =\sum_{S\in \mathcal{C}} \gamma^2_S\int_{\hbox{\bb
R}^{m} }w_C^{2}(x)dx,
\end{equation*}
If at least one cycle is charged then
\begin{equation*}
\mathcal{C}(b,g) =\sum_{S\in \mathcal{C}} \gamma^2_S\int_{\hbox{\bb
R}^{m-1}\times \sS^1 }w_C^{2}(x,\theta)dxd\theta.
\end{equation*}
When several cycles are charged, using expression (\ref{coeff}), we
obtain an exact expression of the limit measure $\mu$.  If
$\mathcal{C}$ denotes the set of charged limit cycles then
\begin{eqnarray*}
\mu= \frac{\sum_{S\in \mathcal{C}}\gamma _{S}^{2}\left( \int_{\hbox{\bb R}%
^{m-1}}z_{S }^{2}(x^{\prime })dx^{\prime }\right) \delta _{x^{\prime
}}\otimes f_{S}^{2}(\theta )d\theta }{\sum_{S\in \mathcal{C}}\gamma
_{S}^{2}\left( \int_{\hbox{\bb R}^{m-1}}z_{S}^{2}(x^{\prime
})dx^{\prime }\right) \int_{0}^{T}f_{S}^{2}(\theta )d\theta },
\end{eqnarray*}
where we have used the result of corollary \ref{scind}. When no
cycles are charged, we have
\begin{eqnarray*}
\mu= \frac{\sum_{P\in \mathcal{C}_{sing}}\gamma _{P}^{2}\left( \int_{\hbox{\bb R}%
^{m}}z_{P }^{2}(x)dx\right) \delta _{P}}{\sum_{P\in
\mathcal{C}_{sing}}\gamma _{P}^{2} \left( \int_{\hbox{\bb
R}^{m}}z_{P}^{2}(x)dx\right)},
\end{eqnarray*}
where $\mathcal{C}_{sing}$ denotes the set of charged critical
points where the topological pressure is achieved. This ends the
proof of theorem \ref{thcf} parts 2-3-4 of theorem \ref{thcy}. \QED


\section{Concentration on two-dimensional torus}

In this section, we shall study  the limit measures of the sequence $%
v^2_{\epsilon} dV_g$, when the recurrent set of the vector field $b$
contains hyperbolic two dimensional torii. We shall not try to
extend our results, when the recurrent sets are of dimension $n\geq
3$, but it is likely that similar results are valid on n dimensional
 manifolds under some  restrictive assumptions.  The characterization of
 the limit measures remains an open problem, even in dimension 3.

 However, we will examine the concentration of eigenfunctions along two dimensional
 torii and show that the limit measures are absolutely continuous with respect to the
probability measure on the torus invariant by the flow of b (see
assumption IV in \ref{notat}: the restriction of the field to the
torus generates an irrational flow).

\subsection{Main result}
\begin{theorem}
\label{torus} On a Riemanian manifold $(V,g)$, let $b$ be a hyperbolic
vector field such that the recurrent set $\mathcal{R}$ consists of points $%
P_{1},...,P_{M}$,  cycles $S_{1}$,.., $S_{N}$ and two-dimensional
irrational torii $\hbox{\bbbis T}_{1},..\hbox{\bbbis T}_{r}$ and
assumptions I-IV of section \ref{notat} are satisfied.

Let $u_{\epsilon }>0$ be the first eigenfunction of $L_{\epsilon }
$, normalized using the $L_2$-norm. We denote by $\mathcal{L}$, a
special Lyapunov function. The set of weak limits of the probability measures $\frac{u_{\epsilon }^{2}e^{-%
\mathcal{L}/\epsilon }dV_{g}}{\int_{V_{n}}u_{\epsilon }^{2}e^{-\mathcal{L}%
/\epsilon }dV_{g}}$  when $\varepsilon$ goes to zero is contained
inside the set
\begin{eqnarray*}
\Lambda _{2} =\{\mu =\sum_{P\in S_{tp}}c_{P}\delta
_{P}+\sum_{\Gamma \in S_{tp}}a_{\Gamma }\delta _{\Gamma }+\sum_{\hbox{\bbbis T}\in S_{tp}}b_{%
\hbox{\bbbis T}}\delta_{\hbox{\bbbis T} }, \,c_{P} \geq 0,a_{\Gamma
}\geq 0,\,b_{\hbox{\bbbis T}}\geq 0
\},
\end{eqnarray*}%
where
\begin{eqnarray*}
\delta_{\hbox{\bbbis T} }(h) =\int_{\hbox{\bbbis T}}h(\theta _{1},\theta
_{2}) f_{\hbox{\bbbis T}}(\theta _{1},\theta _{2})dS_{\hbox{\bbbis T}},
\end{eqnarray*}%
$dS_{\hbox{\bbbis T}}=\frac{d\theta_1 d\theta_2}{\int_{\hbox{\bbbis T}}d\theta_1 d\theta_2}$ is the two-dimensional normalized  measure on $%
\hbox{\bbbis T}$  invariant under the action of the field  $b$ and
$f_{\hbox{\bbbis T}}$ is the unique solution of maximum 1, of the
equation
\begin{eqnarray*}
k_1 \frac{\partial f}{\partial \theta_1}+k_2 \frac{\partial f}{\partial
\theta_2} +cf &=& \mu_2 f \hbox{ on  \bbbis T }, \\
\mu_2&=&\int_{\hbox{\bbbis T}} c dS_{\hbox{\bbbis T}}.
\end{eqnarray*}
The measures $\delta_P$, $\delta_\Gamma$ were defined in the
previous sections.

When the topological pressure is achieved at least on one torus, the
maximum of the sequence $v_{\epsilon }=u_{\epsilon
}e^{-\mathcal{L}/2\epsilon }$ goes to infinity and
\begin{equation*}
\lim_{\epsilon \rightarrow 0}\epsilon ^{n/2-1}\bar{v}_{\epsilon
}^{2}=C(\Omega ,c),
\end{equation*}%
where the constant $C(\Omega ,c)$ is given by
\begin{equation*}
C(\Omega ,c)=\frac{1}{\sum_{\hbox{\bbbis T}\in S^{\prime }}\gamma _{%
\hbox{\bbbis T}}^{2}\int_{\hbox{\bbbis T}}f_{\hbox{\bbbis T}}dS_{%
\hbox{\bbbis
T}}},
\end{equation*}
the sum is extended to all torus where topological pressure is achieved and $%
\gamma _{\hbox{\bbbis T}}$ are the modulating coefficients. If the
topological pressure is not attained on any torus, then the torii do
not contribute to the limit measures.
\end{theorem}

{\noindent \textbf{Remark 1.}} If at  least one torus is charged,
the critical points and limit cycles do not contribute to the limit
measures. On the other hand, if no torii are charged, the
description of the limit measures is the same as when no torii are
present.

\bigskip

{\noindent \textbf{Remark 2.}} Instead of two dimensional torii, we
could have considered more general compact surfaces in V invariant
by the flow of b. But we do not know what assumptions should be made
on the flow restricted to the surface in order to determine the
limit measures. We have restricted ourself to two-dimensional
irrational torii, because we have no knowledge of the properties of
the limit measures on other type of recurrent sets. The case of a
torus is tractable because it has minimal flows \cite{Katok}. Our
assumptions on b imply that its flow on the torus is minimal. On the
other hand, no other compact surface admits minimal flows (the
standard argument is coming from the Euler-Poincar\'e
characteristics).

Also it is conceivable that limit measures on general surfaces for general flows
could be concentrated on proper subsets of the surface.  These subsets could be recurrent
subsets of the restriction of b and this in spite of the fact that the topological pressure
is achieved on the surface.

\subsection{Fundamental propositions}

We have 

\begin{prop}
The topological pressure of a two irrational torus satisfying assumptions IV
is given by
\begin{equation*}
TP=\int_{_{\mathbb{T}}}cdH-trB_{s},\textbf{}
\end{equation*}%
where dH is the unique probability measure on the torus invariant by
the flow of the restriction of b to $\mathbb{T}$, $B_{s}$ is
the stable component of the restriction of the field transverse to
the torus.
\end{prop}

{\noindent \textbf{Proof.}}

\noindent Let us recall the formula \cite{Kifer90}:
\begin{equation*}
TP=\sup \left\{h_{\nu}+ \int (c+\left. \frac{DJ^{s}}{Dt}\right\vert
_{t=0})d\nu |\nu \text{ probability measure on }\mathbb{T}\text{
invariant by the flow }\right\}
\end{equation*}%
where $J^{s}$ is the restriction of the Jacobian to the stable
manifold along $\mathbb{T}$, $h_{\nu}$ is the entropy, equal to zero
\cite{Walter}. \bigskip\ In the present case, since the restriction
of the flow to $\mathbb{T}$ is ergodic, there is only one invariant
probability measure $H$ and $dH=\frac{d\theta _{1}\wedge d\theta _{2}}{\int_{\mathbb{T}}d\theta _{1}\wedge d\theta _{2}}.$ Moreover, using the notation of paragraph \ref{notat} $\left. \frac{DJ^{s}}{Dt}%
\right\vert _{t=0}=-trB_{s}.$ \quad%
\hbox{\hskip 4pt\vrule width 5pt height
6pt depth 1.5pt}

The precise characterization of the concentration of the sequence $%
v_{\epsilon}$ is obtained by studying the renormalized sequence
$w_{\epsilon }$, centered at a maximum sequence point, which
concentrates on a torus. In the normal coordinate system $(x',\theta
_{1},\theta _{2})$ along the torus, defined in section
\ref{notat}-IV, the function $w_{\epsilon }$ is given by
\begin{equation*}
w_{\epsilon }(x^{\prime  },\theta _{1},\theta
_{2})=\frac{v_{\epsilon
}(\sqrt{\epsilon }x^{\prime  },\theta _{1},\theta _{2})}{\overline{%
\bar{v}_{\epsilon }}}.
\end{equation*}
We have 

\begin{prop}
\label{wwp} If a function w is a limit of a sequence $w_{\epsilon }$ as $\epsilon$
tends to zero, then the convergence is uniform on any compact set of
 $\hbox{\bb R}^{m-2}\times \hbox{\bbbis T}_{f}$, where $\hbox{\bbbis T}_{f}$
 is the flat two-dimensional torus.

If $\Delta _{E}^{m-2}$ denotes the Laplacian on $\hbox{\bb R}^{m-2}$,
then $w$ is a weak solution of the following equation
\begin{eqnarray*}
\Delta _{E}^{m-2}w+\sum_{i,j}\Omega _{ij}x^{j}\frac{\partial
w}{\partial x^{i}}+
(\Omega^{//},\nabla w)+\left[ c(0,\theta _{1},\theta _{2})+\frac{\Delta _{g}%
\mathcal{L}}{2}+\Psi (x^{\prime  })\right] w &=&\mu w, \\
0 &<&w\leq 1,
\end{eqnarray*}%
where $\tilde{b}=(k_{1},k_{2})$, $x^{\prime  }$ are the transverse
coordinates, $\Omega _{ij}$ is the transversal part of the field
$\Omega$ and $\mu $ is the first eigenvalue of the operator on the
left hand side.
\end{prop}
The proof will be given later. The most important result is the following
\begin{prop}
\label{pp8}
The function $w$ of proposition \ref{wwp} is in fact regular
and can be written as the product of two functions in the variables $x^{\prime \prime}$ and $%
\theta ^{\prime }$ respectively
\begin{eqnarray*}
w(x)=w_{\hbox{\bbbis T}}(x^{\prime })f_{\hbox{\bbbis T}}(\theta ^{\prime }),
\end{eqnarray*}%
where $w_{\hbox{\bbbis T}}$ and $f$ satisfy
\begin{eqnarray}  \label{eqtorus}
\Delta _{m-2}w_{\hbox{\bbbis T}}+\sum_{i,j}\Omega _{ij}x^{j}\frac{\partial w_{%
\hbox{\bbbis T}}}{\partial x^{i}}+\Psi (x^{\prime })w_{\hbox{\bbbis
T}}=\mu _{1}w_{\hbox{\bbbis T}}\hbox{ on }\hbox{\bb R}^{m-2}
\end{eqnarray}
and
\begin{eqnarray} \label{ftilde}
(\Omega ^{//},\nabla f_{\hbox{\bbbis T}})+c(\theta ^{\prime
})f_{\hbox{\bbbis T}} &=&\mu _{2}f_{\hbox{\bbbis T}}
\label{edd2} \\
\mu _{1}+\mu _{2} &=&\mu, 
\end{eqnarray}%
where
\begin{eqnarray}
\mu _{1}=-tr(B_s)
\end{eqnarray}%
is the first eigenvalue of the operator $\Delta _{m-2}
+\sum_{i,j}\Omega _{ij}x^{j}\frac{\partial }{\partial x^{i}}+\Psi
(x^{\prime })$ and
\begin{eqnarray}
\mu_{2}=\int_{_{\mathbb{T}}}cdH.
\end{eqnarray}%
\end{prop}

Finally, using the blow-up analysis, we can characterize the limit measures along the
torus by

\begin{prop}
\label{pp9} If $\nu$ denotes a limit measure on a torus $\mathbb{T}$, then it is
absolutely continuous with respect to the invariant measure $d\theta_1 d\theta_2$.
If $h_{\mathbb{T}}(\theta )$ is the density of $\nu $ with respect to the probability measure
invariant by the flow, we have
\begin{eqnarray*}
h_{\hbox{\bbbis T}}(\theta )=\frac{\gamma _{_{\mathbb{T}}}^{2}\int_{%
\hbox{\bb R}^{n-2}}w_{_{\mathbb{T}}}^{2}(x^{\prime\prime})dx^{\prime \prime}}{\sum_{%
\hbox{\bbbis
T}^{\prime }\in S^{\prime }}\gamma _{\hbox{\bbbis T}^{\prime }}^{2}\int_{%
\hbox{\bbbis T}^{\prime }}\int_{\hbox{\bb R}^{n-2}}w_{\hbox{\bbbis T}%
^{\prime }}^{2}dxf_{\mathbb{T}}^{2}(\theta )dS_{\hbox{\bbbis T}^{\prime }}}
f_{\mathbb{T}}^{2}(\theta ),
\end{eqnarray*}%
where $w_{_{\mathbb{T}}}$ is the function defined by proposition \ref{wwp}, $%
\gamma _{_{\mathbb{T}}}$ is the modulating coefficient and $S^{\prime }$ is
the subset of torii, where the topological pressure is achieved.
\end{prop}

{\noindent \bf Proof.} By definition of the concentration
coefficient (see definition \ref{concencoeffgeneral}), Proposition
\ref{pp9} is
a consequence of Proposition \ref{pp8}, the $L_{2}$ convergence of the blow up sequence $%
w_{\epsilon }$ and Fubini's theorem. The convergence of the sequence $%
w_{\epsilon }$ in $L_2$ can be proved following the steps of section \ref%
{kupka1}: In a tubular neighborhood $T(\delta )$ of the $\mathbb{T}$, there exists
constants $C>0$ and $C_{0}>0$ , such that
\begin{eqnarray*}
w_{\epsilon }(x^{\prime \prime },\theta _{1},\theta _{2})=\frac{v_{\epsilon }(\epsilon ^{1/2}x^{\prime },\theta_1\theta_2, )}{\bar{%
v_{\epsilon }}}\leq Ce^{-C_{0}dist(x,\mathbb{T} )^2}\text{ for }x\in
T(\delta ).
\end{eqnarray*}
\quad

The limit measure $h_{\hbox{\bbbis T}}(\theta )$ restricted to a torus is computed using a regular solution
of a transport equation that we shall describe now.

\begin{lem}
\label{lemmsys} On an irrational torus $\hbox{\bbbis T}$ endowed
with the coordinate system \ref{notat}, consider a $%
C^\infty$ function $c$ and the field
\begin{eqnarray} \label{omegaii}
\Omega ^{//}= k_{1}\frac{\partial }{\partial \theta _{1}%
}+k_{2}\frac{\partial }{\partial \theta _{2}},
\end{eqnarray}
where $k_{1}$ and $k_{2}$ are defined in \ref{notat}. If the small
divisor assumption (\ref{sd}) is satisfied, then the space of
regular and bounded solutions $f_{\hbox{\bbbis T}}$ of the transport
equation
\begin{eqnarray}
<\Omega ^{//},\nabla f_{\hbox{\bbbis T}}>+c(\theta _{1},\theta
_{2})f_{\hbox{\bbbis T}}=\mu_2 f_{\hbox{\bbbis T}}
\label{transport1}
\end{eqnarray}%
is of dimension one and necessarily
\begin{eqnarray*}
\mu_2 = \int_{\mathbb{T}} c(\theta _{1},\theta_{2}) dH.
\end{eqnarray*}
\end{lem}
\bigskip
\textbf{\noindent Proof.}

\noindent The existence of a solution of \ref{transport1} is
obtained by developing $c$ in Fourier series. Writing $w=e^{h}$,
\begin{eqnarray}
<\Omega ^{//},\nabla h>=\lambda -c(\theta _{1},\theta _{2}).
\label{transportbb}
\end{eqnarray}
We look for h as a Fourier series
\begin{eqnarray*}
\sum_{m_{1},m_{2}}h_{m_{1},m_{2}}e^{i(m_{1}k_{1}+m_{2}k_{2})},
\end{eqnarray*}
Because c is $C^{\infty }$, it can be expanded in Fourier series
\begin{eqnarray*}
c(\theta _{1},\theta
_{2})=\sum_{m_{1},m_{2}}c_{m_{1},m_{2}}e^{i(m_{1}k_{1}+m_{2}k_{2})},
\end{eqnarray*}
where the coefficients $c_{m_{1},m_{2}}$ are rapidly decreasing.
From equation (\ref{transport1}) the coefficients are given by
\begin{eqnarray*}
h_{m_{1},m_{2}}=i\frac{c_{m_{1},m_{2}}}{m_{1}k_{1}+m_{2}k_{2}},
\hbox{ for } (m_1,m_2) \neq (0,0).
\end{eqnarray*}%
The compatibility condition coming from the topological pressure
insures that $\mu_2=c_{0,0}= \int_{\mathbb{T}}c(\theta_1,\theta_2)
dH$, which is exactly the Fourier coefficient. It is well known that
the small divisor condition (\ref{sd}) implies that the sequence
$h_{m_{1},m_{2}}$ is
rapidly decreasing and thus $h$ exists, is a regular function and so is $%
w=e^h $. We remark that since $h$ is defined up to an additive
constant,  the space of $ C^{\infty}$ solutions of equation
(\ref{transport1}) is of dimension of 1 . \bigskip

\noindent \textbf{Remark 1.} The value of $\mu _{2}$ can also be
obtained by
 simple considerations: the function $f_{\hbox{\bbbis T}}$ solution
 of equation (\ref{transport1}) is given by the formula
\begin{eqnarray*}
f_{\hbox{\bbbis T}}(X(t))=f_{\hbox{\bbbis T}}(X(0))e^{\mu
_{2}t-\int_{0}^{t}c(X(s))ds}
\end{eqnarray*}%
along any trajectory $X(t)$ of the restriction of b to the torus.
The function $f_{\hbox{\bbbis T}}$ is positive on the Torus, because
the trajectory $X(t)$ is everywhere dense. Also, there exists a
sequence $t_n\rightarrow +\infty$ such that $X(t_n)\rightarrow
X(0)$, hence
\begin{eqnarray*}
\lim_{ n \rightarrow\infty}e^{\mu
_{2}t_n-\int_{0}^{t_n}c(X(s))ds}=1.
\end{eqnarray*} This implies that
\begin{eqnarray*}
\mu _{2}=\lim_{ n \rightarrow\infty} \frac{1}{t_n}
\int_{0}^{t_n}c(X(s))ds.
\end{eqnarray*}
 Using the ergodic property of the trajectories, we get
\begin{eqnarray*}
\lim_{t\rightarrow \infty }\frac{1}{t}\int_{0}^{t}c(X(s))ds=
\int_{_{\mathbb{T}}}cdH=\mu _{2}.
\end{eqnarray*}%
%
%
%

%

\subsection{Proofs of  Proposition \ref{wwp} and \ref{pp8}}

We have shown on several occasions that the sequence $v_{\epsilon
}^{2}dV_{g} $ concentrates on the set $Z_{\Psi }=\{x\in V, |\Psi _{\mathcal{L%
}}(x)=0\}$. Moreover the sequence $v_{\epsilon }$ converges uniformly to
zero on any compact set $K\subset V-Z_{\Psi }$. 
The blow up function
\begin{eqnarray*}
w_{\epsilon }(x^{\prime \prime },\theta _{1},\theta
_{2})=\frac{v_{\epsilon
}(\sqrt{\epsilon }x^{\prime \prime },\theta _{1},\theta _{2})}{\bar{v}%
_{\epsilon }}.
\end{eqnarray*}%
satisfies in a tubular neighborhood $T_{\hbox{\bbbis T}}(\delta )$
of the torus $\hbox{\bbbis T}$, the following equation
\begin{eqnarray}
\Delta _{\varepsilon}^{m-2}w_{\varepsilon }+<\Omega ^{//},\nabla
w_{\varepsilon }>+\frac{\Omega _{i}(\sqrt{\epsilon }x^{\prime \prime
})}{\sqrt{\epsilon }}\frac{\partial
w_{\epsilon }}{\partial x^{\prime \prime }{}^{i}}+(c+\frac{\Delta _{g}%
\mathcal{L}}{2}+\frac{\Psi _{\mathcal{L}}(\sqrt{\varepsilon }x^{\prime \prime })%
}{\varepsilon })w_{\varepsilon }+R_{\varepsilon }(x^{\prime \prime
},\theta _{1},\theta _{2}) =\lambda _{\varepsilon }w_{\varepsilon },
\label{bust}
\end{eqnarray}
where
\begin{eqnarray*}
R_{\epsilon }=\sqrt{\epsilon }L_{1}(w_{\epsilon })+\epsilon
L_{2}(w_{\epsilon })+\epsilon Q(w_{\epsilon }),
\end{eqnarray*}%
and $\Delta _{\varepsilon}^{m-2}$ is the Laplacian in the $x^{\prime \prime }-$%
variables only, converging uniformly to the standard Euclidean
Laplacian
operator on the class of functions with compact support defined on $\hbox{\bb R}%
^{m-2}$. $L_{1}$ and $L_{2}$ are first order operators containing
Christoffel symbols and $\Omega ^{//}$ is defined in
(\ref{omegaii}).

As $\epsilon $ goes to zero, every limit $w$ of $w_{\epsilon }$
satisfies in the weak sense the following equation (the
argumentation is similar to the one provided for limit cycles)
\begin{eqnarray}
\Delta _{E}^{m-2}w+<\Omega ^{//},\nabla w>+\sum_{i,j}\Omega _{ij}x^{j}\frac{\partial w%
}{\partial x^{i}}+(c(0,\theta _{1},\theta _{2})+ \nonumber \\
\frac{\Delta _{g}\mathcal{%
L\mid }_{\mathbb{T}}}{2}+\psi _{2}(x^{\prime \prime }))w=\mu w\hbox{ on }%
\hbox{\bb R}^{n-2}\times \hbox{\bbbis T}_{f},  \label{totor}
\end{eqnarray}%
where
\begin{eqnarray}
\mu =\lim_{n \rightarrow \infty} \lambda_{\varepsilon}.
\end{eqnarray}%
It is understood that this equation is defined on the universal
covering $\mathbb{R}^{n-2}\times \mathbb{R}^{2}$ of the normal
bundle of  ${\mathbb{T}}$ in $V$. $\psi _{2}$ is polynomial of order
2.
\bigskip Consider $\tilde{w}(x,\theta )=\frac{w(x,\theta
)}{f_{{\mathbb{T}}}(\theta )}$, where the function
$f_{{\mathbb{T}}}$ is defined in Lemma \ref{lemmsys}, then
\begin{eqnarray}  \label{mark1}
\Delta _{E}^{m-2}\tilde{w}+k_{1}\frac{\partial \tilde{w}}{\partial
\theta
_{1}}+k_{2}\frac{\partial \tilde{w}}{\partial \theta _{2}}+\sum_{i,j}\Omega _{ij}x^{j}%
\frac{\partial \tilde{w}}{\partial x^{i}}+(\frac{\Delta _{g}\mathcal{L\mid }%
_{\mathbb{T}}}{2}+\psi _{2}(x^{\prime  }))\tilde{w}=\mu
_{1}\tilde{w},
\end{eqnarray}%
where $\mu _{1}=-trB_{s},$ this relation follows from the definition
of the Topological Pressure. Along the characteristics,
$(\dot{\theta _{1}}= k_1,\dot{\theta_{2}}= k_2)$, we define
\begin{eqnarray}
\tilde{w}(x^{\prime  },t)=\tilde{w}(x^{\prime \prime },\theta
_{1}(t),\theta _{2}(t) )
\end{eqnarray}%
then 
\begin{eqnarray}  \label{bheq}
\Delta _{E}^{m-2}\tilde{w}+\frac{\partial \tilde{w}}{\partial
t}+\sum_{i,j} \Omega_{ij}x^{j}\frac{\partial \tilde{w}}{\partial x^{i}}+(\frac{\Delta _{g}%
\mathcal{L\mid }_{\mathbb{T}}}{2}+\psi _{2}(x^{\prime \prime }))\tilde{w}%
=\mu _{1}\tilde{w}.
\end{eqnarray}%
Because $\tilde{w}$ is a solution of a parabolic equation, by the
regularity theorems (see the case of limit cycles), we conclude that
the solution is regular. Moreover on each compact set of
$\mathbb{R}^{n-2}\times \mathbb{R}^2,$ $w$ is the uniform limit of a
sequence $w_{\epsilon _{n}},$ where $\epsilon _{n}$ goes to 0. To
derive an explicit expression for the function $\tilde{w}$ defined
by equation (\ref{bheq}), we use Kolmogorov's representation
formula. Define
\begin{eqnarray*}
\tilde{z}(x^{\prime  },t) = \tilde{w}(x^{ \prime },t)\exp \left(
\frac{1}{2}<Ax^{\prime },x^{\prime }>_{\mathbb{R}^{m-2}}\right)
\end{eqnarray*}
then
\begin{eqnarray}
\tilde{z}(x^{\prime  },t) =\frac{e^{-t TrB^{s}}}{\left( 4\pi \right) ^{%
\frac{m-1}{2}}\sqrt{\det Q_{t }}}\int_{\mathbb{R}^{m-1}}\widetilde{w}%
(y,\theta_1,\theta_2)e^{-q(x,y,t )}dy. \label{komp}
\end{eqnarray}
This notation have been introduced in lemma \ref{lemako}.
$\widetilde{w}$ is
regular and bounded, thus we conclude as in lemma \ref{lemako} that $\tilde{z%
}$ is well defined and belongs to Tychonoff's class. Because
$\tilde{z}$ is not a periodic function of the variable $t$, we
cannot use lemma \ref{repres} to conclude. We have to modify
slightly the proof of lemma \ref{lemako} to obtain an explicit
expression of the function $\tilde{z}$. Instead of periodicity,  we
use the density of the flow of $\Omega^{//}$ on the torus: We denote
by $\tau(t,\theta_1,\theta_2)$ the flow of $\Omega^{//}$. For any
point $(\theta_1,{\theta_2})$  , there exists a sequence $t_n$
converging to infinity such that $\tau(t_n,\theta_1,{\theta_2})
\rightarrow (\theta_1,{\theta_2})$. For n sufficiently large,
\begin{eqnarray*}
|\tilde{z}(x^{\prime },\theta_1,\theta_2) -\frac{\exp^{ -\left[ {t_n }%
TrB_{s}+\frac{1}{4} <U_{{t_n }}x^{\prime },x^{\prime }>_{\mathbb{R}^{m-2}}%
\right]} }{\left( 4\pi \right) ^{\frac{m-1}{2}}\sqrt{\det Q_{{t_n }}}} \times
\\
\int_{\mathbb{R}^{m-2}}\widetilde{w}(\left( \eta ^{\prime }+R_{{t_n }%
}^{-1}P_{t_n}x^{\prime }\right) ,\theta_1,\theta_2)e^{-\frac{1}{2}%
||R_{s,t_n}\eta _{s}^{\prime }||_{\mathbb{R}^{m-2}}^{2}-\frac{1}{2}||R_{u,{%
t_n }}\eta _{u}^{\prime }||_{\mathbb{R}^{m-1}}^{2}}d\eta ^{\prime }
|\leq \frac{1}{n}.
\end{eqnarray*}
Here, we have used relation (\ref{KKOlm}). Using the continuity of
$\widetilde{w}$ and letting n go to infinity, we get
\begin{eqnarray*}
\tilde{z}(x^{\prime },\theta_1,\theta_2) =\frac{\exp -\left[ \frac{1}{4%
}\left( \int_{0}^{+\infty }e^{tB_{s}}e^{tB_{s}^{\ast }}dt\right)
^{-1}x_{s}^{\prime },x_{s}^{\prime }>_{\mathbb{R}^{m-2}}\right]
}{\left( 4\pi \right) ^{\frac{m-2}{2}}\sqrt{\det \int_{0}^{+\infty
}e^{tB_{s}}e^{tB_{s}^{\ast }}dt}} \tilde{\chi}(\theta _{1},\theta
_{2}),
\end{eqnarray*}
where
\begin{eqnarray*}
\tilde{\chi}(\theta _{1},\theta _{2})=\int_{\mathbb{R}^{m-2}}\widetilde{w}(\eta ^{\prime },\theta_1,\theta_2)e^{-%
\frac{1}{2}||R_{s,\infty }\eta _{s}^{\prime
}||_{\mathbb{R}^{m-2}}^{2}-\frac{ 1}{2}||R_{u,\infty }\eta
_{u}^{\prime }||_{\mathbb{R}^{m-2}}^{2}}d\eta ^{\prime }.
\end{eqnarray*}
This identity proves  that  $\tilde{w}$ is the product of two
functions: an exponential function depending only on the variable
$x_{s}$ and a function $\tilde{\chi}$ of the variable $(\theta
_{1},\theta _{2})$.

Finally, setting
\begin{eqnarray}
M_s =\int_{0}^{+\infty }e^{tB_{s}}e^{tB_{s}^{\ast }}dt,
\end{eqnarray}

\begin{eqnarray} \label{ww}
w(x^{\prime  },\theta _{1},\theta _{2}) = \tilde{\chi}(\theta
_{1},\theta _{2})f_{\hbox{\bbbis T}}(\theta _{1},\theta
_{2})\frac{\exp \left[-\left( \frac{1}{2}<Ax^{\prime
},x^{\prime }>_{\mathbb{R} ^{m-2}} \right)- \frac{1}{4%
}  <M_s ^{-1}x_{s}^{\prime },x_{s}^{\prime
}>_{\mathbb{R}^{m-2}}\right] }{\left( 4\pi \right)
^{\frac{m-2}{2}}\sqrt{\det M_s  }} .
\end{eqnarray}
Because $w$ satisfies equation (\ref{eqtorus}) and $f_{\hbox{\bbbis
T}}$ equation \ref{ftilde}, $\tilde{\chi}$ satisfies the equation
\begin{eqnarray*}
<\Omega^{//}, \nabla \tilde{\chi}>=0.
\end{eqnarray*}
Since $\Omega^{//}$ is ergodic, we conclude that $\tilde{\chi}$ is a
constant. \QED
\bigskip

\noindent {\bf  Proof of proposition \ref{pp9}}

\noindent The proof uses the explicit expression of the function $w$
given by expression (\ref{ww}). Indeed, following the steps of
section \ref{limitmeasure}, it can be proved that there exists a
sequence $w_{{\varepsilon_n}}$ which converges to w in
$L_2(\hbox{\bb R}^{n-2}\times \hbox{\bbbis T})$. Using propositions
\ref{wwp} and \ref{pp8} we have
\begin{eqnarray*}
1=\int_{V_{n}}v_{\epsilon_n }^{2} &=&\sum_{P}\bar{v}_{\epsilon_n
}^{2}\epsilon_n
^{m/2}\int_{B_{P}(\delta /\epsilon_n )}w_{\epsilon_n }^{2}(\sqrt{\epsilon_n }%
x)dV_{g_{\epsilon_n }}+ \bar{v}_{\epsilon_n }^{2}\epsilon_n
^{(m-1)/2}\sum_{\Gamma }\int_{T_{\Gamma }(\delta /\epsilon_n
)}w_{\epsilon_n }^{2}(\sqrt{\epsilon_n }x^{\prime },\theta
)dV_{g_{\epsilon_n }}+\\&&\bar{v}_{\epsilon_n }^{2}\epsilon ^{(m-2)/2}\sum_{%
\hbox{\bbbis T}}\int_{T_{\hbox{\bbbis T}}(\delta /\epsilon_n
)}w_{\epsilon_n }^{2}(\sqrt{\epsilon_n }x^{\prime  },\theta
_{1},\theta _{2})dV_{g_{\epsilon_n }}.
\end{eqnarray*}%
Now suppose that a torus is charged,
\begin{eqnarray*}
\lim_{\epsilon_n \rightarrow 0}\bar{v}_{\epsilon_n }^{2}\epsilon_n
^{(m-2)/2}=C>0
\end{eqnarray*}%
and in that case,
\begin{eqnarray*}
\lim_{\epsilon_n \rightarrow 0}\bar{v}_{\epsilon_n }^{2}\epsilon_n
^{(m-1)/2}=\lim_{\epsilon_n \rightarrow 0}\bar{v}_{\epsilon_n
}^{2}\epsilon_n ^{(m)/2}=0,
\end{eqnarray*}%
thus no cycles or points can contribute to the limit measure. Using
the exponential decay of the sequence $w_{\varepsilon_n }$ and the
strong convergence in $L_2(\hbox{\bb R}^{n-2}\times \hbox{\bbbis
T})$ to the function $w$, we obtain the following expression for the
constant $C$:
\begin{eqnarray*}
1=C\sum_{\hbox{\bbbis T}}\int_{\hbox{\bbbis T}}\int_{\hbox{\bb R}%
^{n-2}}\gamma _{\hbox{\bbbis T}}^{2}w_{\hbox{\bbbis T}}(x^{\prime  })^{2}f_{%
\hbox{\bbbis T}}(\theta _{1},\theta _{2})dx^{\prime  }d\theta
_{1}d\theta _{2},
\end{eqnarray*}%
where the sum is extended to the charged torii, where the
topological pressure is attained. $\gamma _{\hbox{\bbbis T}}$ is a
modulating coefficient. When no torii are charged, the limit
measures have been studied in the section \ref{lcs} on limit cycles.
\quad\hbox{\hskip 4pt\vrule width 5pt height 6pt depth 1.5pt}


\section{Conjectures and open questions }

In this section, we offer some conjectures and state some open
problems.

An interesting extension of the present work would be to study the
blow up function on a component of the recurrent set that is a
manifold $M_k$ of dimension k ($\geq 2$) normally hyperbolic.
Probably to obtain substantial results, one has to make more
assumptions on the hyperbolic structure and the restriction of the
field to the manifold, such as assuming that the restriction is
ergodic (with respect to some invariant measure on the manifold).

For example, it could happen that with some unspecified assumptions
on  the field and on $M_k$ , we could have a variable-separation
phenomenon for the limit of the blown up sequence as in lemma
\ref{lemmsys} for the case of a torus and in proposition \ref{prop3}
for a cycle.

In a different direction, if the restriction of the field to the
manifold $M_k$ has a nonzero entropy, it would be interesting to
study the limit of the blown up function, as we did here in lemma
\ref{lemmsys} when the entropy is zero. In particular, one can
expect a new characterization of the entropy $\mu$ as follow:
\begin{eqnarray} \mu = \lim_{T\rightarrow \infty} \frac{1}{T} \ln
w_{M_k}(X(T)), \end{eqnarray} where $X(T)$ is the flow of the field
restricted to $M_k$, $w_{M_k}$ is a nonzero solution of equation
\begin{eqnarray}
<\Omega ^{//},\nabla w_{M_k}>+cw_{M_k}=\mu_2 w_{M_k},
\label{transport}
\end{eqnarray}
$\Omega ^{//}$ is the restriction of the field to $M_k$ and
\begin{eqnarray}
\mu_2 = \mu +\int cdH.
\end{eqnarray}
H is the unique invariant and ergodic measure with respect to
$\Omega ^{//}$.


\subsection{Anisotropic concentration when  $\Psi $ vanishes to order 4 and more.}

In the present work, we have assumed that the special Lyapunov
function $\Psi_{\mathcal{L}}$ vanishes at order 2 on the recurrent
set of the field. However, when $\Psi_{\mathcal{L}}$ vanishes to a
higher order (four or more),  the analysis used here to study the
concentration process of the eigenfunction sequence does not apply
anymore: when $\Psi_{\mathcal{L}}$ vanishes at order 4, the limit
measures are not necessarily concentrated on the components of the
recurrent set on which the topological pressure is attained. In
fact, the sequence $w_{\epsilon}$ may not converge in $L_2$ because
in equation (\ref{ww}), the function $\psi_2$ is identically zero.
To analyze furthermore this case, we consider a small ball centered
at a saddle point $S$ where the dimension of the stable and unstable
spaces satisfies $m_s+m_u=m$. Consider a canonical cartesian
structure in the neighborhood of a saddle point, defined by the
stable and unstable manifolds \cite{Robin}. In that cartesian
structure ($x_s$ (stable), $x_u$ (unstable)), define the following
cones: for small $\delta>0$,
\begin{eqnarray*}
C_{s}(\delta)=\{ x=(x_{u},x_{s}) \hbox{ such that } \mid x_{u} \mid
\leq \mid x_{s} \mid \leq\delta \}
\end{eqnarray*}
and
\begin{eqnarray*}
C_{u}(\delta)=\{ x=(x_{u},x_{s}) \hbox{ such that } \mid x_{s} \mid
\leq \mid x_{u} \mid \leq \delta \}.
\end{eqnarray*}
We have the following: for small $\delta>0$,
\begin{eqnarray}  \label{ratios}
\lim_{\epsilon \rightarrow
0}\frac{\int_{C_{s}(\delta)}v^{2}_{\epsilon} }{
\int_{C_{u}(\delta)}v^{2}_{\epsilon}}=0.
\end{eqnarray}
This shows that concentration does not take place in all the
neighborhood of the saddle point. Rather, it occurs along the
unstable manifold. This is a totally anisotropic concentration.

We are now going to prove formula \ref{ratios}. Because the sequence
$w_{\epsilon_n}$  converges to a function $w_s$ and all these
functions decay exponentially in the variable $x_s$ only (in the
neighborhood of a stable manifold), we have in the cartesian
coordinates,

\begin{eqnarray*}
 \lim_{ \epsilon \rightarrow 0} \frac{1 }{\epsilon^{m/2} \bar{v}^2 _{\epsilon}}
\int_{C_{s}} v^{2}_{\epsilon} dx= \lim_{ \epsilon \rightarrow 0}
\int_{B_{s}(0,\delta/\sqrt{\epsilon})} w^{2}_{s}(x_s)
 \int_{B_{u}(0, ||x_s||)} dx^{u}dx^{s} <\infty,
\end{eqnarray*}
where the exterior integral converges because $\int_{B_{u}(0,
||x_s||)} dx^{u}$ is bounded by a polynomial in $||x_s||$ and
$w^{2}_{s}$ decays exponentially.

Now,
\begin{eqnarray*}
  \frac{1 }{\epsilon^{m/2} \bar{v}^2 _{\epsilon}}
 \int_{C_{u}} v^{2}_{\epsilon} &=&
\int_{B_{u}(0,\delta/\sqrt{\epsilon})}dx^{u}
 \int_{B_{s}(0, ||x_u||)} w^{2}_{s}(x_s) dx^{s} \\
&\geq& C\epsilon^{-n_u/2}\delta^{n_u}.
\end{eqnarray*}
where $C>0$ is a constant. Thus the ratio in equation \ref{ratios}
converges to zero.


\subsection{Final remarks}



\subsubsection{Concentration near polycycle}

When the recurrent set of the field contains polycycles, (that is a
close curve which is the union of critical points and separatrices
joining these points). In that case, we expect that the
concentration will not occur uniformly on the polycycle, but rather
at the critical points. It is unclear what is the effect on the
concentration phenomena of the discontinuity of the tangent vectors
at the critical points.


\subsubsection{The case of Hamiltonian systems}

We have developed here a method for hyperbolic field, but for
Hamiltonian systems, the present method cannot be applied and a new
approach has to be introduced. In particular it is not clear what is
the support of the limit measures.

\subsubsection{Study the spectrum}

The characterization of the set of possible limit measures of
eigenfunctions associated to other eigenvalues remains an open
problem. Such a study is feasible because for a Morse-Smale drift,
the blow up analysis near the points and cycles leads to an
eigenfunction problem of the Ornstein -Ulhenbeck operator and the
solution can be expressed explicitly by Hermite functions.


\subsubsection{Asymptotic computation}

The existence of a asymptotic expansion,
\begin{equation*}
\lambda _{\epsilon }\approx \text{topological
pressure}+I_{1}\epsilon ^{1/2}+I_{2}\epsilon +...
\end{equation*}%
of the first eigenvalue $\lambda _{\epsilon }$ as a function of
$\epsilon $ is an open problem and so is the determination of
  the coefficients $I_{k}$.


\subsection{Appendix}

1)The notations are those of the section "Proofs of the theorems" except we
denote the function $w$ by $w_{0}$ so as to avoid confusions. First let us
show that $w_{0}$ is $C^{\infty }.$Recall that $w_{0}$ belongs to $%
L_{loc}^{2} $ and that it is a weak solution of equation%
\begin{eqnarray*}
-\sum_{i=1}^{m-1}\frac{\partial ^{2}w}{\left( \partial x^{i}\right) ^{2}}%
+\sum_{i,j=1}^{m-1}\Omega _{j}^{i}x^{j}\frac{\partial w}{\partial x^{i}}+%
\frac{\partial w}{\partial \theta }+((c+\frac{\Delta _{g}\mathcal{L}}{2}%
)(0,\theta )+\psi _{2}(x^{\prime }))w=\lambda w
\end{eqnarray*}

The theorems 13.4.1,page 191,vol.II and 4.4.1,page110,vol I of
(\cite{hor}) show that any weak solution of such a parabolic
operator is $C^{\infty }.$

\bigskip 2)Recall the operator:%
\begin{eqnarray*}
L_{\varepsilon }=L_{2\varepsilon }+L_{1\epsilon }+L_{0\varepsilon }
\end{eqnarray*}

where:

\begin{eqnarray*}
L_{2\varepsilon }=-\sum_{i,j=1}^{m-1}g_{\varepsilon }^{ij}\frac{\partial ^{2}%
}{\partial x^{i}\partial x^{j}}-2\sqrt{\varepsilon }\sum_{i=1}^{m-1}g_{%
\varepsilon }^{im}\frac{\partial ^{2}}{\partial x^{i}\partial \theta }%
-\varepsilon g_{\varepsilon }^{mm}\frac{\partial ^{2}}{\partial \theta ^{2}}
\end{eqnarray*}

\begin{eqnarray*}
L_{1\epsilon }=\sum_{k=1}^{m-1}\left( \frac{\Omega ^{k}}{\sqrt{\varepsilon }}%
-\sqrt{\varepsilon }\sum_{i,j=1}^{m}g_{\varepsilon }^{ij}\Gamma
_{\varepsilon ij}^{k}\right) \frac{\partial }{\partial x^{k}}+\left( \Omega
^{m}-\varepsilon \sum_{i,j=1}^{m}g_{\varepsilon }^{ij}\Gamma _{\varepsilon
ij}^{m}\right) \frac{\partial }{\partial \theta }
\end{eqnarray*}

\begin{eqnarray*}
L_{0\varepsilon }=c_{\varepsilon }+\frac{\left( \Delta _{g}\mathcal{L}%
\right) _{\varepsilon }}{2}+\frac{\Psi _{\mathcal{L\varepsilon }}}{%
\varepsilon }
\end{eqnarray*}

and $g_{\varepsilon }^{ij}=g^{ij}(x^{\prime }\sqrt{\varepsilon },\theta
),1\leq i,j\leq m,$ $\Gamma _{\varepsilon ij}^{k}=\Gamma _{ij}^{k}(x^{\prime
}\sqrt{\varepsilon },\theta ),$ $\left( \Delta _{g}\mathcal{L}\right)
_{\varepsilon }=\Delta _{g}\mathcal{L}(x^{\prime }\sqrt{\varepsilon },\theta
),$ $c_{\varepsilon }=c(x^{\prime }\sqrt{\varepsilon },\theta ),\Psi _{%
\mathcal{L\varepsilon }}=\Psi _{\mathcal{L}}(x^{\prime }\sqrt{\varepsilon }%
,\theta ).$Using Schweins'formulas we can write:

\begin{eqnarray*}
L_{\varepsilon }=L_{2\varepsilon }^{^{\prime }}+L_{1\epsilon
}^{^{\prime }}+L_{0\varepsilon }
\end{eqnarray*}

\begin{eqnarray*}
L_{2\varepsilon }^{^{\prime }}=\sum_{k=1}^{m}X_{k}^{\ast }X_{k}
\end{eqnarray*}

\begin{eqnarray*}
L_{1\epsilon }^{^{\prime }}=\sum_{k=1}^{m-1}\left( \frac{\Omega ^{k}}{\sqrt{%
\varepsilon }}+\sqrt{\varepsilon }\sum_{i,j=1}^{m}g_{\varepsilon
}^{ik}\Gamma _{\varepsilon ji}^{j}\right) \frac{\partial }{\partial x^{k}}%
+\left( \Omega ^{m}+\varepsilon \sum_{i,j=1}^{m}g_{\varepsilon
}^{im}\Gamma _{\varepsilon ji}^{j}\right) \frac{\partial }{\partial
\theta }
\end{eqnarray*}

where:%
\begin{eqnarray*}
X_{k}=\sum_{i=1}^{k}a_k^{i}\frac{\partial }{\partial x^{i}}
\end{eqnarray*}
for k=1,..,m-1,

\begin{eqnarray*}
X_{m}=\sum_{i=1}^{m-1}a_k^{i}\frac{\partial }{\partial x^{i}}
+\sqrt{\varepsilon }a_k^{m}\frac{\partial }{\partial \theta }.
\end{eqnarray*}
$X_{k}^{\ast }$ is the formal adjoint of $X_{k}$%
\begin{eqnarray*}
X_{k}^{\ast }=-\sum_{i=1}^{k}\frac{\partial }{\partial
x^{i}}a^{i}_{k},
\end{eqnarray*}
for k=1,..,m-1, and
\begin{eqnarray*}
X_{m}^{\ast }=-\sum_{i=1}^{m-1}\frac{\partial }{\partial x^{i}}a^{i}_{k}-\sqrt{%
\varepsilon }\frac{\partial }{\partial \theta }a^{m}_{k},
\end{eqnarray*}
where $1\leq i\leq j\leq m$:%
\begin{eqnarray*}
a^{i}_j=\frac{G(i j)}{\sqrt{G(j j)G(j+1 j+1)}},
\end{eqnarray*}
and
\begin{eqnarray*}
G(j j)&=&det\{ g_{\varepsilon }^{\alpha \beta };j\leq \alpha,
\beta\leq m\}\\
G(i j)&=&det\{ g_{\varepsilon }^{\alpha \beta };i\leq \alpha\leq m,
j\leq \beta\leq n\}.
\end{eqnarray*}
\begin{eqnarray*}
g_{\varepsilon }^{\alpha \beta}(x',\theta)=g_{\varepsilon }^{\alpha
\beta}(\sqrt{\varepsilon}x',\theta)
\end{eqnarray*}
When $\varepsilon \rightarrow 0+,L_{\varepsilon }\rightarrow
L_{0}=L_{20}^{\prime }+L_{10}^{\prime }+L_{00}$ where:%
\begin{eqnarray*}
L_{20}^{\prime }=-\sum_{i=1}^{m-1}\frac{\partial ^{2}}{\left(
\partial x^{i}\right) ^{2}}
\end{eqnarray*}

\begin{eqnarray*}
L_{10}^{\prime }=\sum_{i,j=1}^{m-1}\Omega _{j}^{i}x^{j}\frac{\partial }{%
\partial x^{i}}+\frac{\partial }{\partial \theta }
\end{eqnarray*}

Recall that by our choice of coordinates along a cycle $\Omega ^{m}(0,\theta
)=1.$Also $X_{k}\rightarrow \frac{\partial }{\partial x^{k}}$ as $%
\varepsilon \rightarrow 0,$if $1\leq k\leq m-1,$and $X_{m}\rightarrow 0.$

We know that $w$ is a weak solution of the equation:%
\begin{eqnarray}
L_{0}w=\lambda w  \label{limeq}
\end{eqnarray}

where $\lambda =\underset{\varepsilon \rightarrow 0}{\lim }\lambda
_{\varepsilon }.$\ Hence it follows from Theorem 22.2.1,page 353, in volume
274 \ of (\cite{hor}) \ that $w$ is a $C^{\infty }$ solution of equation (%
\ref{limeq}).

On the other hand for any $\varepsilon >0,w_{\varepsilon }$satisfies the
relation%
\begin{eqnarray*}
L_{\varepsilon }w_{\varepsilon }=\lambda _{\varepsilon }w_{\varepsilon }%
\text{ on }\mathbb{R\times }B_{0}^{m-1}(\delta /\sqrt{\varepsilon })
\end{eqnarray*}

Let $\ \delta _{\varepsilon }=w_{\varepsilon }-w_{0}.$ Then $\delta
_{\varepsilon }$ satisfies the following equation on $\mathbb{R\times }%
B_{0}^{m-1}(\delta /\sqrt{\varepsilon })$%
\begin{eqnarray}
L_{\varepsilon }\delta _{\varepsilon }=\lambda _{\varepsilon }\delta
_{\varepsilon }+\left( L_{0}-L_{\varepsilon }\right) w_{0}+\left(
\lambda _{\varepsilon }-\lambda \text{ }\right) w_{0}
\label{auuxeq}
\end{eqnarray}

It is easy to see that for any compact $\mathcal{K\subset }\mathbb{R}^{m-1},$
there is a $\varepsilon _{\mathcal{K}}>0,$such that $\mathcal{K\subset }%
B_{0}^{m-1}(\delta /\sqrt{\varepsilon })$ if $\ 0<\varepsilon \leq
\varepsilon _{\mathcal{K}}.$ Moreover there exists a $C^{\infty }$
differential operator $P_{\varepsilon }$ defined on $\mathbb{R\times }%
B_{0}^{m-1}(\delta /\sqrt{\varepsilon _{\mathcal{K}}})$ such that
\begin{eqnarray*}
L_{\varepsilon }-L_{0}=\sqrt{\varepsilon }P_{\varepsilon }
\end{eqnarray*}

Taking a compact set $\mathcal{K}_{1},$ two functions $\phi _{1},\phi _{2}$
in $C_{c}^{\infty }(\mathbb{R}^{m-1};[0,1])$ such that $\phi _{1}=1$ on $%
\mathcal{K}_{1}$ and that $\phi _{2}=1$ on the support of $\phi _{1},$ for
any $s>0,$ there are constants $\sigma =\sigma (\mathcal{K}_{2})>0,\tau
<0,C_{s}=C(s,\mathcal{\phi }_{1},\phi _{2})$ such that for all $\varepsilon
, $ $0\leq \varepsilon \leq \varepsilon _{\mathcal{K}_{2}}$ where $\mathcal{K%
}_{2}$ is the support of $\phi _{2},$%
\begin{eqnarray}
\left\vert \left\vert \phi _{1}\delta _{\varepsilon }\right\vert
\right\vert _{s+\sigma }\leq C_{s}\left( \varepsilon \left\vert
\left\vert \phi _{2}w_{0}\right\vert \right\vert _{s+2}+\left\vert
\lambda _{\varepsilon }-\lambda \right\vert \left\vert \left\vert
\phi _{2}w_{0}\right\vert \right\vert _{s}+\left\vert \left\vert
\phi _{2}\delta _{\varepsilon }\right\vert \right\vert _{\tau
}\right)  \label{horm}
\end{eqnarray}

This follows from Lemmas 22.2.4, page 356 and 22.2.5,page 357 in vol.III of (%
\cite{hor}). Because in our case the coefficients of the operator $%
L_{\varepsilon }$ and all their derivatives are continuous functions of the
variables ($x^{\prime }$,$\theta $) and also $\varepsilon \in \lbrack 0,1],$%
it is easy to check that all the constant appearing in the proofs of
section 22.2 in vol. III of (\cite{hor}) can be taken independent of
$\varepsilon .$ $\phi _{2}w_{0}$ is the strong limit of the
sequence \{$\phi _{2}w_{\varepsilon _{n}}|n\in \mathbb{N}$\}in any space $%
H_{\rho }$ \ with $\rho <0.$

The equation (\ref{horm}) implies that the sequence $\{\delta _{\varepsilon
_{n}}|n\in \mathbb{N}\}$ and all its derived sequences tend to 0 uniformly.

\subsection{Appendix II: estimation of $E_x \{ \chi _{3}(X_{\varepsilon
}) e^{-\frac{\int_0^t\Psi_L(X_\varepsilon)(s)ds}{\varepsilon}} \}$}

$\chi _{3}=\chi _{E_{3}},$ $E_{3}=\ \{\gamma |d_{\infty }(\gamma
,\gamma _{x})\leq \eta \}.$ \ \ Evaluate:\ $E_{x}[\chi
_{3}(X_{\varepsilon }(t))\exp
\left( -\int_{0}^{t}\frac{\Psi _{\mathcal{L}}(X_{\varepsilon }(s))ds}{%
\varepsilon }\right) ]$
\begin{eqnarray}
dX_{\varepsilon }(t)=-\Omega _{\varepsilon }(X_{\varepsilon
}(t))dt+\sqrt{2\varepsilon }\sigma (X_{\varepsilon }(t))dw(t),
\end{eqnarray}
where
\begin{eqnarray}
\Omega _{\varepsilon }(x)=\Omega (x)+\varepsilon \widehat{\Omega
}(x), \hbox{ where } \\
\widehat{\Omega }^{k}=\sum_{ij=1}^{m}g^{ij}\Gamma _{ij}^{k},1\leq
k\leq m.
\end{eqnarray}
Define
\begin{eqnarray}
 Y_{\varepsilon }&=&X_{\varepsilon }-\gamma _{x},\\
 dY_{\varepsilon }(t)&=&-\left( \gamma _{x}^{\prime }+\Omega
_{\varepsilon }(Y_{\varepsilon }(t)+\gamma _{x}(t))\right)
dt+\sqrt{2\varepsilon }\sigma (Y_{\varepsilon }(t)+\gamma
_{x}(t))dw(t)
\end{eqnarray}
Define
\begin{eqnarray}
dZ_{\varepsilon }(t)=\left( \Omega (\gamma_x(t))-\Omega
_{\varepsilon }(Z_{\varepsilon }(t)+\gamma _{x}(t))\right) dt+\sqrt{
2\varepsilon }\sigma (Z_{\varepsilon }(t)+\gamma _{x}(t))dw(t)
\end{eqnarray}
Applying Girsanov's formula between $Y_{\varepsilon }$ and
$Z_{\varepsilon}$, we get

\begin{eqnarray*}
\frac{dP_{Y_{\varepsilon }}}{dP_{Z_{\varepsilon }}}=\exp \{-
\int_{0}^{t}<\gamma _{x}^{\prime }(s)+\Omega _{\varepsilon }(\gamma
_{x}(s)),\ dZ_{\varepsilon }(s)-\Omega (Z_{\varepsilon }(s))+\Omega
_{\varepsilon }(Z_{\varepsilon }(s)+\gamma
_{x}(s))>_{g(Z_{\varepsilon }(s)+\gamma
_{x}(s))}\\
+\frac{1}{2}\int_{0}^{t}||\gamma _{x}^{\prime }(s)+\Omega
_{\varepsilon }(\gamma _{x}(s))||_{_{g(Z_{\varepsilon }(s)+\gamma
_{x}(s))}}^{2} \}
\end{eqnarray*}

$E_{x}[\chi _{3}(X_{\varepsilon }(t))\exp \left( -\int_{0}^{t}\frac{\Psi _{%
\mathcal{L}}(X_{\varepsilon }(s))ds}{\varepsilon }\right)
=E_{x}^{P_{Z_{\varepsilon }}}[\chi _{3}(Z_{\varepsilon }(t)+\gamma
_{x}(t))\exp \left( -I(t)\right) ]$

\begin{eqnarray*}
I(t)&=&\frac{1}{\varepsilon }\int_{0}^{t} \Psi _{\mathcal{L}%
}(Z_{\varepsilon }(s)+\gamma _{x}(s))+<\gamma _{x}^{\prime
}(s)+\Omega _{\varepsilon }(\gamma _{x}(s)),\sqrt{2\varepsilon
}\sigma (Z_{\varepsilon
}(s)+\gamma _{x}(s))dw(s)\ >_{g(Z_{\varepsilon }(t)+\gamma _{x})}+\\ & &\frac{1}{2\varepsilon}%
\int_{0}^{t}||\gamma _{x}^{\prime }(s)+\Omega (\gamma
_{x}(s))||_{_{g(Z_{\varepsilon }(s)+\gamma _{x}(s))}}^{2} ds
\end{eqnarray*}

$Z_{\varepsilon }=\sqrt{\varepsilon }z_{1}+\varepsilon z_{2}+\varepsilon ^{%
\frac{3}{2}}R_{3,\varepsilon }=\sqrt{\varepsilon }z_{1}+\varepsilon
R_{2,\varepsilon }=\sqrt{\varepsilon }R_{1,\varepsilon }$ \ \ \
\begin{eqnarray}
dz_{1}=-\frac{\partial \Omega }{\partial x}(\gamma _{x})z_{1}dt+\sqrt{2}%
\sigma (\gamma _{x})dw,
\end{eqnarray}
\begin{eqnarray}
dz_{2}=-\left[ \frac{\partial \Omega }{\partial x%
}(\gamma _{x})z_{2}+\frac{1}{2}\frac{\partial ^{2}\Omega }{\left(
\partial
x\right) ^{2}}(\gamma _{x})\left[ z_{1},z_{1}\right] +\widehat{\Omega }%
(\gamma _{x})\right] dt+\sqrt{2}\frac{\partial \sigma }{\partial
x}(\gamma _{x})\left[ z_{1}\right] dw,
\end{eqnarray}
where $\left[ z_{1},z_{1}\right]$ mean the tensorial product. We
define the following notation: if A and B denote two n-dimensional
vectors, $A \odot B$ is a matrix of coordinates:
\begin{eqnarray*}
(A \odot B)_{kq}= \frac{1}{2}(A_kB_q+ A_qB_k)
\end{eqnarray*}

\begin{eqnarray*}
dR_{3,\varepsilon }&=& \{-\frac{\partial \Omega }{\partial x}(\gamma
_{x})R_{3,\varepsilon }-\frac{\partial ^{2}\Omega
}{\left( \partial x\right) ^{2}}(\gamma _{x})\left[ z_{1}\odot R_{2,\varepsilon }+%
\sqrt{\varepsilon }R_{2,\varepsilon }\odot R_{2,\varepsilon }\right]
\\ &&-\int_{0}^{1}\frac{(1-\theta )^{2}}{2}\frac{\partial ^{3}\Omega
}{\left( \partial x\right) ^{3}}(\gamma _{x}+\theta Z_{\varepsilon
})[R_{1,\varepsilon }\odot R_{1,\varepsilon }\odot R_{1,\varepsilon
} ]d\theta + \int\limits_{0}^{1}\frac{\partial \widehat{\Omega
}}{\partial x}(\gamma _{x}+\theta Z_{\varepsilon })[R_{1,\varepsilon
}]d\theta\} dt\\
&&+\sqrt{2}\frac{\partial \sigma }{%
\partial x}(\gamma _{x})\left[ R_{2,\varepsilon }\right] dw+\int_{0}^{1}(1-%
\theta )\frac{\partial ^{2}\sigma }{\left( \partial x\right)
^{2}}(\gamma _{x}+\theta Z_{\varepsilon })\left[ R_{1,\varepsilon
}\odot R_{1,\varepsilon }\right] dw d\theta
\end{eqnarray*}

\begin{eqnarray*}dR_{2,\varepsilon }&=& \{-\frac{\partial \Omega }{\partial
x}(\gamma _{x})R_{2,\varepsilon }-\frac{1}{2}\frac{\partial
^{2}\Omega }{\left(
\partial x\right) ^{2}}(\gamma _{x})\left[ R_{1,\varepsilon }\odot
R_{1,\varepsilon }\right] -\\&&\sqrt{\varepsilon
}\int_{0}^{1}\frac{(1-\theta )^{2}}{2}\frac{\partial ^{3}\Omega
}{\left( \partial x\right) ^{3}}(\gamma _{x}+\theta Z_{\varepsilon
})[R_{1,\varepsilon }\odot R_{1,\varepsilon }\odot R_{1,\varepsilon
}]d\theta +\int\limits_{0}^{1}\frac{\partial \widehat{\Omega
}}{\partial x}(\gamma _{x}+\theta Z_{\varepsilon
})[R_{1,\varepsilon }]d\theta \} dt\\
&& +\sqrt{2}\frac{\partial \sigma }{%
\partial x}(\gamma _{x})\left[ R_{2,\varepsilon }\right] dw+\int_{0}^{1}(1-%
\theta )\frac{\partial ^{2}\sigma }{\left( \partial x\right)
^{2}}(\gamma _{x}+\theta Z_{\varepsilon })\left[ R_{1,\varepsilon
}\odot R_{1,\varepsilon }\right] d\theta dw
\end{eqnarray*}

Let us decompose
\begin{eqnarray}
I(t)=\frac{1}{\varepsilon}I_{0}(t)+I_{1}(t)+I_{2}(t)+I_{3}(t),
\end{eqnarray}
 where
\begin{eqnarray}
 I_{0}(t)=\int_{0}^{t}\left[ \left[ \Psi _{\mathcal{L}}(\gamma _{x}(s))\right] +%
\frac{1}{2}||\gamma _{x}^{\prime }(s)+\Omega (\gamma
_{x}(s))||_{_{g(\gamma _{x}(s))}}^{2}\right]ds,
\end{eqnarray}

\begin{eqnarray*}
I_{1}(t)&=&\frac{1}{\sqrt{\varepsilon }}\int_{0}^{t}\left[ \sum_{k=1}^{m}%
\frac{\partial \Psi _{\mathcal{L}}(\gamma _{x})}{\partial x^{k}}+%
\frac{1}{2}\sum_{ijk=1}^{m}\frac{\partial g_{ij}(\gamma _{x})}{\partial x^{k}%
}(\gamma _{x}^{\prime ,i}+\Omega ^{i}(\gamma _{x}))(\gamma
_{x}^{\prime,j}+\Omega ^{j}(\gamma _{x}))\right] z_{1}^{k}ds+\\
&&\frac{\sqrt{2}}{\sqrt{%
\varepsilon }}\int_{0}^{t}<\gamma _{x}^{\prime }+\Omega
_{\varepsilon }(\gamma _{x}),\sigma (\gamma _{x})dw>_{g(\gamma
_{x})}
\end{eqnarray*}

\begin{eqnarray*}
I_{2}(t)&=&\int_{0}^{t} { \sum_{k=1}^{m}\frac{\partial \Psi _{\mathcal{L}%
}(\gamma _{x})}{\partial x^{k}}z_{2}^{k}+\frac{1}{2}\sum_{ijk=1}^{m}\frac{%
\partial g_{ij}(\gamma _{x})}{\partial x^{k}}(\gamma _{x}^{\prime ,i}+\Omega
^{i}(\gamma _{x}))(\gamma _{x}^{\prime ,j}+\Omega ^{j}(\gamma_{x}))z_{2}^{k}}\\
&& \frac{1}{4}\sum_{ijkl=1}^{m}\frac{\partial
^{2}g_{ij}(\gamma_{x})}{\partial x^{k}\partial x^l}(\gamma
_{x}^{\prime ,i}+\Omega^{i}(\gamma _{x}))(\gamma _{x}^{\prime
,j}+\Omega ^{j}(\gamma _{x}))z_{1}^{k}z_{1}^{l}
+\frac{1}{2}\sum_{kl=1}^{m}\frac{\partial^{2}
\Psi _{\mathcal{L}}(\gamma _{x})}{\partial x^{k}\partial x^{l}}z_{1}^{k}z_{1}^{l}
+  \\
& &  \sqrt{2}\int_{0}^{t} \sum_{ijkl=1}^{m}%
\left[\frac{\partial g_{ij}(\gamma _{x})}{\partial
x^{k}}(\gamma_{x}^{\prime ,i}+\Omega ^{i}(\gamma _{x}))\sigma
^{jl}(\gamma _{x}) +g_{ij}(\gamma _{x})(\gamma _{x}^{\prime
,i}+\Omega ^{i}(\gamma _{x}))\frac{\partial \sigma ^{jl}(\gamma
_{x})}{\partial x^{k}}\right]z_{1}^{k}dw_{l}
\end{eqnarray*}

\begin{eqnarray*}
I_{3}(t)=I_{31}(t)+I_{32}(t)+I_{33}(t)+I_{34}(t)
\end{eqnarray*}

\begin{eqnarray*}
I_{31}(t)=\sqrt{\varepsilon }\int_{0}^{t} \sum_{k=1}^{m}\left( \frac{%
\partial \Psi _{\mathcal{L}}(\gamma _{x})}{\partial x^{k}}+\frac{1}{2}%
\sum_{ij=1}^{m}\frac{\partial g_{ij}(\gamma _{x})}{\partial
x^{k}}(\gamma _{x}^{\prime ,i}+\Omega ^{i}(\gamma _{x})(\gamma
_{x}^{\prime ,j}+\Omega ^{j}(\gamma _{x})\right) R_{3,\varepsilon
}^{k}ds
\end{eqnarray*}

\begin{eqnarray*}
I_{32}(t)&=&\sqrt{\varepsilon
}\int_{0}^{t}\frac{1}{2}\sum_{kl=1}^{m}\left( \frac{\partial
^{2}\Psi _{\mathcal{L}}(\gamma _{x})}{\partial x^{k}\partial
x^{l}}+\frac{1}{2}\sum_{ij=1}^{m}\frac{\partial ^{2}g_{ij}(\gamma _{x})}{%
\partial x^{k}\partial x^{l}}(\gamma _{x}^{\prime ,i}+\Omega ^{i}(\gamma
_{x}))(\gamma _{x}^{\prime ,j}+\Omega ^{j}(\gamma _{x}))\right)
\times
\\&&\left(
z_{1}^{k}R_{2,\varepsilon }^{l}+z_{1}^{l}R_{2,\varepsilon }^{k}+\sqrt{%
\varepsilon }R_{2,\varepsilon }^{k}R_{2,\varepsilon }^{l}\right) ds
\end{eqnarray*}

\begin{eqnarray*}
I_{33}(t)&=&\sqrt{\varepsilon }\int_{0}^{t}\int_{0}^{1}\frac{(1-\theta )^{2}}{%
2}\sum_{kl,n=1}^{m}( \frac{\partial ^{3}\Psi _{\mathcal{L}}(\gamma
_{x}+\theta Z_{\varepsilon })}{\partial x^{k}\partial x^{l}\partial
x^{n}}+ \\& & \frac{1}{2}\sum_{ijkl,n=1}^{m}\frac{\partial
^{3}g_{ij}(\gamma _{x}+\theta Z_{\varepsilon })}{\partial
x^{k}\partial x^{l}l\partial x^{n}}(\gamma _{x}^{\prime ,i}+\Omega
^{i}(\gamma _{x}))(\gamma _{x}^{\prime ,j}+\Omega ^{j}(\gamma
_{x}))) \times \\& & R_{1,\varepsilon }^{k}R_{1,\varepsilon
}^{l}R_{1,\varepsilon }^{n}d\theta ds
\end{eqnarray*}

\begin{eqnarray*}
I_{34}(t)&=&\sqrt{\frac{\varepsilon
}{2}}\int_{0}^{1}\int_{0}^{t}(1-\theta )\sum_{ijkl,n=1}^{m } [
\frac{\partial ^{2}g_{ij}(\gamma _{x}+\theta Z_{\varepsilon
})}{\partial x^{k}\partial x^{l}}(\gamma _{x}^{\prime ,i}+\Omega
^{i}(\gamma _{x})\sigma ^{jn}(\gamma _{x}+\theta Z_{\varepsilon
})+\\
&& 2\frac{\partial g_{ij}(\gamma _{x}+\theta Z_{\varepsilon
})}{\partial x^{k}}(\gamma _{x}^{\prime ,i}+\Omega ^{i}(\gamma
_{x})\frac{\partial \sigma ^{jn}(\gamma _{x}+\theta Z_{\varepsilon
})}{\partial x^{l}} \\
& & +g_{ij}(\gamma _{x}+\theta
Z_{\varepsilon })(\gamma _{x}^{\prime ,i}+\Omega ^{i}(\gamma _{x})\frac{%
\partial ^{2}\sigma ^{jn}(\gamma _{x}+\theta Z_{\varepsilon })}{\partial
x^{k}\partial x^{l}} ] R_{1,\varepsilon }^{k}R_{1,\varepsilon
}^{l}dw_{n}d\theta
\end{eqnarray*}

\emph{Estimate of $I_{1}(t).$ }
 The Euler-Lagrange  equations of the minimization problem imply that if
 we set, for $1\leq k\leq m$
\begin{eqnarray*}
p_{k}=\sum_{i=1}^{m}g_{ik}(\gamma _{x})(\gamma _{x}^{\prime
,i}+\Omega ^{i}(\gamma _{x})) \end{eqnarray*}

we have using equations \ref{equa1}, \ref{equa2} of the Hamiltonian
system,
\begin{eqnarray*}
\frac{dp_{k}}{ds}=\frac{\partial \Psi _{\mathcal{L}}(\gamma
_{x})}{\partial
x^{k}}-\frac{1}{2}\sum_{ij=1}^{m}\frac{\partial g^{ij}(\gamma _{x})}{%
\partial x^{k}}p_{i}p_{j}+\sum_{j=1}^{m}p_{j}\frac{\partial \Omega
^{j}(\gamma _{x})}{\partial x^{k}}
\end{eqnarray*}
Hence:
\begin{eqnarray*}
\frac{dp_{k}}{ds}=\frac{\partial \Psi _{\mathcal{L}}(\gamma
_{x})}{\partial
x^{k}}+\frac{1}{2}\sum_{ij=1}^{m}\frac{\partial g_{ij}(\gamma _{x})}{%
\partial x^{k}}(\gamma _{x}^{\prime ,i}+\Omega ^{i}(\gamma _{x})(\gamma
_{x}^{\prime ,j}+\Omega ^{j}(\gamma
_{x})+\sum_{j=1}^{m}p_{j}\frac{\partial \Omega ^{j}(\gamma
_{x})}{\partial x^{k}}
\end{eqnarray*}
This implies:%
\begin{eqnarray*}
I_{1}(t)=\frac{1}{\sqrt{\varepsilon }}\int_{0}^{t}\left[ \sum_{k=1}^{m}%
\left( \frac{dp_{k}}{ds}-\sum_{j=1}^{m}p_{j}\frac{\partial \Omega
^{j}(\gamma _{x})}{\partial x^{k}}\right)
z_{1}^{k}ds+\sqrt{2}<\gamma _{x}^{\prime }+\Omega _{\varepsilon
}(\gamma _{x}),\sigma (\gamma _{x})dw>_{g(\gamma _{x})}\right]
\end{eqnarray*}

Using the stochastic equation for $z_{1}:$

\[
I_{1}(t)=\frac{1}{\sqrt{\varepsilon }}\int_{0}^{t}\left[ \sum_{k=1}^{m}%
\left( \frac{dp_{k}}{ds}-\sum_{j=1}^{m}p_{j}\frac{\partial \Omega
^{j}(\gamma _{x})}{\partial x^{k}}\right)
z_{1}^{k}ds+\sum_{j=1}^{m}p_{k}\left( dz_{1}^{k}+\sum_{j=1}^{m}\frac{%
\partial \Omega ^{k}(\gamma _{x})}{\partial x^{j}}z_{1}^{j}ds\right) \right]
\]

\begin{eqnarray*}
I_{1}(t)=\frac{1}{\sqrt{\varepsilon }}\int_{0}^{t}\left[ \sum_{k=1}^{m}\frac{%
dp_{k}}{ds}z_{1}^{k}ds+\sum_{j=1}^{m}p_{k}dz_{1}^{k}\right] =\left.
\sum_{j=1}^{m}p_{k}z_{1}^{k}\right\vert _{0}^{t}
\end{eqnarray*}
But, $z_{1}^{k}(0)=0,1\leq k\leq m$ and according to the boundary
conditions for the optimization problem $p_{k}(t)=0,1\leq k\leq m$,
thus
\begin{eqnarray*}
I_{1}(t)=0.
\end{eqnarray*}

\bigskip

\emph{Estimate of }$I_{2}(t).$ Proceeding as above:%

\begin{eqnarray*}
&& \frac{1}{\sqrt{\varepsilon }}\int_{0}^{t}\left[
\sum_{k=1}^{m}\frac{\partial
\Psi _{\mathcal{L}}(\gamma _{x})}{\partial x^{k}}z_{2}^{k}+\frac{1}{2}%
\sum_{ijk=1}^{m}\frac{\partial g_{ij}(\gamma _{x})}{\partial
x^{k}}(\gamma _{x}^{\prime ,i}+\Omega ^{i}(\gamma _{x})(\gamma
_{x}^{\prime ,j}+\Omega
^{j}(\gamma _{x})z_{2}^{k}\right] ds\\
&=&\frac{1}{\sqrt{\varepsilon }}%
\int_{0}^{t}\sum_{k=1}^{m}\left( \frac{dp_{k}}{ds}-\sum_{j=1}^{m}p_{j}\frac{%
\partial \Omega ^{j}(\gamma _{x})}{\partial x^{k}}\right) z_{2}^{k}dt
\end{eqnarray*}

and using the stochastic equation for $z_{2}:$%
\begin{eqnarray*}
& &\sqrt{2}\int_{0}^{t}\sum_{ijkl=1}^{m}g_{ij}(\gamma _{x})(\gamma
_{x}^{\prime
,i}+\Omega ^{i}(\gamma _{x}))\frac{\partial \sigma ^{jl}(\gamma _{x})}{%
\partial x^{k}}z_{1}^{k}dw_{l}=\\
&& \int_{0}^{t}\sum_{k=1}^{m}p_{k}\left(
dz_{2}^{k}+\left[ \sum_{j=1}^{m}\frac{\partial \Omega ^{k}}{\partial x^{j}}%
(\gamma _{x})z_{2}^{j}+\frac{1}{2}\sum_{j,l=1}^{m}\frac{\partial
^{2}\Omega
^{k}}{\partial x^{j}\partial x^{l}}(\gamma _{x})z_{1}^{j}z_{1}^{l}+\widehat{%
\Omega ^{k}}(\gamma _{x})\right] \right) ds
\end{eqnarray*}

Then inserting these expressions in $I_{2}(t):$
\begin{eqnarray*}
I_{2}(t) &=&\int_{0}^{t}\left(
\frac{1}{2}\sum_{kl=1}^{m}\frac{\partial
^{2}\Psi _{\mathcal{L}}(\gamma _{x})}{\partial x^{k}\partial x^{l}}%
z_{1}^{k}z_{1}^{l}+\frac{1}{4}\sum_{ijkl=1}^{m}\frac{\partial
^{2}g_{ij}(\gamma _{x})}{\partial x^{k}\partial xl}(\gamma
_{x}^{\prime ,i}+\Omega ^{i}(\gamma _{x})(\gamma _{x}^{\prime
,j}+\Omega ^{j}(\gamma
_{x})z_{1}^{k}z_{1}^{l}+\right.  \\
&&\left. \frac{1}{2}\sum_{jk,l=1}^{m}p_{k}\frac{\partial ^{2}\Omega ^{k}}{%
\partial x^{j}\partial x^{l}}(\gamma _{x})z_{1}^{j}z_{1}^{l}+\sum_{k=1}^{m}%
p_{k} \widehat{\Omega ^{k}}(\gamma _{x})\right) ds+\\
&&\sqrt{2}\int_{0}^{t}%
\sum_{ijkl=1}^{m}\frac{\partial g_{ij}(\gamma _{x})}{\partial
x^{k}}(\gamma _{x}^{\prime ,i}+\Omega ^{i}(\gamma _{x}))\sigma
^{jl}(\gamma _{x})z_{1}^{k}dw_{l}
\end{eqnarray*}

\bigskip

The second variation:%
\begin{eqnarray*}
&&\frac{1}{2}\mathcal{V}(\gamma_x )[z,z] \\
&=&\int_{0}^{t}\left( \frac{1}{2}%
\sum_{kl=1}^{m}\left[ \frac{\partial ^{2}\Psi _{\mathcal{L}}(\gamma _{x})}{%
\partial x^{k}\partial x^{l}}z+\frac{1}{2}\sum_{ij=1}^{m}\frac{\partial
^{2}g_{ij}(\gamma _{x})}{\partial x^{k}\partial x^{l}}(\gamma
_{x}^{\prime ,i}+\Omega ^{i}(\gamma _{x}))(\gamma _{x}^{\prime
,j}+\Omega ^{j}(\gamma _{x}))+\sum_{j=1}^{m}p_{j}\frac{\partial
^{2}\Omega ^{j}}{\partial
x^{k}\partial x^{l}}(\gamma _{x})\right] \right.  \\
&&z^{k}z^{l}+\left. \sum_{ij=1}^{m}g_{ij}(\gamma _{x})(z^{\prime ,i}+\sum_{l=1}^{m}%
\frac{\partial \Omega ^{i}}{\partial x^{l}}(\gamma
_{x})z^{l})(z^{\prime ,j}+\sum_{k=1}^{m}\frac{\partial \Omega
^{j}}{\partial x^{n}}(\gamma _{x})z^{k})\right)
ds\\
&&+\int_{0}^{t}\sum_{ijk=1}^{m}\frac{\partial g_{ij}(\gamma
_{x})}{\partial x^{k}}(\gamma _{x}^{\prime ,i}+\Omega
^{i}(\gamma _{x}))\left( dz^{j}+\sum_{l=1}^{m}\frac{\partial \Omega ^{j}}{%
\partial x^{l}}(\gamma _{x})z^{l}ds\right)
\end{eqnarray*}
Using the stochastic equation for $z_{1}$, we get
\begin{eqnarray*}
\sqrt{2}\int_{0}^{t}\sum_{ijkl=1}^{m}\frac{\partial g_{ij}(\gamma _{x})}{%
\partial x^{k}}(\gamma _{x}^{\prime ,i}+\Omega ^{i}(\gamma _{x}))\sigma
^{jl}(\gamma
_{x})z_{1}^{k}dw_{l}&=&\\
\int_{0}^{t}\sum_{ijk=1}^{m}\frac{\partial
g_{ij}(\gamma _{x})}{\partial x^{k}}(\gamma _{x}^{\prime ,i}+\Omega
^{i}(\gamma _{x}))\left( dz_{1}^{j}+\sum_{l=1}^{m}\frac{\partial \Omega ^{j}%
}{\partial x^{l}}(\gamma _{x})z_{1}^{l}ds\right)
\end{eqnarray*}
\begin{eqnarray*}
& &\frac{1}{2}\mathcal{V}(\gamma
)[z,z]=I_{2}(t;z)+\\
& &\int_{0}^{t}\sum_{ij=1}^{m}g_{ij}(\gamma _{x})(z^{\prime
,i}+\sum_{l=1}^{m}\frac{\partial \Omega ^{i}}{\partial x^{l}}(\gamma
_{x})z^{l})(z^{\prime ,j}+\sum_{k=1}^{m}\frac{\partial \Omega
^{j}}{\partial x^{n}}(\gamma
_{x})z^{k})ds-\int_{0}^{t}\sum_{k=1}^{m}p_{k}\widehat{\Omega
^{k}}(\gamma _{x})ds
\end{eqnarray*}
Using considerations of paragraph 7.8 of \cite{Azencott} p274-275,
we conclude that
\begin{eqnarray*}
Pr \{ -I_2(t) \geq r \} \leq e^{-cr},
\end{eqnarray*}
where $c>1$. This implies that there exists a $\beta>0$ such that
\begin{eqnarray*}
E_x \left(\exp-(1+\beta)I_2(t) \right)<+\infty.
\end{eqnarray*}

\bigskip

\emph{Estimate of }$I_{3}(t)$%

We have
\begin{eqnarray*}
I_{31}(t)=O(\sqrt{\varepsilon }\overline{R_{3,\varepsilon }}),
\end{eqnarray*}
\begin{eqnarray*}
I_{32}(t)=O(\sqrt{\varepsilon }\overline{R_{2,\varepsilon }}\overline{z_{1}}%
+(\sqrt{\varepsilon }\left( \overline{R_{2,\varepsilon }}\right)
^{2}),
\end{eqnarray*}
\begin{eqnarray*}
I_{33}(t)=O(\sqrt{\varepsilon }\left( \overline{R_{1,\varepsilon
}}\right) ^{3}),
\end{eqnarray*}
\begin{eqnarray*}
I_{34}(t)=\int_{0}^{t}O(\sqrt{\varepsilon }\left(
\overline{R_{1,\varepsilon }}\right) ^{2})dw,
\end{eqnarray*}
where if $X_{s},Y_{s}$, $s\in \lbrack 0,T]$ are two processes, we say that $%
Y_{s}=O(\overline{X})$ for $s\in \lbrack 0,t]$ if $\overline{Y}\leq K%
\overline{X},$ for some positive constant $K,$ and where $\overline{X}=%
\underset{s\in \lbrack 0,t]}{\sup }||X_{s}||$.  Using the same
consideration as in \cite{Azencott} p 270-271, paragraph 7.8 , we
conclude that there exists a function $\rho(\alpha)$, defined for
all positive $\alpha$ small enough, such that if the radius $\eta$
of the ball $E_3$ is smaller then $\rho(\alpha)$, then
\begin{eqnarray*}
E_x \left(e^{(1+\alpha)I_3(t)} \right)\leq C.
\end{eqnarray*}
We conclude that there exists a constant $C>0$ such that
\begin{eqnarray*}
E_x \{ \chi _{3}(X_{\varepsilon })
e^{-\frac{\int_0^t\Psi_L(X_\varepsilon)(s)ds}{\varepsilon}} \} \leq
C\exp\{-\frac{I_0(t)}{\varepsilon}\}.
\end{eqnarray*}

\end{document}